\documentclass[preprint,12pt]{elsarticle}
\usepackage{lineno,hyperref,graphicx,enumerate,color,multirow,multirow}
\modulolinenumbers[5]
\usepackage[margin= .5 in]{geometry}
\usepackage{amsmath,amsthm,amssymb,url}
\usepackage{amsfonts}
\usepackage{amssymb}
\usepackage{caption}
\usepackage{subfigure}
\usepackage[table]{xcolor}
\usepackage{subcaption}
\usepackage{algorithm}
\usepackage{algpseudocode}
\usepackage{float}
\usepackage{color}
\usepackage{tikz}
\usepackage{bm}
\usepackage{setspace}
\usepackage{diagbox}
\usepackage{mwe}    

\definecolor{cyan}{rgb}{0.0, 1.0, 1.0}
\definecolor{electricultramarine}{rgb}{0, 0.0, 1.0}
\definecolor{coralred}{rgb}{1.0, 0.25, 0.25}

\definecolor{darkgreen}{rgb}{0.0, 0.5, 0.0}

\newcommand{\darkgreenline}{\raisebox{2pt}{\tikz{\draw[-,darkgreen,solid,line width = 1.5pt](0,0) -- (10mm,0);}}}

\definecolor{electricgreen}{rgb}{0.0, 1.0, 0.0}

\newcommand{\coralredline}{\raisebox{2pt}{\tikz{\draw[-,coralred,solid,line width = 1.5pt](0,0) -- (5mm,0);}}}
\newcommand{\bluedashedline}{\raisebox{2pt}{\tikz{\draw[-,electricultramarine,dashed,line width = 1.5pt](0,0) -- (10mm,0);}}}

\definecolor{lightblue}{rgb}{0.68, 0.85, 0.9}

\definecolor{grayfill}{rgb}{0.7, 0.7, 0.7}
\newcommand{\grayfill}{\raisebox{2pt}{\tikz{\fill[grayfill] (0,0) rectangle (10mm,2mm);}}}

\definecolor{blueoutline}{rgb}{0.0, 0.0, 0.5}

\definecolor{translucentblue}{rgb}{0.0, 0.0, 1.0} 
\newcommand{\translucentbluefill}{\raisebox{2pt}{\tikz{\fill[translucentblue,opacity=0.2] (0,0) rectangle (10mm,2mm);}}}
\newcommand{\InputLayer}{\bm{\mathcal{ X}}}
\newcommand{\OutputLayer}{\bm{\mathcal{ Y}}}
\newcommand{\Rcal}{\mathbb{R}}
\newcommand{\SLIDEWindow}{{\mathrm{t}}_{\text{d}}}
\newcommand{\kSteps}{{\mathrm{k}}}
\newcommand{\iStep}{{\mathrm{i}}}
\newcommand{\ix}{\bm{x}}
\newcommand{\iy}{\bm{y}}
\newcommand{\nTrain}{{\mathrm{n}}_{\mathrm{train}}}
\newcommand{\NN}{{\bm{\mathcal{N}}}}
\newcommand{\NNPara}{{\bm{\Psi}}}
\newcommand{\Loss}{{\mathcal{L}}}
\newcommand{\nVal}{{\mathrm{n}}_{\mathrm{val}}}
\newcommand{\neVal}{{\mathrm{n}}_{\mathrm{eval}}}

\newcommand{\Mm}{\mathbf{M}}
\newcommand{\qv}{\mathbf{q}}
\newcommand{\Qv}{\mathbf{Q}}
\newcommand{\Gv}{\mathbf{G}}
\newcommand{\uv}{\mathbf{u}}
\newcommand{\lm}{\bm{\lambda}}
\newcommand{\ODEy}{\mathbf{y}}
\newcommand{\ODE}{\mathrm{ODE}_{1}}
\newcommand{\Angle}{\mathbf{\theta}}
\newcommand{\rv}{\mathbf{r}}
\newcommand{\cv}{\mathbf{c}}

\journal{}

\begin{document}

\begin{frontmatter}

\title{Real-Time Structural Deflection Estimation in Hydraulically Actuated Systems Using {3D} Flexible Multibody Simulation and DNNs}

\author[adres]{Qasim Khadim\corref{mycorrespondingauthor}}
\cortext[mycorrespondingauthor]{qasim.khadim@oulu.fi}
\author[adres1]{Peter Manzl}
\author[adres]{Emil Kurvinen}
\author[adres2]{Aki Mikkola}
\author[adres2]{Grzegorz Orzechowski}
\author[adres1]{Johannes Gerstmayr}

\address[adres]{University of Oulu, Finland}
\address[adres1]{University of Innsbruck, Austria}
\address[adres2]{Lappeenranta-Lahti University of Technology LUT, Finland}
\begin{abstract}
The precision, stability, and performance of lightweight high-strength steel structures in heavy machinery is affected by their highly nonlinear dynamics. This, in turn, makes control more difficult, simulation more computationally intensive, and achieving real-time autonomy, using standard approaches, impossible. Machine learning through data-driven, physics-informed and physics-inspired networks, however, promises more computationally efficient and accurate solutions to nonlinear dynamic problems. This study proposes a novel framework that has been developed to estimate real-time structural deflection in hydraulically actuated three-dimensional systems~\footnote{GitHub source, \url{https://github.com/qkhadim22/SLIDE-EstimatingStructuralDeflection.git}}. It is based on SLIDE, a machine-learning-based method to estimate dynamic responses of mechanical systems subjected to forced excitations.~Further, an algorithm is introduced for the data acquisition from a hydraulically actuated system using randomized initial configurations and hydraulic pressures.~The new framework was tested on a hydraulically actuated flexible boom with various sensor combinations and lifting various payloads. The neural network was successfully trained in less time using standard parameters from PyTorch, ADAM optimizer, the various sensor inputs, and minimal output data. The SLIDE-trained neural network accelerated deflection estimation solutions by a factor of $10^7$ in reference to flexible multibody simulation batches and provided reasonable accuracy. These results support the studies goal of providing robust, real-time solutions for control, robotic manipulators, structural health monitoring, and automation problems.
 
\end{abstract}
\end{frontmatter}
\section{Introduction}
\subsection{Research background: Structural deflection—Impacting automation in heavy machines}
Heavy machinery is important for primary production and transportation to enable a modern lifestyle \cite{bissadu2024society}. Most of these machines are human-operated and typically tailored for specific hydraulic actuation \cite{khadim2020targeting}. However, there is a growing trend toward increasing automation to enhance operational safety, productivity, and efficiency~\cite{billard2019trends, sanchez2020innovation, mann2023benign}. Increasing the level of automation, leading eventually to autonomous operation, requires accurate information on physical dynamics, particularly structural dynamics during operation~\cite{physicsbasedDT, haggerty2023control}. Flexible behavior, a result of material properties, geometry and other factors~\cite{wagg2010nonlinear}, can significantly affect the precision and accuracy needed to achieve autonomous operation~\cite{sarkhel2023robust, 10438059, cui2020trajectory}.

\subsection{Research motivation: Electrification and its impact on structural deflections in heavy machinery}
Additionally, emerging trends in electrification are driving engineers to innovate lighter structural designs to compensate for the weight of batteries and other heavy components~\cite{lajunen2018overview, czerwinski2021current}. This goal can be achieved using lightweight materials like modern metal alloys, \textit{e.g.}, ultra-high-strength steel or aluminum, composites, and sandwich structures~\cite{czerwinski2021current}. However, this shift towards lighter structures comes with new challenges. As the machinery becomes lighter, its structural components also become more flexible, which can affect stability and performance, especially under dynamic conditions~\cite{sarkhel2023robust, rigatos2018robotic,uyar2023implementation}.

\subsection{State of art methods and their limitations}
The increased flexibility makes it more difficult to maintain precise control over the machine, particularly when it experiences varying or sudden loads~\cite{ rigatos2018robotic,uyar2023implementation}. State-of-the-art methods for interpreting structural flexibility include direct measurement~\cite{avci2021review, malekloo2022machine}, dynamic models~\cite{sarkhel2023robust, dwivedy2006dynamic, lee2020critical}, and Machine Learning~(ML)~\cite{aldakheel2021feed, jensen2022online, rouvinen1997deflection}. 
The direct measurement methods—such as digital image correlation~\cite{pezeshki2023state}, fiber optic sensors~\cite{fernandez2021long}, wireless sensor networks~\cite{grundkotter2022precision}, ultrasonic nondestructive testing~\cite{kot2021recent}, laser scanning, and LIDAR~\cite{chen2023experimental, rasmussen2022non}—are commonly used in civil engineering applications. Real-time processing demands and implementation constraints are limitations of these methods.

In heavy machines conventional strain sensors can interpret structural deflections at specific locations on the structure. However, these sensors have several disadvantages, such as high cost, insufficient reliability, limited measuring points, and poor real-time capability due to challenging working conditions. To overcome these challenges, alternative approaches such as dynamic modeling and ML offer promising solutions for the accurate collection of deflection data.

\subsection{Research gaps: Conventional dynamic modeling and ML approaches in controlling flexibility}
In flexible systems, dynamic modeling involves links and joints or both~\cite{10438059}.~The Assumed Mode Method~(AMM)\cite{sun2018fuzzy}, Lumped Parameter Model~(LPM)\cite{sarkhel2023robust}, Transfer Matrix Method~(TMM)\cite{shi2023dynamics}, and Finite Elements Method (FEM)~\cite{uyar2023implementation} are commonly used to model flexible links. Broadly speaking, these methods demonstrate limitations in real-time control due to high computational cost, complex-geometry handling~\cite{lochan2017robust}, many Degrees Of Freedom (DOFs)~\cite{uyar2023implementation}, and improper boundary conditions and modes~\cite{10438059, dwivedy2006dynamic, meng2021motion}. A detailed description of flexible systems modeling methods and their pitfalls can be found in~\cite{10438059}. 

In flexible multibody systems, the Floating Frame of Reference Formulation (FFRF) is one of the most widely used methods in engineering applications involving large translations and rotations~\cite{shabana2020dynamics}. Across several FFRF versions, generalized Component Mode Synthesis (CMS) has been proposed as a suitable way to analyze 3D flexible systems~\cite{zwolfer2019co, zwolfer2020concise}. The Equations Of Motion~(EOMs) in CMS are further simplified using modal reduction techniques based on eigenmodes, which yields reasonable computational efficiency and accuracy~\cite{zwolfer2021nodal}. Nevertheless, the real-time processing of structural deflections in flexible multibody simulation remains challenging due to model complexity and the handling of hydraulics and electric components, contacts, and friction.    

ML offers computationally efficient and accurate solutions in diverse fields such as image processing~\cite{2012_krizhevsky_AlexnetPaper, 2016_He_deepResidualLearningForImageRecognition}, playing Atari games~\cite{2013_Mnih_PlayingAtari_DQN}, natural language processing~\cite{2017_Vaswani_attentionIsAllYouNeed}, fluid mechanics~\cite{2021_Cai_PINNsForFluidDynamics}, heat transfer~\cite{2021_Cai_PINNsHeatTransfer}, and the real-time estimation of rigid~\cite{hashemiMultibodyDynamicsControl2023,gerstmayrMultibodyModelsGenerated2024} and flexible multibody dynamics~\cite{goRapidlyTrainedDNN2024}. 
In ML, neural networks often act as black-box models that effectively map input-output relationships~\cite{1989_Hornik_FFNareUniversalFunApproximators}.~For flexible systems, neural networks have been used to estimate structural deflections~\cite{ jensen2022online, rouvinen1997deflection, omisore2021kinematics}, develop physics-inspired~\cite{lutter2023combining, khadim2024simulation} and physics-informed~\cite{2021_Cai_PINNsHeatTransfer,lutter2023combining} networks, and replace flexible multibody simulation~\cite{slimak2025machine,goRapidlyTrainedDNN2024, manzl2024slide}.

As universal approximators, neural networks often overlook the underlying system mechanics. The studies in~\cite{2021_Angeli_DeepLearningMinimalCoordinates, 2021_Choi_dataDrivenSimulation_DeepNerualNetworks, 2024_Slimak_overviewDesignConsiderationDataDrivenTimeStepping, 2021_Han_DNNFlexible} utilize neural networks within a time-stepping scheme, which involves estimating a state vector at $T^{th}$ step. The solution of the previous step is used autoregressively, which significantly affects training data size and computational training resources. To accelerate supervised learning, Principal Component Analysis~(PCA) was used to reduce the output vector and data size in flexible multibody systems~\cite{goRapidlyTrainedDNN2024}. However, in flexible machines, the acquisition of data presents several challenges for neural network applications~\cite{ jensen2022online, rouvinen1997deflection, omisore2021kinematics}. These include data collection, extensive preprocessing, identifying exact input-output, and data quality~\cite{2017_Vaswani_attentionIsAllYouNeed}.

\subsection{Research outline}
Physics-informed networks~\cite{roehrl2020modeling, lutter2023combining} combine physical laws with ML to affectively reduce computational training cost.~A physics-inspired network, the SLiding-window Initially-truncated Dynamic-response Estimator (SLIDE) method was introduced~\cite{manzl2024slide} to efficiently train the introduced networks on system dynamics using the SLIDE window size $\SLIDEWindow$. In a damped systems,~$\SLIDEWindow$ demonstrates the time taken by a system to reach the steady-state condition under forced excitation~\cite{manzl2024slide}.~The $\SLIDEWindow$ approximation is based on the complex eigenvalues of the system’s linearized EOMs~\cite{manzl2024slide}. It plays a key role in defining input-output relationships for neural networks.

Without building physics-informed networks, the SLIDE approach offers computational advantages in terms of data size and training time, and it enables real-time single-step and multi-steps estimations. However, this approach has not yet been studied with hydraulically actuated flexible systems including nonlinear friction forces for real-time structural deflection estimation applications. In particular, the calculation of $\SLIDEWindow$ might be challenging due to the highly nonlinear dynamics and discontinuities. This study proposes a statistics-based method to compute $\SLIDEWindow$ in hydraulically actuated flexible systems.

This study introduces a real-time structural deflection estimation framework based on a novel SLIDE approach for hydraulically actuated 3D flexible multibody systems. The framework learns the behavior of structural deflections at arbitrary location using inputs within $\SLIDEWindow$ that are based on its full-scale {3D} finite element model subjected to forced excitations. The proposed approach was tested on a hydraulically actuated flexible boom with various combinations of sensors and lifting various payloads. Simulation data for the flexible boom was obtained by combining the modally reduced CMS and lumped fluid theory in the Exudyn environment.~The trained neural network enabled the estimate of structural deflection with a mean-absolute-percentage error~(MAPE) less than 1.5\% compared to the reference solution.  The scientific contributions of this study are as follows.

\begin{itemize}
\item {A real-time structural deflection estimation framework for hydraulically actuated 3D flexible multibody systems has been developed that can be used for single-step and {multi-step} estimations in heavy machinery.} 

\item {A computationally efficient SLIDE data acquisition approach for hydraulically actuated flexible multibody systems and a method to compute $\SLIDEWindow$ without the EOMs have been introduced.} 

\item{The estimation accuracy and computational efficiency of proposed framework is verified with respect to the reference solutions for various payloads and sensor combinations.} 

\end{itemize}
The final trained neural network model enables the real-time estimate of structural deflections, which is particularly beneficial in applications where quick decision-making is critical, such as control, robotic manipulators, structural health monitoring, and automation. 

\section{Deep Neural Network (DNN) Model}
\label{DataSurrogate}
\begin{figure}[h]
    \centering
\includegraphics[width=0.5\textwidth]{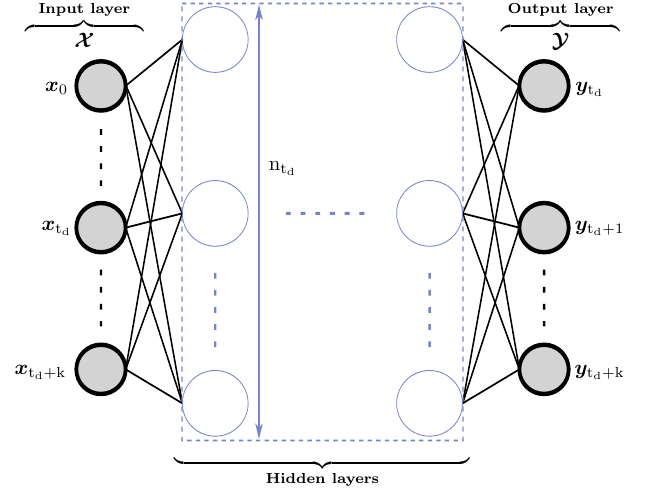}
        \label{SLIDE}
    \caption{\textbf{Proposed method}—A general framework of SLIDE-neural network model.}
    \label{FFN_Arch}
\end{figure}

The general framework of the SLIDE approach is illustrated in Fig.~(\ref{FFN_Arch}). The proposed neural network model learns the nature of $\OutputLayer$ through regression tasks using an input layer ${\InputLayer}$ using $\SLIDEWindow$.

The composition of  $\InputLayer$ and $\OutputLayer$ for a damped mechanical system can be described as
\vspace{-0.50cm}
\begin{equation}\label{InputLayer}
   \InputLayer = \begin{bmatrix}
        \ix_0 & \ix_1 & \hdots & \ix_{\SLIDEWindow}& \hdots &{\ix}_{\SLIDEWindow+\kSteps}
    \end{bmatrix}^{{\top}}, \quad~\text{and}
\end{equation}
\vspace{-1.0cm}
\begin{equation}\label{OutputLayer}
   { \OutputLayer = \begin{bmatrix}
        \iy_{\SLIDEWindow} &\iy_{\SLIDEWindow+1} & \hdots & \iy_{\SLIDEWindow+\kSteps}
    \end{bmatrix}^{\top}},
\end{equation}

where ${\ix}_\iStep$ and ${\iy}_\iStep$ are $i^{th}$ input and target vectors. In~\cite{manzl2024slide}, $\SLIDEWindow$ is approximated using the complex eigenvalues of the system’s linearized EOMs. A statistics-based method in Sec.~\ref{ComputeSlide} describes the computing of $\SLIDEWindow$ without the EOMs. 
The regression relationship in ${\InputLayer}$ and ${{\OutputLayer}}$ can be formulated as follows~\cite{bergstra2012random}. 
\vspace{-0.50cm}
\begin{equation} \label{DNN_Reconstruction_Model}
{\widehat{\OutputLayer}}_{0:\nTrain} = \NN ({\InputLayer}_{0:\nTrain}; \NNPara),
\end{equation}
where ${\widehat{\OutputLayer}}_{0:\nTrain}$ is the estimated output layer, $\NN$ is the neural network model, $\NNPara$ is the vector of trainable parameters, and $\nTrain$ is the number of training samples. In supervised learning, the objective of the neural network model is to minimize the loss function ${\Loss}$ and keep it below a threshold value ${\Loss}_{\text{min}}$ using the optimal training parameters. It can be described as
\vspace{-0.50cm}
\begin{equation} 
\label{MinimizeLoss}
\text{argmin} \, \, \, \, \Loss   \quad \text{subject to} \quad \Loss \leq \Loss_{\text{min}} 
\end{equation}
The loss function is computed between the estimated output layer ${\widehat{\OutputLayer}}$ and the ground truth output layer ${{\OutputLayer}}$ using the Mean Square Error (MSE) as follows.
\vspace{-0.50cm}
\begin{equation} 
\label{LossDefinition}
\Loss  = \frac{1}{{\nTrain}} \sum_{i=1}^{\nTrain} (\widehat{\OutputLayer}_i - \OutputLayer_i)^2
\end{equation}
Adam optimizer~\cite{kingma2014adam} minimizes the loss function in $\NN$ training.~{The hidden layers size is according to the number of steps $\SLIDEWindow$, highlighted in Fig.~(\ref{FFN_Arch}) as $\mathrm{n}_{\SLIDEWindow}$. The standard neural network parameters ${\NNPara}^{*}$ are selected to achieve optimum training performance.}
In the evaluation phase, the output layer is estimated in a continuous-time frame with the trained neural network as \vspace{-0.50cm}
{
\begin{equation} \label{DNN_Prediction_Model}{\widehat{\OutputLayer}}^{\neVal}_{\SLIDEWindow:\mathrm{t}_\mathrm{f}} = \NN ({\InputLayer}^{\neVal}_{i:{\SLIDEWindow}+i}),
\end{equation}}

where $i = 0,1,\dots {\mathrm{t}_\mathrm{f}}-\SLIDEWindow$, and ${\mathrm{t}_\mathrm{f}}$ is the final time in the evaluation data $\neVal$. 
\section{General Purpose Flexible Multibody Dynamics}
\label{DynamicModel}
The equations of motion for a constrained mechanical system can be described as 
\vspace{-0.50cm}
\begin{equation} \label{Eq01: GeneralEOM}
    {\Mm ({{\qv}})}{\ddot{\qv}} + { ({\Gv}_{{\qv}})^{\top} \lm}  = {{\Qv ({{\qv}}, {\dot{\qv}}, {{\uv}}, t)  }},
\end{equation}
\begin{equation} \label{Eq01: GeneralEOM1}
    {{\Gv} ({{\qv}},t)} = {\mathbf{0}}, \quad~\text{and}
\end{equation}
\begin{equation} \label{Eq03: GeneralEOM1}
 \dot{\ODEy} + \frac{\partial \Gv}{{\partial {\ODEy} }^{\top}} \lm = {{\mathbf{f}_{\ODE}} ({{\qv}}, {\dot{\qv}}, {{\uv}}, t)}, 
\end{equation}
where ${{\qv}} \in \Rcal^{n}$, ${\dot{\qv}} \in \Rcal^{n}$ , and ${\ddot{\qv}} \in \Rcal^{n}$ are the position, velocity, and acceleration vectors of the generalized coordinates, respectively. ${\Mm} \in \Rcal^{n \times n }$ is the system mass matrix, ${\Qv \in \Rcal^{n}}$ is generalized force vector, ${{\Gv}}$ represents the {holonomic} constraint equations, and ${{\mathbf{f}_{\ODE}}}$ is the vector of first-order differential equations. Here, $n$ is the number of coordinates in the system. 
\begin{figure}[h]
\centering
\includegraphics[width=0.58\textwidth]{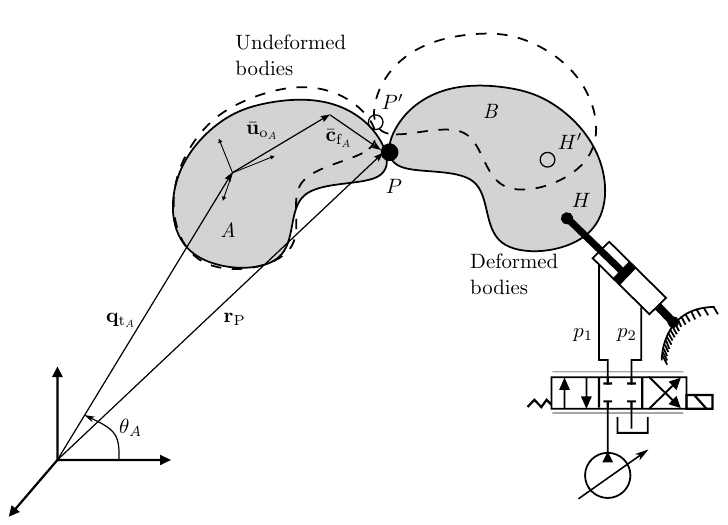}
\caption{\textbf{Flexible system} -- A system consisting of two flexible bodies $\text{A}$ and $\text{B}$ connected at the joint $P$, controlled by hydraulic actuation -- The position of the joint is ${\rv}_{\mathrm{P}} = \qv_{{\mathrm{t}}_{\text{A}}}+{{\mathbf{A}}_{\mathrm{A}}}({\bar{\uv}}_{{\mathrm{o}}_{\text{A}}}+ {\bar{\cv}}_{{\mathrm{f}}_\text{A}})$, where $\qv_{{\mathrm{t}}_{\text{A}}} \in \Rcal^{3 \times 1}$ is the translational coordinates vector, ${\mathbf{A}}_\text{A} = {\mathbf{A}} (\Angle_{\text{A}}) $ is the rotation matrix, $\Angle_{\text{A}} \in \Rcal^{n_{\mathrm{r}} \times 1}$ is the rotational parametrization vector, ${\bar{\uv}}_{{\mathrm{o}}_{\text{A}}}$ is the local coordinate vector, and    ${\bar{\cv}}_{{\mathrm{f}}_\text{A}} \in \Rcal^{3 \times 1}$ is the local flexible coordinates vector in body $\text{A}$. Additionally, ${n_{\mathrm{r}}}$ is the rotational DOFs.  }
\label{fig:FlexibleSystem}
\end{figure} 
To explain deformable bodies, Fig.~(\ref{fig:FlexibleSystem}) describes a hydraulically actuated flexible multibody system, where a flexible body $\text{A}$ is defined with the generalized coordinate vector $\qv_{{\text{A}}} = {\begin{bmatrix}
    \qv_{{\mathrm{t}}_{\text{A}}}^{\top} &
    {\Angle_{\text{A}}}^{\top} &
    {\bar{\cv}}_{{\mathrm{f}}_\text{A}}^{\top} 
\end{bmatrix}}^{\top}$~\cite{shabana2020dynamics}. The nodal-based FFRF can FE-discretize a {3D} $\text{A}$ body into $n_{\text{n}}$ nodes, with the resulting local flexible coordinates vector described as ${\bar{\cv}}_{{\mathrm{f}}_\text{A}} \in \Rcal^{3n_{\text{n}} \times 1}$~\cite{zwolfer2020concise}.~The vector of modal coordinates ${{\mathbf{\zeta}}_\text{A}} \in \Rcal^{3n_{\text{m}} \times 1}$ further reduces the local flexible coordinates vector as ${\bar{\cv}}_{{\mathrm{f}}_\text{A}} \approx {\overline{\NNPara}_\text{A}} {{\mathbf{\zeta}}_\text{A}},~\text{with}~n_{\text{m}}=\dim({{\mathbf{\zeta}}_\text{A}}) \ll \dim({\bar{\cv}}_{{\mathrm{f}}_\text{A}}) = 3n_{\text{n}}$ in CMS~\cite{zwolfer2021nodal}. The variable~$n_{\text{m}}$ represents the modal coordinates, and ${\overline{\NNPara}}_\text{A}  \in \Rcal^{3n_{\text{n}} \times n_{\text{m}}}$ is the column-wise modes reduction-basis. The definition of joint and constraint equations for the multiple interconnected flexible bodies is further detailed in~\cite{Johannes2024} using the modally reduced CMS. The components of Eq.~\eqref{Eq01: GeneralEOM} for a flexible multibody system are as follows~\cite{zwolfer2021nodal}.
\vspace{-0.2cm}
\begin{equation} \label{Eq03_04: ConstraintsAndForces1}
{\Mm} = {\begin{bmatrix}
    {{\mathbf{\widehat{M}_{\text{tt}}}}}  & {{\mathbf{\widehat{M}_{\text{tr}}}}} & {{\mathbf{\widehat{M}_{\text{tf}}}}}\\
    & {{\mathbf{\widehat{M}_{\text{rr}}}}} & {{\mathbf{\widehat{M}_{\text{rf}}}}} \\
    \text{sym.} &  & {{\mathbf{\widehat{M}_{\text{ff}}}}}
\end{bmatrix}},  
\end{equation}
\begin{equation} \label{Eq03_04: ConstraintsAndForces2}
{ ({\Gv}_{{\qv}})^{\top} \lm} = \underbrace{\begin{bmatrix}
     \mathbf{\widehat{Q}_{\text{c}_{\text{t}}}} \\
     \mathbf{\widehat{Q}_{\text{c}_{\text{r}}}} \\
     \mathbf{\widehat{Q}_{\text{c}_{\text{f}}}} 
\end{bmatrix}}_{\substack{\text{constraint force} \\ \text{vector}}}, \quad~\text{and}
\end{equation}
\begin{equation} \label{Eq03_04: ConstraintsAndForces}
\Qv = \underbrace{\begin{bmatrix}
     \mathbf{\widehat{Q}_{\text{e}_{\text{t}}}} \\
     \mathbf{\widehat{Q}_{\text{e}_{\text{r}}}} \\
     \mathbf{\widehat{Q}_{\text{e}_{\text{f}}}} 
\end{bmatrix}}_{\substack{\text{elastic force} \\ \text{vector}}} + \underbrace{\begin{bmatrix}
     \mathbf{\widehat{Q}_{\text{v}_{\text{t}}}} \\
     \mathbf{\widehat{Q}_{\text{v}_{\text{r}}}} \\
     \mathbf{\widehat{Q}_{\text{v}_{\text{f}}}} 
\end{bmatrix}}_{\substack{\text{quadratic velocity} \\ \text{vector}}}+ \underbrace{\begin{bmatrix}
     \mathbf{\widehat{Q}_{\text{a}_{\text{t}}}} \\
     \mathbf{\widehat{Q}_{\text{a}_{\text{r}}}} \\
     \mathbf{\widehat{Q}_{\text{a}_{\text{f}}}} 
\end{bmatrix}}_{\substack{\text{applied force} \\ \text{vector}}}.
\end{equation}

Further details of terms in ${\Mm}$ and ${\Qv}$ can be found in~\cite{zwolfer2021nodal}. 
The components of applied force vector are 
\vspace{-0.50cm}
\begin{equation}\label{Eq03: AppliedForce1}
 {{\mathbf{\widehat{Q}_{\text{a}_{\text{t}}}}}} = \displaystyle\sum_{i=1}^{{n_{\text{n}}}} {{\mathbf{f}}_{\mathrm{a}}}^{(i)}, \quad {{\mathbf{f}}_{\mathrm{a}}}^{(i)} \neq 0, 
\end{equation}
\begin{equation}\label{Eq03: AppliedForce2}
{{\mathbf{\widehat{Q}_{\text{a}_{\text{r}}}}}} = {\overline {\Gv}}^{\top} \displaystyle\sum_{i=1}^{{n_{\text{n}}}}
\begin{bmatrix}
{\widetilde{\overline{\ix}}}^{(i)} + \displaystyle\sum_{\text{m}=1}^{{n_{\text{m}}}}{\widetilde{\overline{\NNPara}}_{\text{m}}}^{(i)} \mathbf{\zeta}_\text{m} \end{bmatrix}
\mathbf{A}^{\top} {{\mathbf{f}}_{\mathrm{a}}}^{(i)}, \quad {{\mathbf{f}}_{\mathrm{a}}}^{(i)} \neq 0, \quad~\text{and}
\end{equation}
\begin{equation}\label{Eq03: AppliedForce3}
{{\mathbf{\widehat{Q}_{\text{a}_{\text{f}}}}}} = \displaystyle\sum_{i=1}^{{n_{\text{n}}}} 
\left[ 
\begin{array}{c}
\overline{\NNPara}_1^{(i)^{\top}} \\
\vdots \\
\overline{\NNPara}_{{n_{\text{m}}}}^{(i)^{\top}}
\end{array}
\right] 
\mathbf{A}^{\top} {{\mathbf{f}}_{\mathrm{a}}}^{(i)}, \quad {{\mathbf{f}}_{\mathrm{a}}}^{(i)} \neq 0, 
\end{equation}
where ${{\mathbf{f}}_{\mathrm{a}}}^{(i)}$ is the applied force vector at an arbitrary node $i$, and {${\widetilde{\overline{\ix}}}^{(i)}$ is the undeformed (reference) nodal
coordinates. In hydraulic actuation, ${{\mathbf{f}}_{\mathrm{a}}}^{(i)}$ can be computed from the hydraulic force vector ${{\mathbf{F}}_{\mathrm{h}}}$ as follows.
\vspace{-0.50cm}
\begin{equation}\label{Eq04: NodalHydraulicForce}
    {{\mathbf{f}}_{\mathrm{a}}}^{(i)} = {w^{(i)}} {{\mathbf{F}}_{\mathrm{h}}} ,
\end{equation}
where ${w^{(i)}} = \frac{1}{3A_B} \sum_{\text{j}} A_{\text{j}}, \quad \text{and} \quad \sum_i {w^{(i)}} = 1$ is the nodal weighting. The variable $A_{\text{j}}$ is the area of each triangle ${\text{j}}$, and $A_B$ is the total boundary area. This weighting ensures nearly constant strain distribution with equally distributed axial forces. The hydraulic force produced by the cylinder is calculated as $F_h = p_1A_1-p_2A_2-F_{\mu}$, where $p_1$ and $p_2$ correspond to the cylinder pressures on cylinder areas $A_1$ and $A_2$, respectively. Friction force in the cylinder, $F_{\mu}$, is calculated according to~\cite{brown2016continuous}. The hydraulic pressure, ${p}_{\mathrm{h}}$, in a volume ${V}_{\mathrm{h}}$, can be modeled using lumped fluid theory~\cite{watton2009fundamentals} in differential form as follows.
\vspace{-0.50cm}
\begin{equation} \label{DiffPressure}
\dot{p}_{\mathrm{h}}  = \frac{B_{{\mathrm{e}}_{\mathrm{h}}}}{V_{\mathrm{h}}}  \left(- \frac{d {V_{\mathrm{h}}}}{dt} + Q_{{\mathrm{s}}}  \right) \quad~\text{and}
\end{equation}
\begin{equation} \label{EffBulk}
\begin{aligned}
B_{{\mathrm{e}}_{\mathrm{h}}}  = \left(\frac{1}{B_{\mathrm{o}}} + \sum_{c=1}^{n_{\mathrm{h}}} \frac{V_{\mathrm{c}}}{{V_{\mathrm{h}}}{B_{\mathrm{c}}}} \right)^{-1}, \\[6pt]
\end{aligned}
\end{equation}
where $B_{{\mathrm{e}}_{\mathrm{h}}}$ is the effective bulk modulus, $Q_{\mathrm{s}}$ is the sum of the incoming and outgoing flows, $B_{\mathrm{o}}$ is the oil bulk modulus, $n_{\mathrm{h}}$ is the total number of hydraulic volumes, $V_{\mathrm{c}}$ denotes the sub-volume, and $B_{\mathrm{c}}$ is the bulk modulus of the sub-volume. A semi-empirical method can be used to compute the flow rate using the relative spool position $U$~\cite{watton2009fundamentals}.
\begin{figure}[h]
\centering
\includegraphics[width=0.8\textwidth]{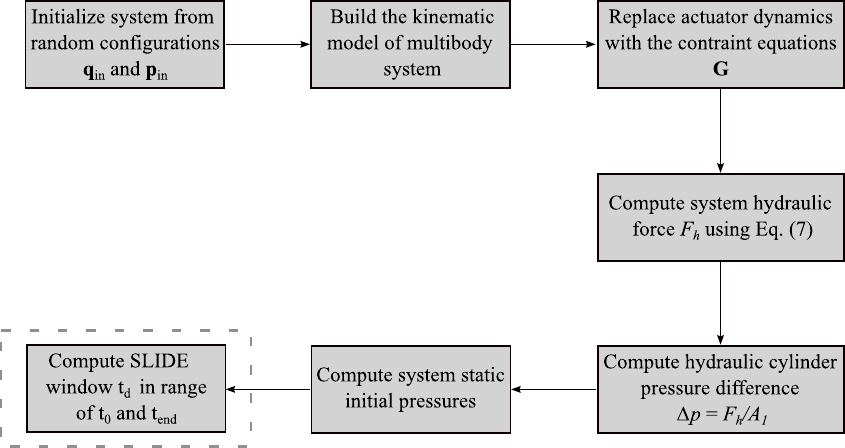}
\caption{{\textbf{SLIDE data acquisition} -- A method for generating data from hydraulically actuated mechanical systems at random initial configurations -- Initial pressures are calculated based on the static equilibrium force for each system configuration. The SLIDE window $\SLIDEWindow$ is also determined during the data acquisition, and the data is arranged accordingly.}}
\label{fig:Eq034}
\end{figure}
The DNN model needs training data from various initial configurations of the system.  Specifically for hydraulic systems, this results in computational challenges because each system's initial configuration must reach static equilibrium. To this end, the algorithm in Fig.~\ref{fig:Eq034} is introduced to achieve static equilibrium at random system configuration ${\qv}_\mathrm{in}$ and initial pressure ${\mathbf{p}}_\mathrm{in}$. 

For a random system configuration ${\qv}_\mathrm{in}$, the actuator dynamics in the system are replaced with the equivalent constraint equations $\Gv$ in a closed kinematic loop. At static equilibrium, the system's EOMs described in Eq.~\eqref{Eq01: GeneralEOM} are reduced into $ { ({\Gv}_{{\qv}})^{\top} \lm}  = {{\Qv ({{\qv}}, {\dot{\qv}}, {{\uv}}, t_0)  }}$ and ${{\Gv} ({{\qv}},t_0)} = {\mathbf{0}}$, which results in the computation of external hydraulic force~$F_h$. The pressure difference~$\Delta p$, computed from $F_h$, enables the computation of initial pressures in the hydraulic cylinder. Further, the proposed algorithm facilitates computing $\SLIDEWindow$. 

\section{Flexible Simulation Setup}
\label{FFNN_application}
Application of the proposed method is demonstrated by estimating the tip deflections in a hydraulically actuated flexible boom. The simulation setup of the flexible boom is described in Fig.~(\ref{CaseExample}a). 
\begin{figure}[h] 
\centerline{\includegraphics[width=\linewidth]{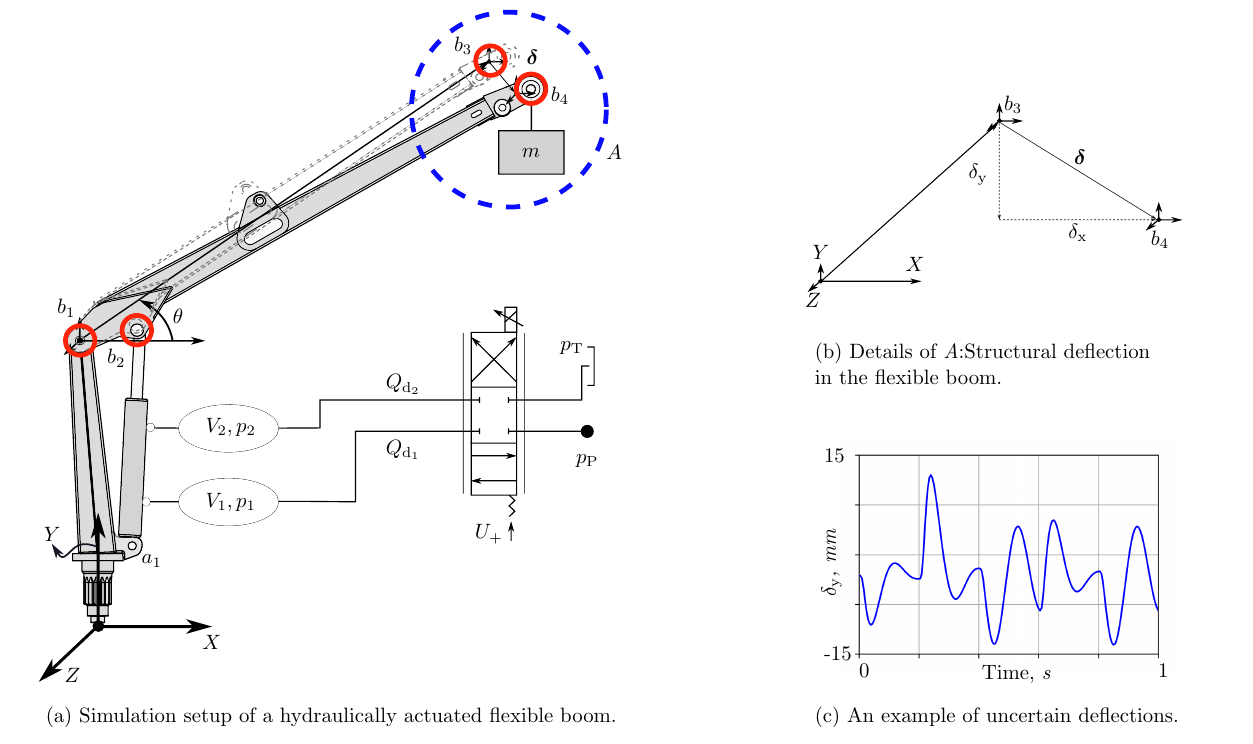}}
\caption{\textbf{Simulation setup} -- A hydraulically actuated flexible boom lifting a payload $m$ -- The flexible boom is attached to rigid pillar at node $b_1$. It is actuated by a hydraulic cylinder, which has been added between the nodes $a_1$ and $b_2$. The known payload $m$ is connected to the flexible boom at node $b_4$, and tip deflection is measured at node $b_3$. }
\label{CaseExample}
\end{figure}
It comprises a rigid pillar, flexible boom, known payload ~$m$, and the hydraulics. Tip structural deformation $\bm{\delta}$ of the deformed boom is defined in Fig.~(\ref{CaseExample}b). It is measured relative to the global coordinate system~${XYZ}$. The components of structural deflection are represented by ${\delta}_{\mathrm{x}}$, ${\delta}_{\mathrm{y}}$ and ${\delta}_{\mathrm{z}}$. In this study, only~$\delta_{\mathrm{y}}$ is estimated due to the payload forces acting along $Y$-axis, nonlinear friction force $F_\mu$  in the hydraulic cylinder, and the highly nonlinear dynamics. Fig.~(\ref{CaseExample}c) further describes the uncertain and non-linear nature of tip deflection $\delta_{\mathrm{y}}$ during a working cycle. 

\subsection{Flexible multibody system}
The simulation setup was modeled in the open-source software Exudyn~\cite{gerstmayr2023exudyn}. Exudyn\footnote{Version 1.8, \url{https://github.com/jgerstmayr/EXUDYN}} is a general-purpose flexible multibody dynamics systems modelling software in the Python environment. It includes various versions of the FFRF, including the modally reduced CMS and lumped fluid theory. 
\begin{figure}[h]
\centering
\includegraphics[width=0.8\textwidth]{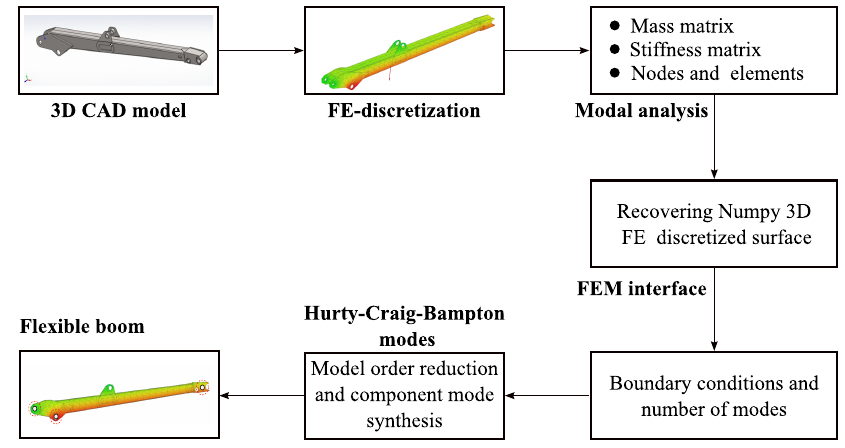}
\caption{Workflow for modeling flexible boom in the Exudyn software~\cite{gerstmayr2023exudyn} -- The  commercial Abaqus\textsuperscript{\textregistered} environment is used to perform FE discretization of the three-dimensional lift boom model. The nodes $b_1, b_2$ and $b_3$ on the flexible boom describe the boundary conditions in the flexible multibody system. }
\label{fig:FlexibleBoom}
\end{figure} 
The {3D} boom geometry is discretized into 11495~{first-order tetrahedral elements ({C3D10})}~in the commercial FE software Abaqus\textsuperscript{\textregistered}. Fig.~(\ref{fig:FlexibleBoom}) shows the meshed flexible boom in Abaqus\textsuperscript{\textregistered}. The material properties set in the Abaqus\textsuperscript{\textregistered} software include Young’s modulus $E= 2.1 \times 10^{11}~\text{Pa}$, Poisson’s ratio $\nu = 0.3$, and density $\rho = 7850~\text{kg}/\text{m}^3$. The mass and stiffness matrices, required in Eq.~\eqref{Eq03_04: ConstraintsAndForces} for Exudyn, were exported from Abaqus\textsuperscript{\textregistered}. The nodal points $b_1$, $b_2$ and $b_3$ describe the location of joints and forces on the flexible boom. Eight Hurty Craig-Bampton modes are computed in software Exudyn using the specified boundary conditions. The eigen-frequencies of first bending and normal modes are 29 Hz and 31 Hz, respectively. 

The accuracy of $\delta_{\mathrm{y}}$ in Exudyn was confirmed with Abaqus\textsuperscript{\textregistered} during a static analysis with an approximated error of 0.5\%. The flexible boom connects to the pillar at node $b_1$ via a revolute joint through an RBE2 standard interface. The nodal points $b_2$
and $b_4$ on the flexible boom defines the locations of hydraulics and payload. The flexible boom was hydraulically actuated by a 4/3 directional control valve through hydraulic volumes $V_1$ and $V_2$, constant pressure sources from pump, and tank $p_\mathrm{P}$ and $p_\mathrm{T}$. The details of the hydraulic parameters used in this example can be found in Table~(\ref{Tab.patuHydraulic}) and~\cite{KHADIM2023105405}. 
\begin{table}[!htb]
\centering
\caption{Hydraulic parameters of the lift boom}
\begin{tabular}{|c|c|c|c|}
\hline
\rowcolor{gray!30} \textbf{Parameter} & \textbf{Value} & \textbf{Parameter} & \textbf{Value} \\
\hline
Pump pressure & 140~$ \text {bar}$ & Tank pressure & 1~$ \text {bar}$ \\
\rowcolor{gray!10}
Cylinder diameter & $100~\text{mm}$ & Piston diameter & $56~\text{mm}$ \\
Cylinder length~$(l_{\mathrm{cyl}})$ & $535~\text{mm}$ & Cylinder stroke~$(l_{\mathrm{pist}})$ & 820~$\text{mm}$ \\
\hline
\end{tabular}
\label{Tab.patuHydraulic}
\end{table}
\subsection{SLIDE data acquisition}
In Exudyn, the coupled differential equations of the hydraulically actuated flexible boom are solved using the 
generalized-alpha integration scheme.~The coupled system is solved for 1~\text{s} simulation time with a fixed time step of 5 \text{ms}, comprising each simulation of 200 steps. To generate varied training data, the flexible  boom is actuated between $\theta_{\text{min}} = -10^{\circ}$ and $\theta_{\text{max}}= +50^{\circ}$, which are determined according to the minimum and maximum actuator lengths. The randomized initial angle ${{\theta}_\mathrm{in}}$ for each simulation is computed as follows.
\vspace{-0.50cm}
\begin{equation}
    {{\theta}_\mathrm{in}} = \text{rand}^{*}(\theta_{\min}, \theta_{\max})+\theta_{min},
\end{equation}

where $\text{rand}^{*}$ is the NumPy function $\texttt{numpy.random.rand()}$~\cite{harris2020array}, which populates random samples using a uniform distribution. For each ${{\theta}_\mathrm{in}}$, the hydraulic pressures~$p_1$ and $p_2$  are computed  through the algorithm introduced in Fig.~(\ref{fig:Eq034}) during a static analysis.~The relative tolerances of $1~\times~10^{-7}$~\text{m} at the position level were used in the Newton-Raphson solver for the static analysis. The simulations were run in batches of 80, 160, 320, 640, 1280 and 2560, with the resulting data saved in the standard Numpy format. The simulation batches include 80\% training data and 20\% validation data.

\subsection{Designing the control signal}
In each simulation, the hydraulic valve signal was actuated using a randomly generated control signal. The random control signal varies between the minimum spool position $-1$, the neutral spool position $0$, and maximum spool position~$+1$.
\begin{figure}[h]
    \centering
    \subfigure{
        \includegraphics[width=0.23\textwidth]{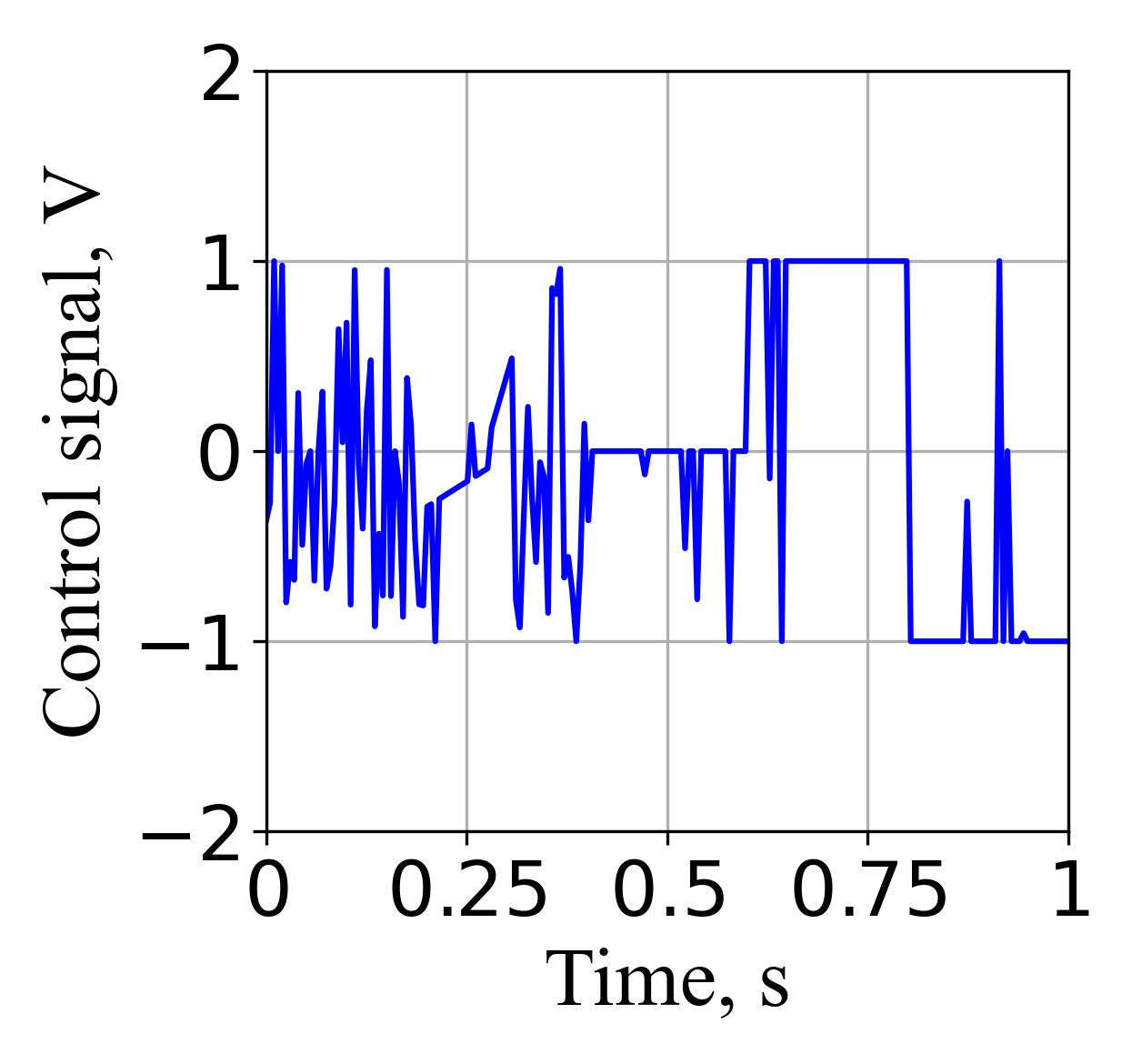}
        \label{signal1}}
        \subfigure{
        \includegraphics[width=0.23\textwidth]{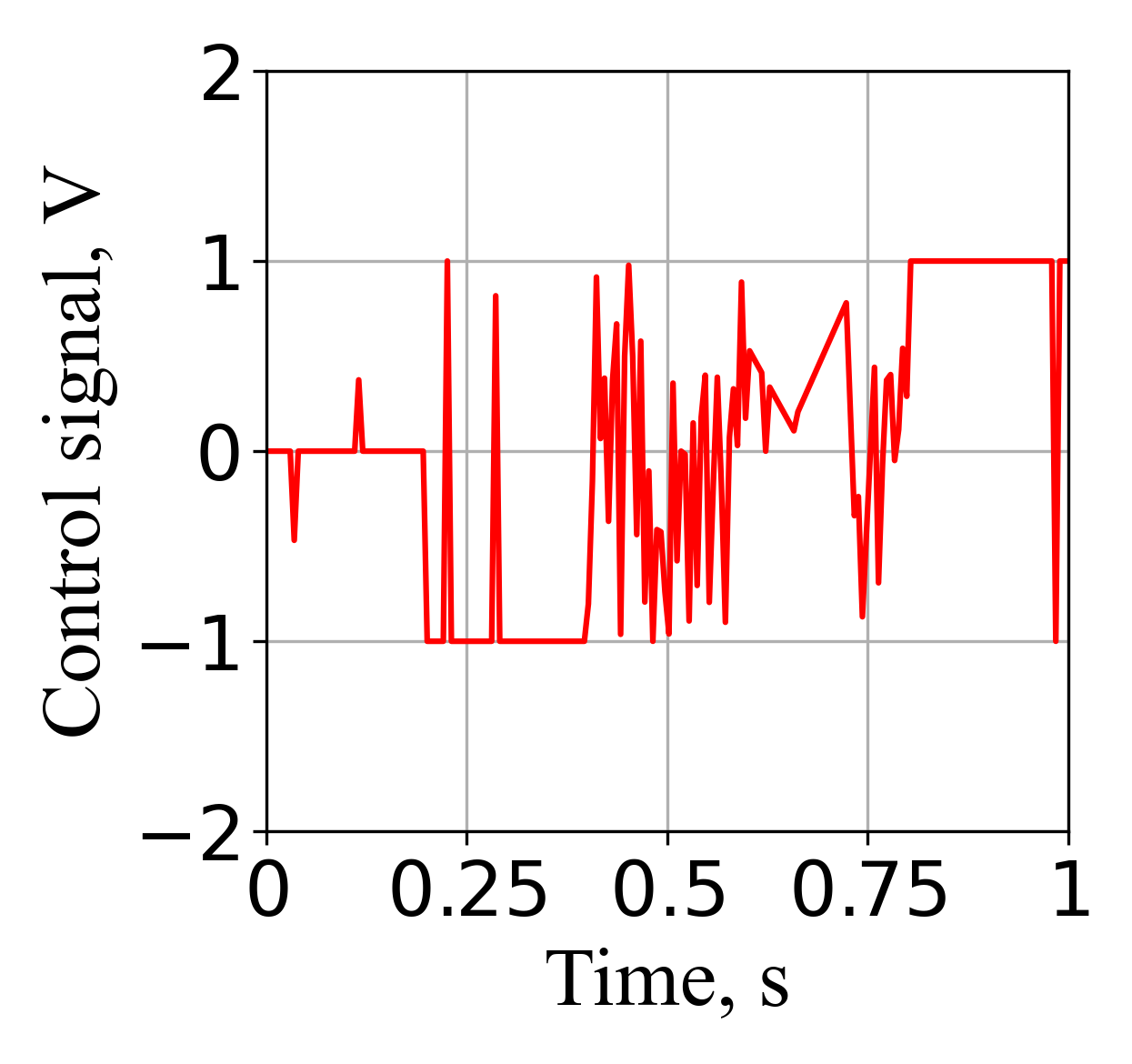}
        \label{signal2}}
        \subfigure{
        \includegraphics[width=0.23\textwidth]{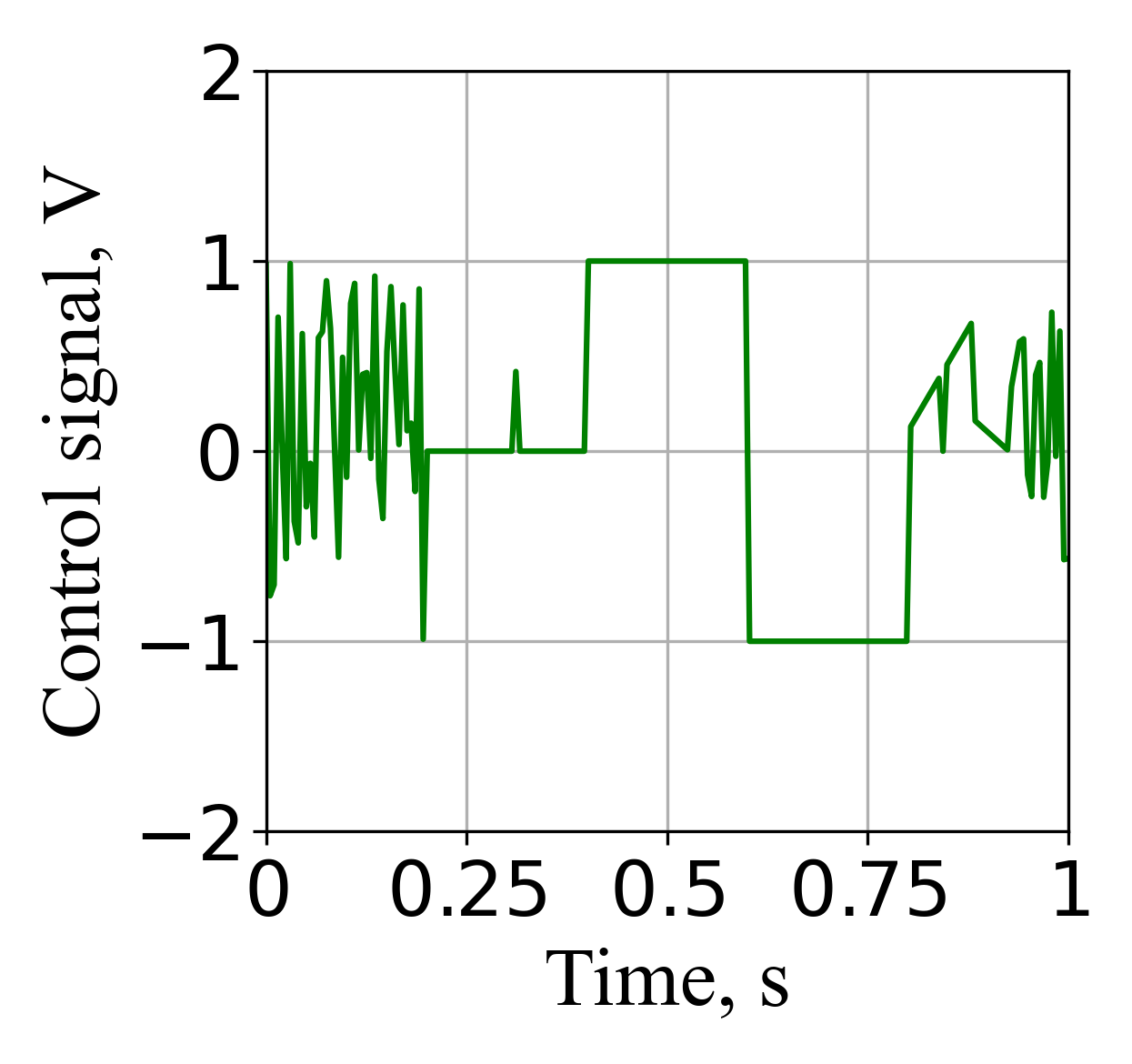}
        \label{signal3}}
        \subfigure{
        \includegraphics[width=0.23\textwidth]{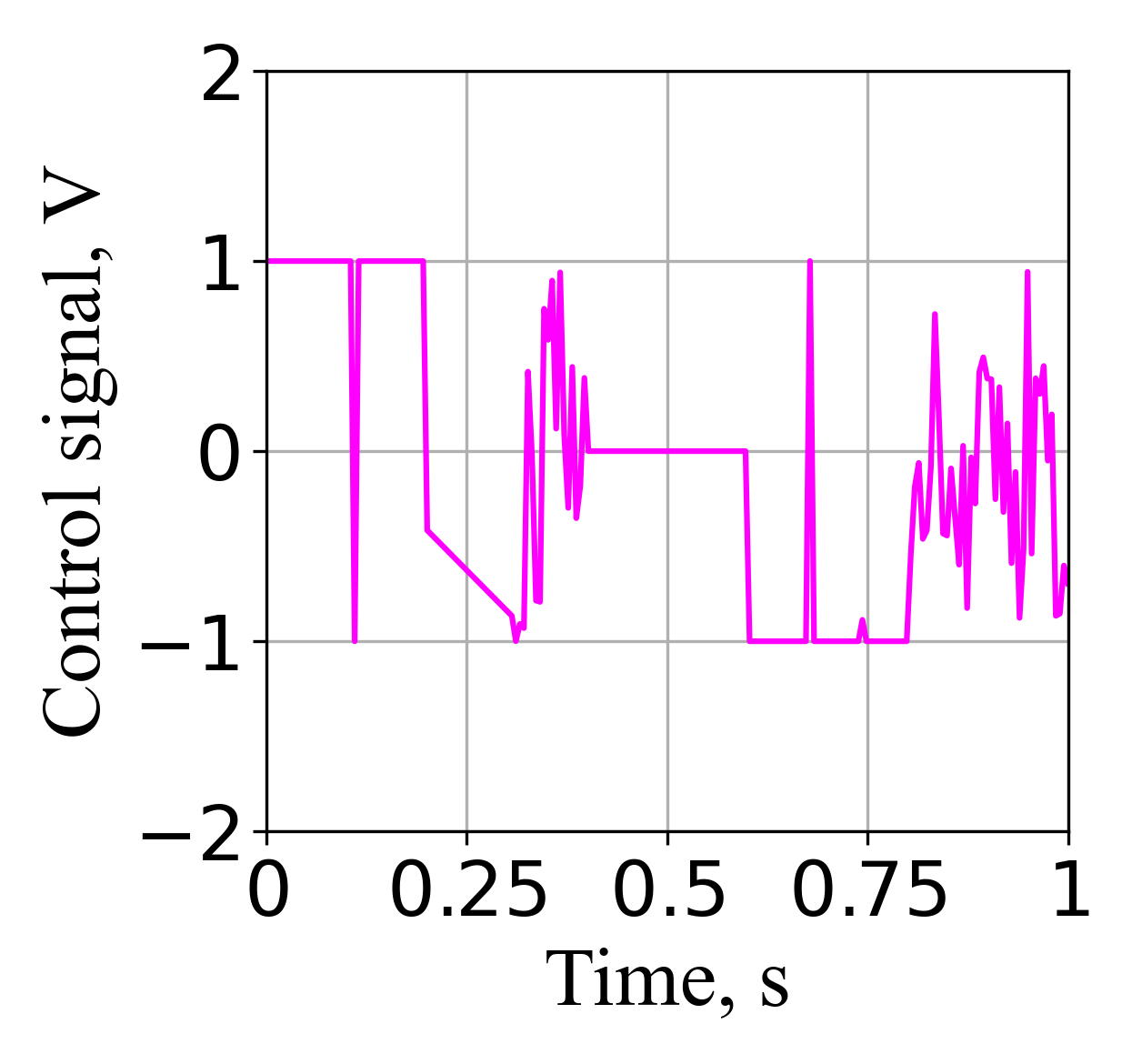}
        \label{signal4}}
        \vfill
            \subfigure{
        \includegraphics[width=0.23\textwidth]{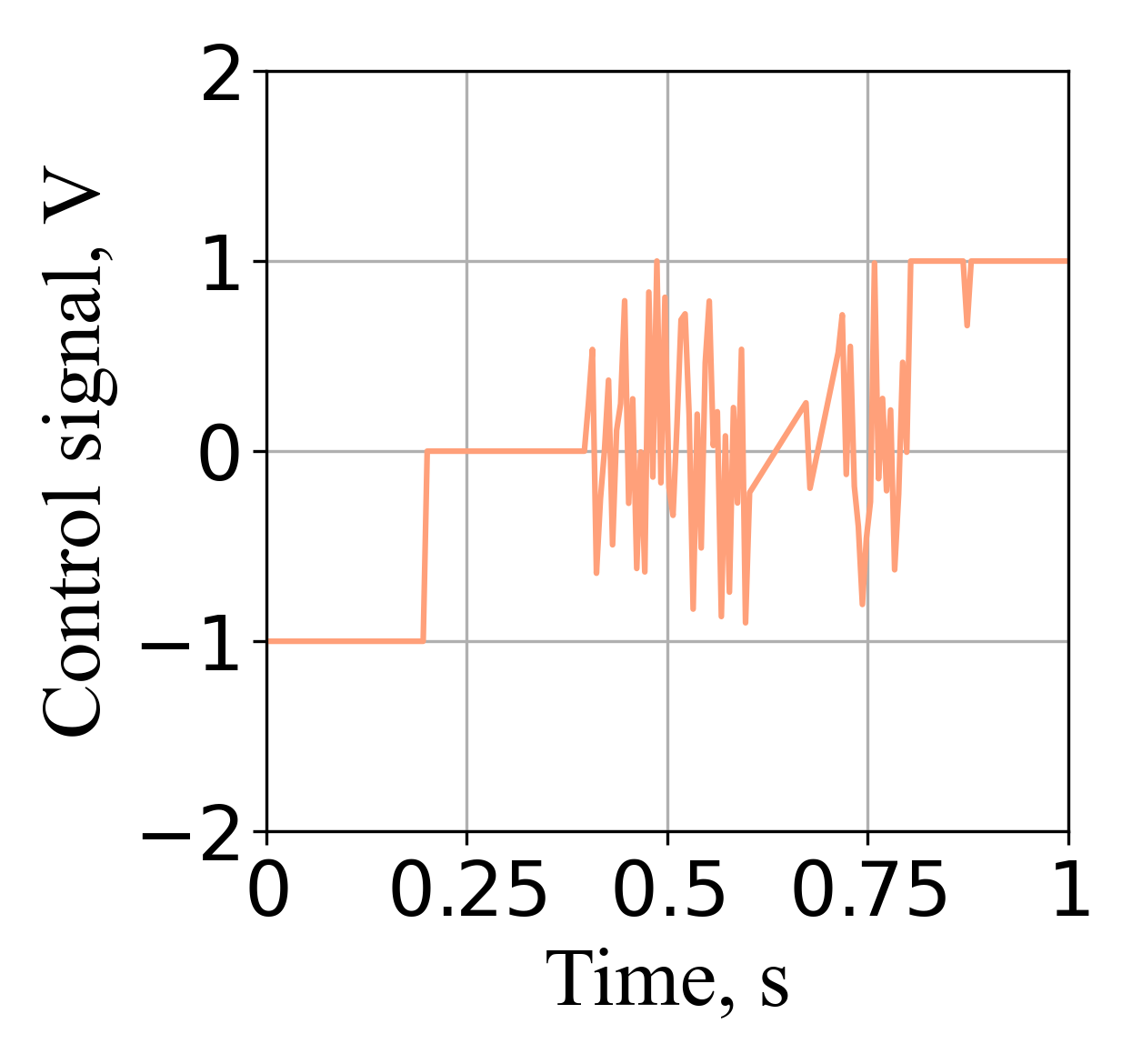}
        \label{signal5}}
        \subfigure{
        \includegraphics[width=0.23\textwidth]{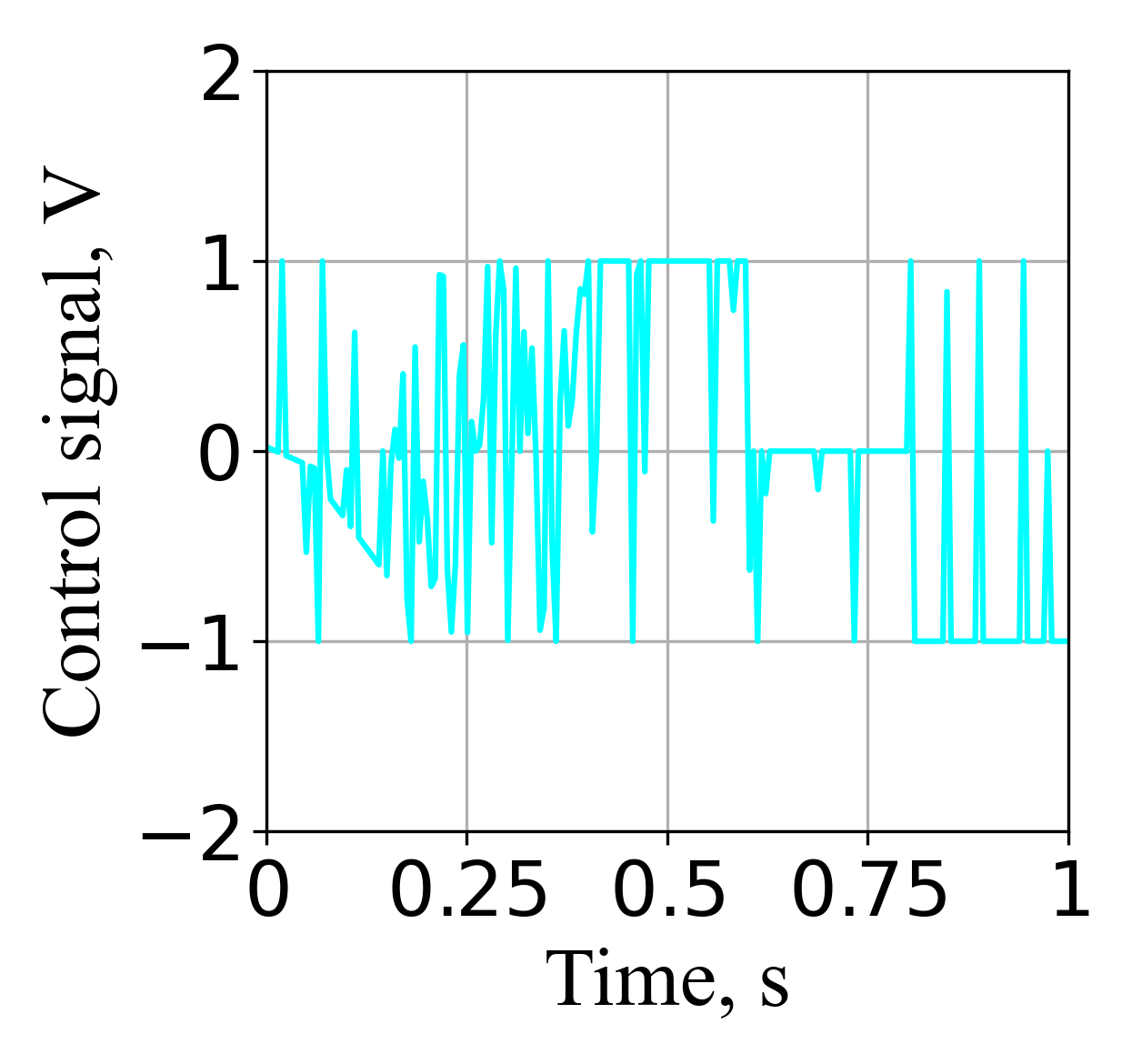}
        \label{signal6}}
        \subfigure{
        \includegraphics[width=0.23\textwidth]{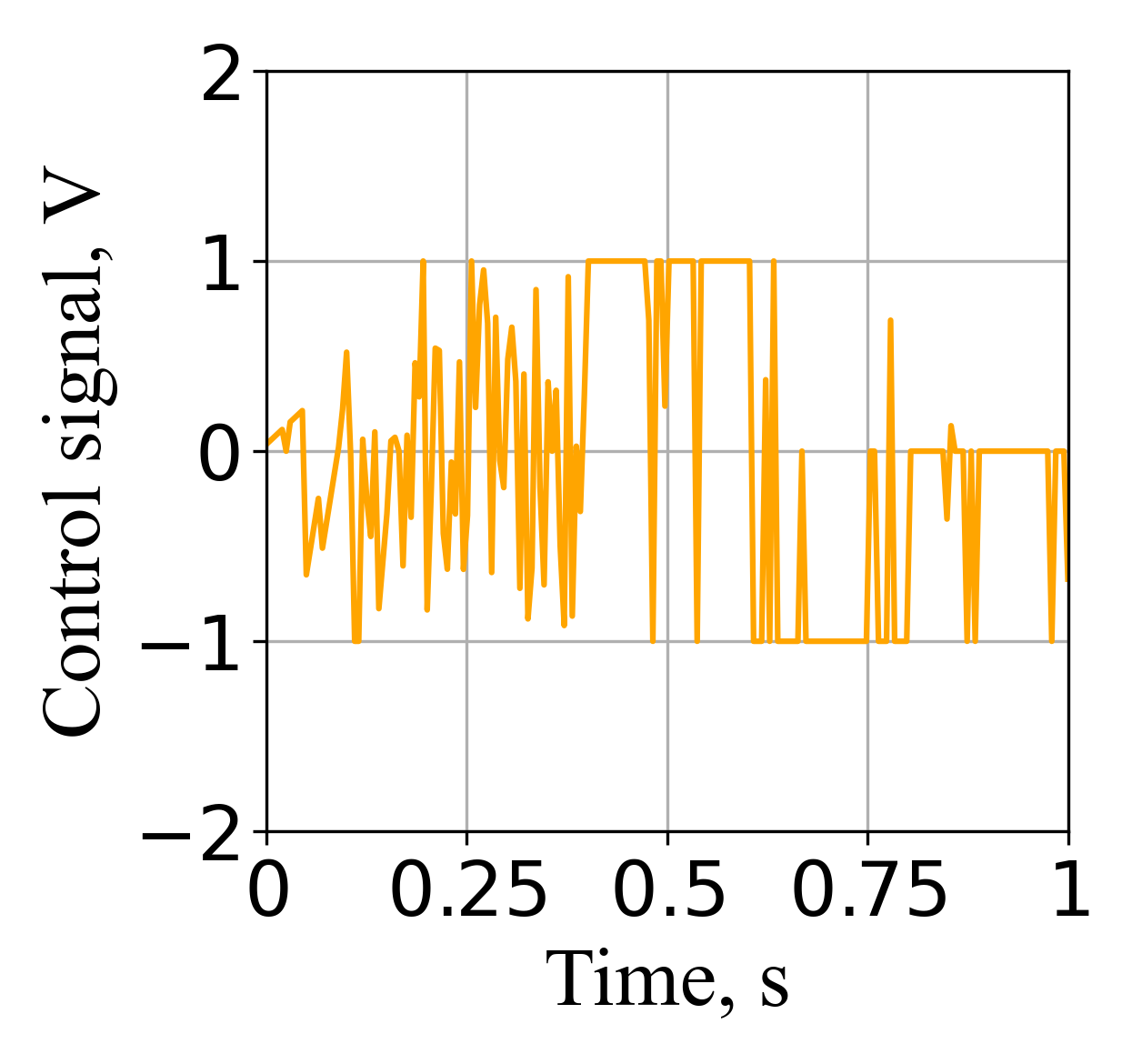}
        \label{signal7}}
        \subfigure{
        \includegraphics[width=0.23\textwidth]{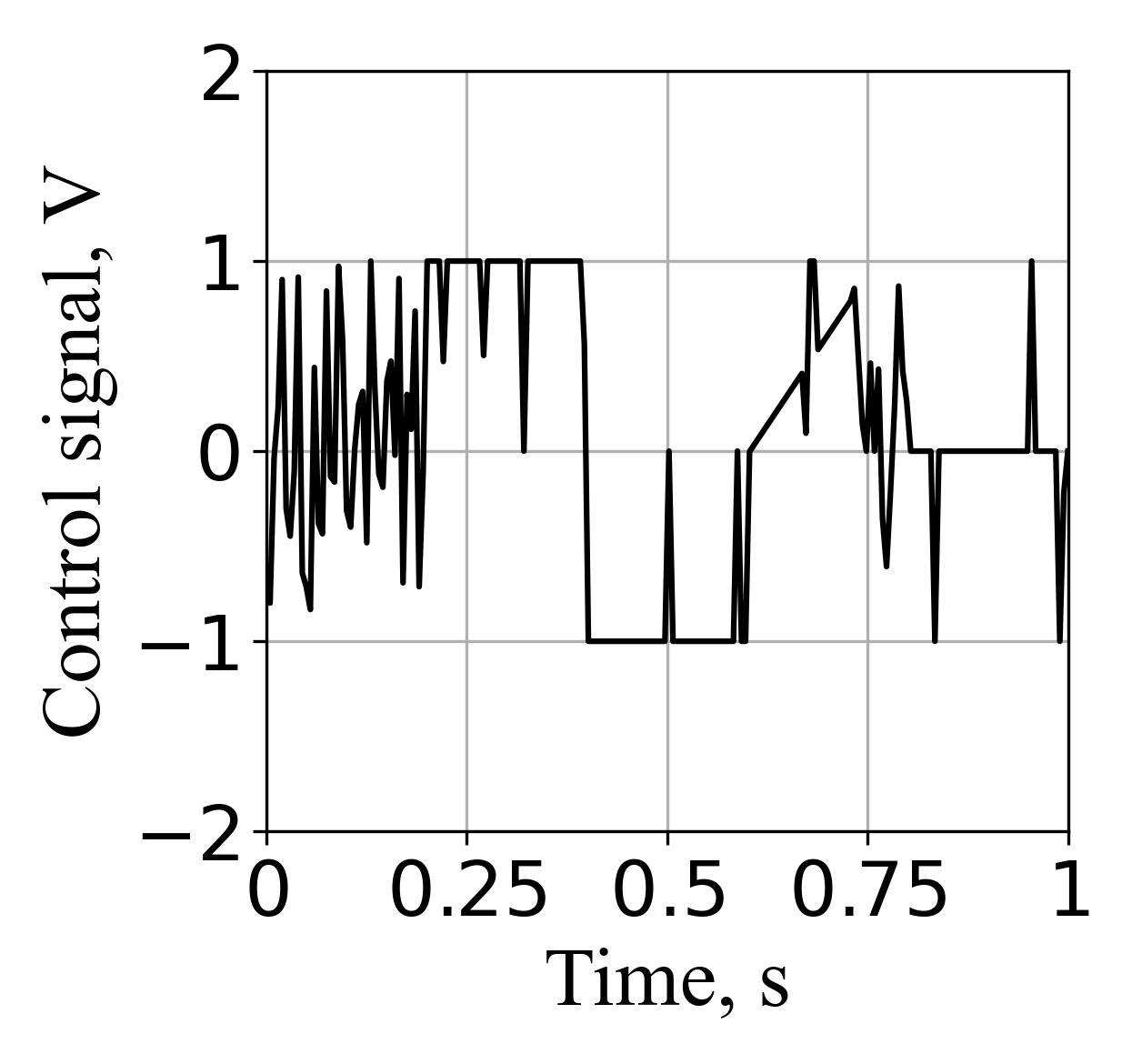}
        \label{signal8}}
    \caption{\textbf{Random control signal}—An example of control signal in the training data }
    \label{ControlSignal}
\end{figure}
It comprises 20\% values at 0, 20\% values at +1, 20\% values at -1, 20\% linear ramped values between -1 and +1 and 20\% random values between -1 and +1. The length of each ramp signal is randomly determined between the points to ensure that the total ramped signal contains 20\% of total simulation steps. The different segments of the control signal are randomly shuffled to formulate the control signal, as described in Fig.~\ref{ControlSignal}.

\subsection{Calculating SLIDE window}
\label{ComputeSlide}
SLIDE data acquisition arranges data according to $\SLIDEWindow$. The variable ${\SLIDEWindow}^{*}$ represents the SLIDE window during a training sample. To calculate ${\SLIDEWindow}^{*}$, the flexible boom is actuated from a random configuration $\theta_\mathrm{in}$ using the control signal described in Fig.~(\ref{TestControl}).
\begin{figure}[h]
    \centering
    \subfigure[Actuator valve control signal ]{
        \includegraphics[width=0.48\textwidth]{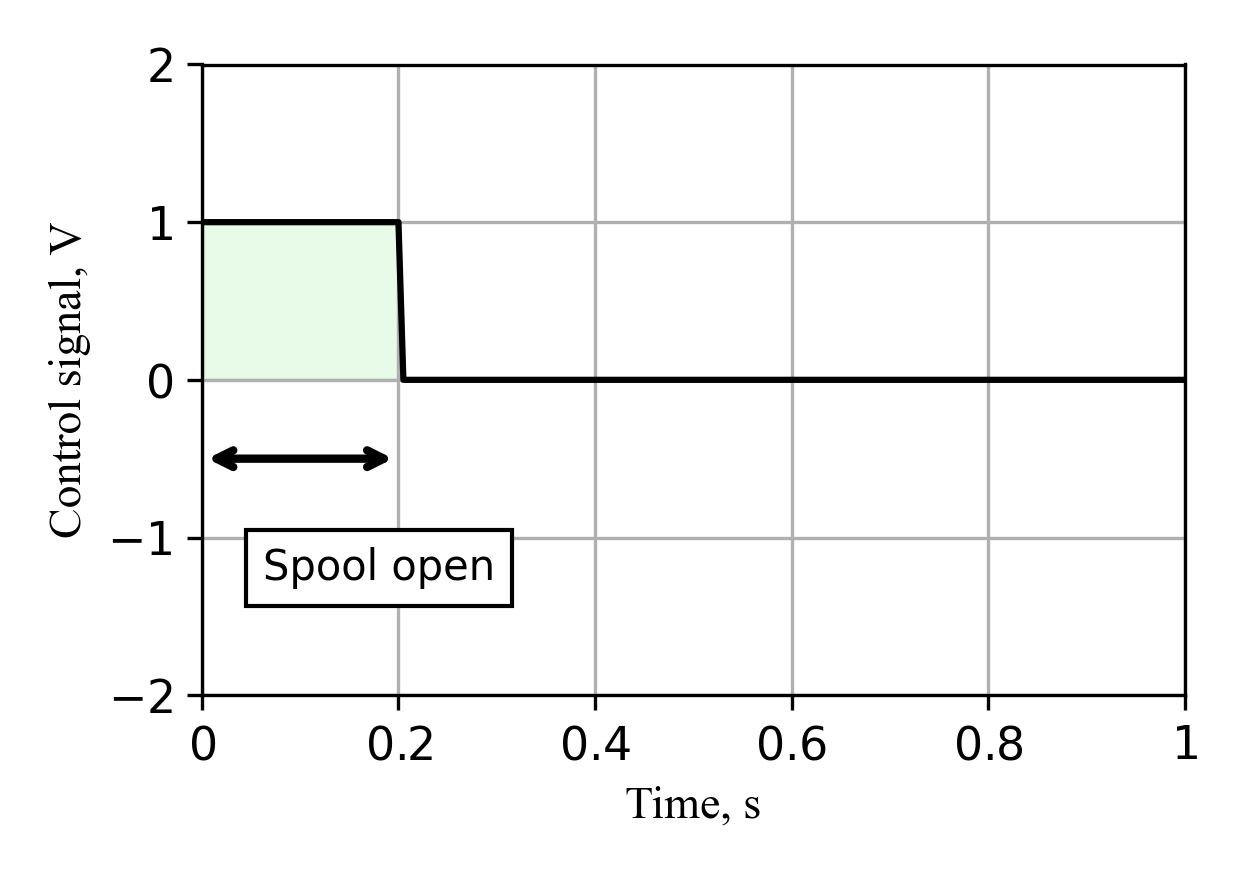}
        \label{TestControl}}
        \subfigure[The SLIDE window is the time in computing~${{\delta}_{\mathrm{y}}}^{*}$ ]{
        \includegraphics[width=0.48\textwidth]{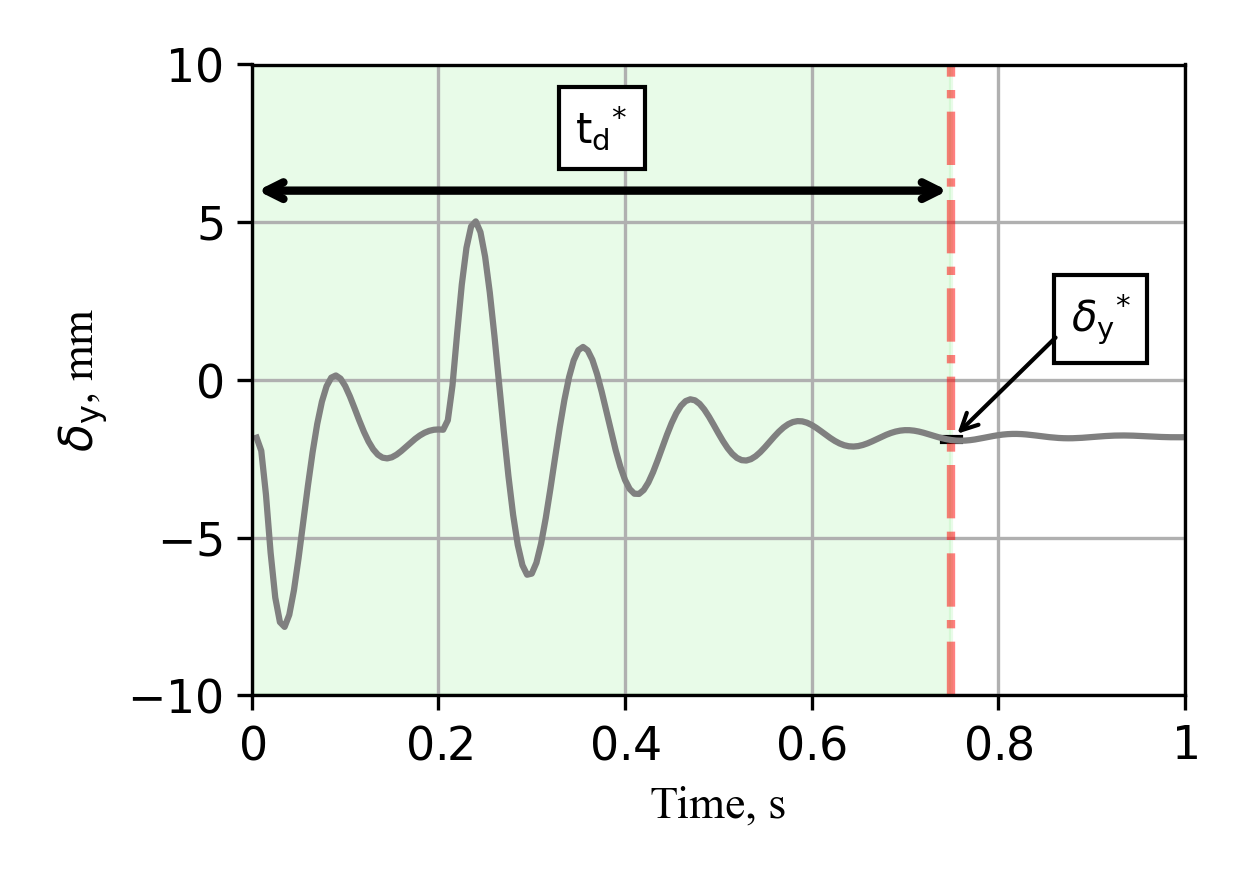}
        \label{TestDeflection}}
    \caption{\textbf{Computing SLIDE window} -- ${\SLIDEWindow}^{*}$ is computed from the threshold deflection~${{\delta}_{\mathrm{y}}}^{*}$ by actuating the hydraulic valve with a control signal. The threshold deflection is 5\% more than the mean of ${\delta}_{\mathrm{y}}$, presented in (b). }
    \label{Damped_Test}
\end{figure}
The control signal involves opening the hydraulic valve for 20\% of simulation time and closing. $\nabla$ is the ${\delta}_{\mathrm{y}}$ data in discrete-time format from Fig.~(\ref{TestDeflection}).
\vspace{-0.50cm}
\begin{equation}
    \nabla = \begin{bmatrix}
        t_0 & t_1 & t_2 & \cdots & t_{\mathrm{end}} \\
        \delta_{{\mathrm{y}}_0} & \delta_{{\mathrm{y}}_1}  & \delta_{{\mathrm{y}}_2}& \cdots & \delta_{{\mathrm{y}}_\mathrm{end}} 
    \end{bmatrix}^{{\top}}
    \label{dampedData}
\end{equation}
\vspace{-0.5cm}
The threshold deflection~${{\delta}_{\mathrm{y}}}^{*}$ is calculated from the mean structural deflection as
\begin{equation}
    {{\delta}_{\mathrm{y}}}^{*} = 1.05 \, \text{mean}(\delta_{\mathrm{y}}).
    \label{threshold}
\end{equation}
\vspace{-0.50cm}
The size of~${\SLIDEWindow}^{*}$ is determined by finding $n^{*}$ step, where~$\delta_{\mathrm{y}} < {{\delta}_{\mathrm{y}}}^{*}$ holds for  
$n^{*} = {n_{end}}/10$ steps, as follows.
\begin{equation}
\left| \delta_{\mathrm{y}_\kSteps} \right|, \left| \delta_{\mathrm{y}_{\kSteps+1}} \right|, \left| \delta_{\mathrm{y}_{\kSteps+2}} \right|, \hdots, \left| \delta_{\mathrm{y}_{\kSteps+\text{n}^{*}-1}} \right| < {{\delta}_{\mathrm{y}}}^{*}
,
\label{DampedSteps}
\end{equation}
where $\kSteps$ is the first step, for~$\delta_{\mathrm{y}} < {{\delta}_{\mathrm{y}}}^{*}$ in $\nabla$.  For $\nTrain$ training samples, $\SLIDEWindow$ is computed as
\vspace{-0.50cm}
\begin{equation}
    \SLIDEWindow = \frac{\displaystyle\sum_{\text{n}=1}^{\nTrain }{\SLIDEWindow}^{*}}{\nTrain}.
\end{equation}

The SLIDE window is computed as ${\SLIDEWindow}^{*} = \kSteps+\mathrm{n}^{*}-1 $. The size of~${\SLIDEWindow}^{*}$ depends on the payload and the Rayleigh damping factor used in the simulation. See Fig.~(\ref{TestDef_loads}) for more details. Note that this method is performed without the EOMs, which demonstrates its industrial applications with the experimental data. 

\subsection{Feature scaling}
The virtual sensors recorded actuator position~$s$, actuator velocity~$\dot{s}$, pressures~$p_1$ and $p_2$, and structural deflection $\delta_{\mathrm{y}}$ from the simulation setup.~The sensor values are scaled for ${\InputLayer}$ as follows. 
\vspace{-0.50cm}
\begin{equation}\label{inputVec}
{\InputLayer} = \frac{1}{{{S}}_{{\InputLayer}}}
\begin{bmatrix}
\ix_{0}&\ix_{1}  & \hdots & \ix_{\SLIDEWindow} & \hdots & \ix_{\SLIDEWindow+\kSteps} 
\end{bmatrix}^{\top}, \quad~\text{and}
\end{equation}
\begin{equation}\label{outputVec}
{\OutputLayer} = \frac{1}{{{S}}_{{\OutputLayer}}}
\begin{bmatrix}
  \iy_{{\SLIDEWindow}} & \iy_{\SLIDEWindow+1} & \hdots & \hdots &\iy_{\SLIDEWindow+\kSteps} 
\end{bmatrix}^{\top},
\end{equation}
where ${\ix}_\iStep=\begin{bmatrix}
     U_\iStep & s_\iStep & {\dot{s}}_\iStep & p_{1_\iStep} & p_{2_\iStep} 
\end{bmatrix}$ is the input vector, and ${\iy}_\iStep=\begin{bmatrix}
    {\delta_{y}}_\iStep
\end{bmatrix}$ is the target vector at the $i^{th}$ step. In Eqs.~\eqref{inputVec}--\eqref{outputVec}, ${{{S}}_{{\InputLayer}}} = \begin{bmatrix}
{{S}}_{\InputLayer_0} & {{S}}_{{{{\InputLayer}}}_1} & \hdots & {{S}}_{{{{\InputLayer}}}_{\SLIDEWindow}}&\hdots & {{S}}_{{{{\InputLayer}}}_{\SLIDEWindow+\kSteps}}
\end{bmatrix}$ and ${{{S}}_{{\OutputLayer}}} = \begin{bmatrix}
{S}_{{\OutputLayer}_{\SLIDEWindow}} & {S}_{{\OutputLayer}_{\SLIDEWindow+1}} & \hdots & \hdots & {S}_{{\OutputLayer}_{\SLIDEWindow+\kSteps}} 
\end{bmatrix}$ are the scaling matrices for $\InputLayer$ and $\OutputLayer$. 
The scaling factor for actuator position is calculated using $l_{\mathrm{cyl}}+l_{\mathrm{pist}}$.  The actuator velocity and hydraulic pressures are scaled according to the maximum velocity, 6.67~\text{m}/\text{s}, and maximum pressures, 200~\text{bar},  produced by the hydraulic cylinder. Similarly, the output scaling factor is determined based on the maximum capacity of the strain measurement sensor, which is ${30~\text{mm}}$.

\subsection{Data distribution}
 The quality of the training data is further accessed by analyzing $\InputLayer$ $(\protect \translucentbluefill)$ and $\OutputLayer$ $(\protect \grayfill)$ on the normal probability distribution curves. See Fig.~(\ref{DataStats}). The x-axis in these plots demonstrates the range of scaled measurements, and the y-axis is the probability density of the fitted normal distribution.
\begin{figure}[h]
\centering
\includegraphics[width=0.95\textwidth]{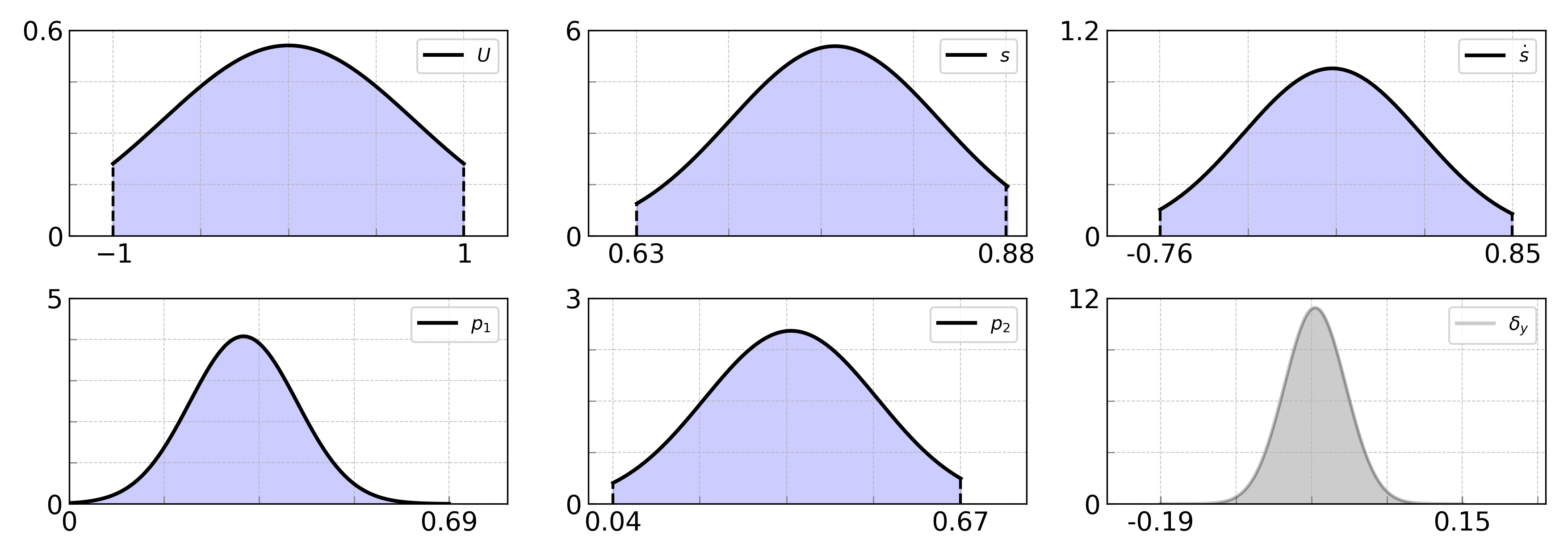}
\caption{\textbf{Data distribution} -- Representation of $\InputLayer$ $(\protect \translucentbluefill)$ and $\OutputLayer$ $(\protect \grayfill)$ on the normal probability distribution curves -- Each curve is calculated using the mean and standard deviation.}
\label{DataStats}
\end{figure}
As shown, the specified scaling factors scale $s$ between $+0.63$ and $+0.88$,~$\dot{s}$ between $-0.76$  and $+0.85$, pressures between $0$ and $+0.7$, and $\delta_{\mathrm{y}}$ between $-0.19$ and $+0.15$. Positive pressures in the hydraulic cylinder are ensured during the data acquisition process.
 
\subsection{Multi-step estimation}
{Using $\SLIDEWindow$, the SLIDE-neural networks can be used for both single-step and multi-step estimations. Fig.~(\ref{Fig:SingleStep})--(\ref{Fig:MultiStep}) illustrate the data arrangement of $\InputLayer$ and $\OutputLayer$ for both estimation approaches.} 
\begin{figure}[h]
    \centering
    \subfigure[Data arrangement in the single-step estimator ]{\includegraphics[width=0.482\textwidth]{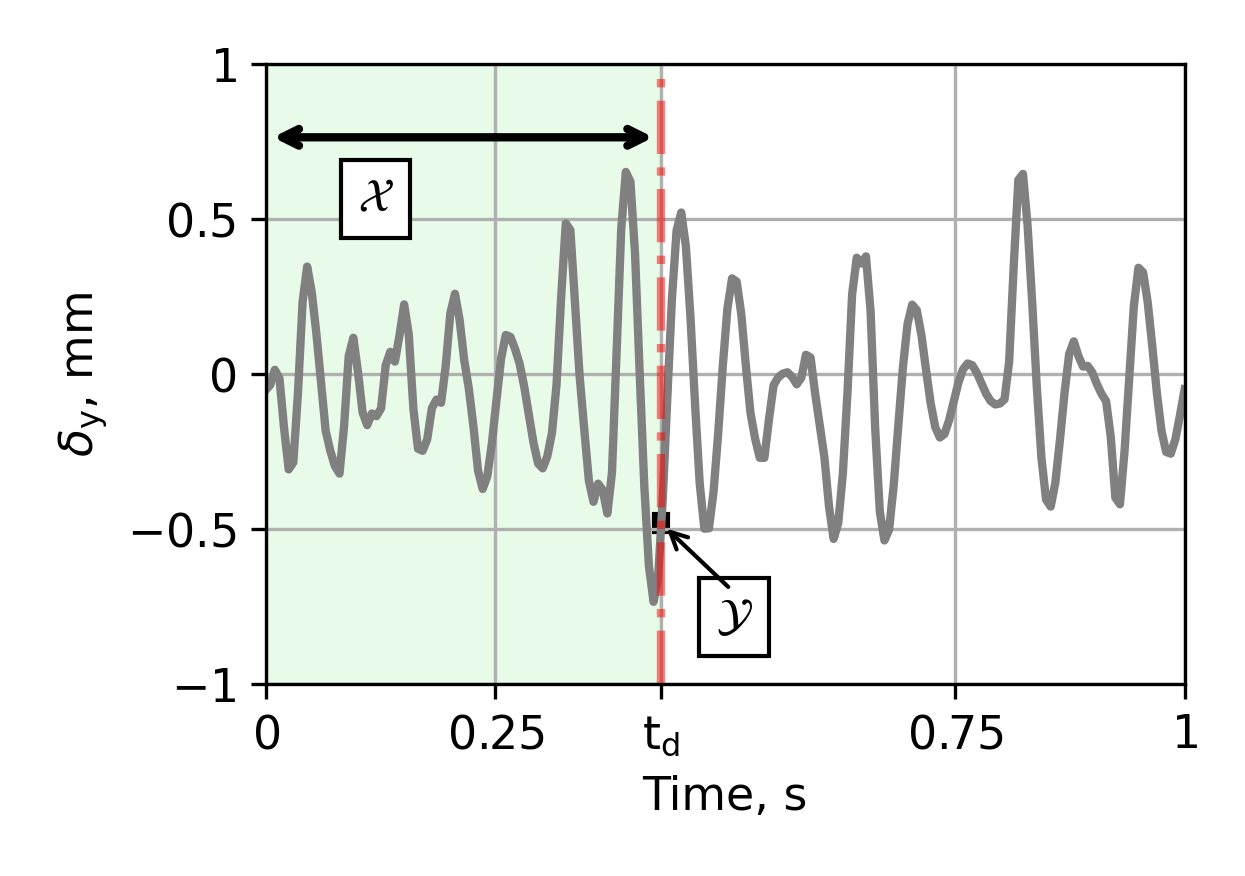}
        \label{Fig:SingleStep}}
        \subfigure[Data arrangement in the multi-step estimator ]{\includegraphics[width=0.482\textwidth]{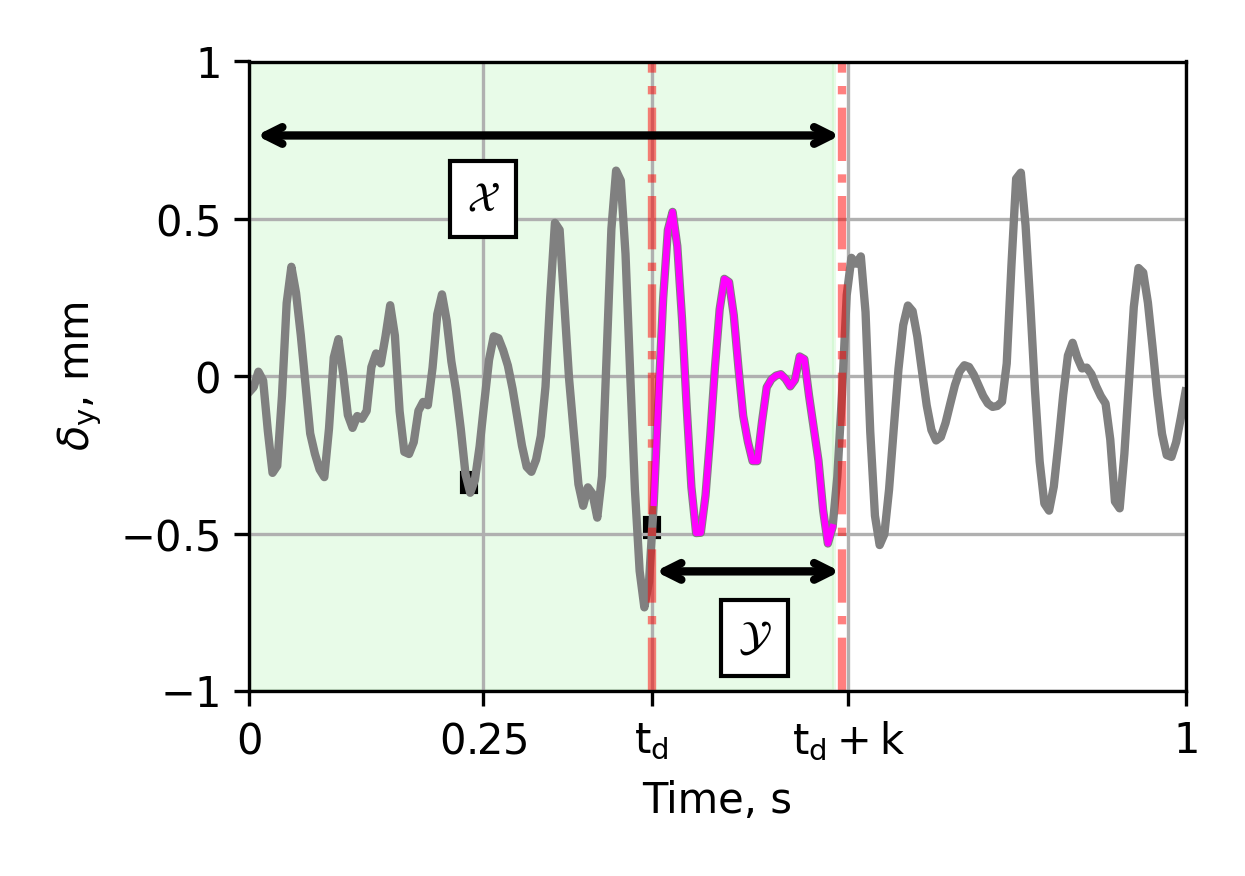}
        \label{Fig:MultiStep}}
    \caption{{\textbf{SLIDE data arrangement} -- Arrangement of $\InputLayer$ and $\OutputLayer$ in the single-step and multi-step estimation schemes}  }
    \label{fig:Steppredictor}
\end{figure}
{The network model learns the dynamics of $\OutputLayer$ from $\InputLayer$ in $\SLIDEWindow$, which is representing the steady-state condition under the forced excitations. In single-step estimation, the output layer ${\OutputLayer}$ contains the target measurements at~$\SLIDEWindow$. However, the multi-step estimator generates estimations starting from $\SLIDEWindow$ for $\kSteps$ steps forward. The applications of multi-step estimation can also be found in the reference study~\cite{manzl2024slide}.} 
\subsection{DNN training parameters}
The proposed DNN model is implemented using the ML PyTorch\footnote{Version 2.3, \url{https://github.com/pytorch/pytorch}} library~\cite{2024_Ansel_Pytorch2}. The model computations are accelerated with the CUDA toolkit\footnote{Version 12.4, \url{https://developer.nvidia.com/cuda-downloads}}. 
\begin{table}[!htb]
\centering
\caption{Standard parameters from PyTorch are used for the ADAM optimizer.}
\begin{tabular}{|c|c|c|c|}
\hline 
\rowcolor{gray!30} \textbf{Parameter} & \textbf{Value} & \textbf{Parameter} & \textbf{Value} \\
\hline
Optimizer & ADAM~\cite{kingma2014adam} & Variable type & float32 \\
\rowcolor{gray!10}
Learning rate & $1~\times~10^{-4}$ & Batch size & $\nTrain / 8$ \\
Training set size $n_{\text{train}}$ & 80 ... 2560 & Validation set size $\nVal$ & 20\% of $\nTrain$ \\
\rowcolor{gray!10}
Validation frequency & every 20 epochs & ${\Loss}_{\text{min}}$ & $5~\times~10^{-6}$ \\
\hline
\end{tabular}
\label{NN_Parameters}
\end{table}
The parameters of the DNN are listed in Table~(\ref{NN_Parameters}). The size of the hidden layers is determined by multiplying $\mathrm{n}_{\SLIDEWindow}$. The training  network comprises the linear layer~(L), sigmoid layer~(S), tangent hyperbolic layer~(T), and ReLU layer~(R). 
\section{DNN training and its evaluation performance}
\label{Results}
The DNN model was trained on a computer with an $11^{th}$ generation Intel\textsuperscript{\textregistered} Core\texttrademark{} i5-11500H CPU running at a base speed of 2.92GHz, complemented by 32 GB RAM. This machine has 6 operating cores and 12 logical processors. It also has an NVIDIA T1200 GPU. The operating system is 64-bit Windows 11. The Exudyn parameter variation function is used to facilitate supervised learning in various hidden layer combinations, leveraging parallel computing power. Following paragraphs describe the DNN training and evaluation performance in detail.
\vspace{-0.50cm}
\begin{figure}[h]
    \centering
    \subfigure[Training loss with single-step estimation. ]{
\includegraphics[width=0.482\textwidth]{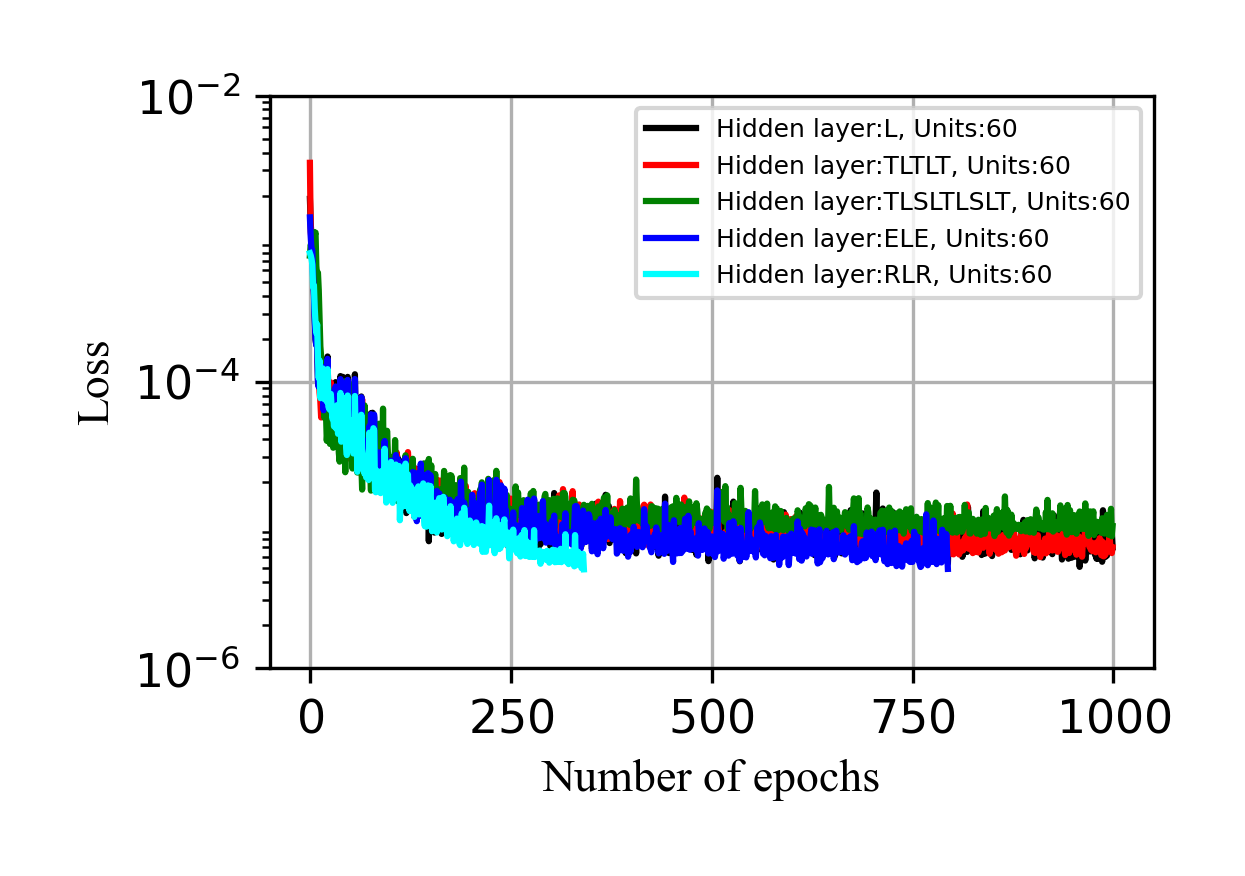}
        \label{Training_5Sensor}}
        \hfill
        \subfigure[Training loss with multi-steps estimation. ]{
        \includegraphics[width=0.482\textwidth]{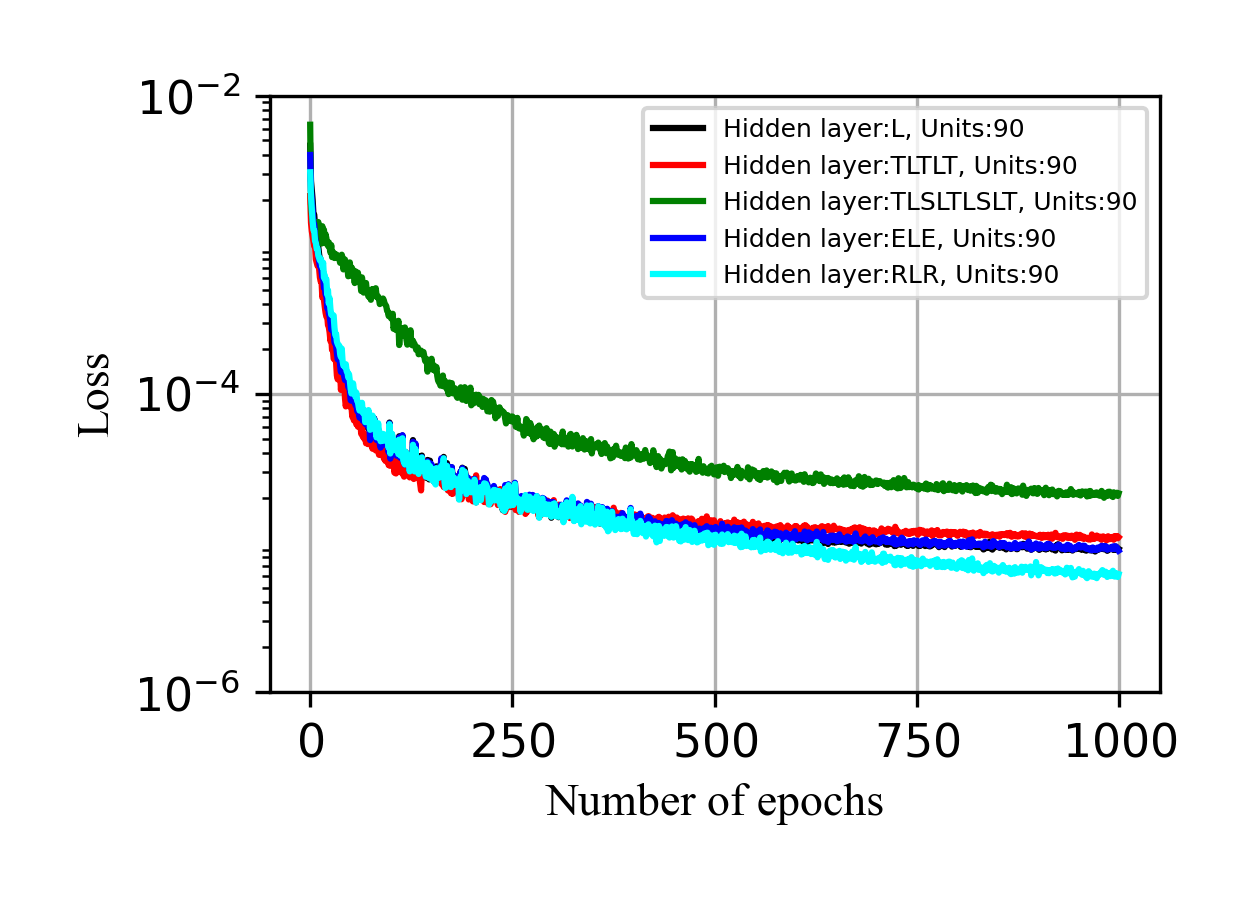}
        \label{Training_30STeps}}
    \caption{\textbf{Training performance}—Mean squared error (MSE) loss for the single-step and multi-steps SLIDE-neural network models evaluated with five sensors and hidden layers combinations without payload. }
    \label{Fig:Training}
\end{figure}

\subsection{Training performance}
In supervised learning, various combinations of data size, input-output arrangement and hidden layer size are explored. Fig.~(\ref{Fig:Training}) shows the training performance of the single-step and multi-step SLIDE-networks, evaluated with five sensors and hidden layer combinations without payload. The different hidden layer combinations include 'L', 'TLTLT', 'TLSLTLSLT','ELE', and 'RLR'. The five sensors providing data were —$U$, $p_1$, $s$, $p_2$ and $\dot{s}$.~Two sensors include $U$ and $s$ in $\InputLayer$, which is used in later estimation tasks. SLIDE-network 'L' shows good training results, assisted by the availability of all data in $\InputLayer$ for the force computation.  The hidden layer 'RLR' and 'ELE' meet the ${\Loss}_{\text{min}}$ criteria for five sensors combinations. Other hidden layer combinations reached 1000 epochs during the supervised learning. In Fig.~(\ref{Training_30STeps}), 30 steps forward in $\SLIDEWindow$ was considered during the training process. This results in the smooth supervised learning of the SLIDE-networks. However, none of hidden layer combinations meet the ${\Loss}_{\text{min}}$ criteria.

\subsection{Evaluation performance}
In the evaluation phase, the performance of estimation for the final single-step DNN models is tested $\delta_{\mathrm{y}}$ with an unseen control signal.~The multi-step estimations are demonstrated in Fig.~(\ref{fig:MultistepEstimations}).~The $\delta_{\mathrm{y}}$ from the reference simulation is represented by~$(\protect \coralredline)$, and~$(\protect \bluedashedline)$ corresponds to the DNN model estimates.
\begin{figure}[h]
    \centering
        \subfigure[\textbf{No payload:} Estimating $\delta_{\mathrm{y}}$ using 'L' network with two sensors. ]{\includegraphics[width=0.75\textwidth]{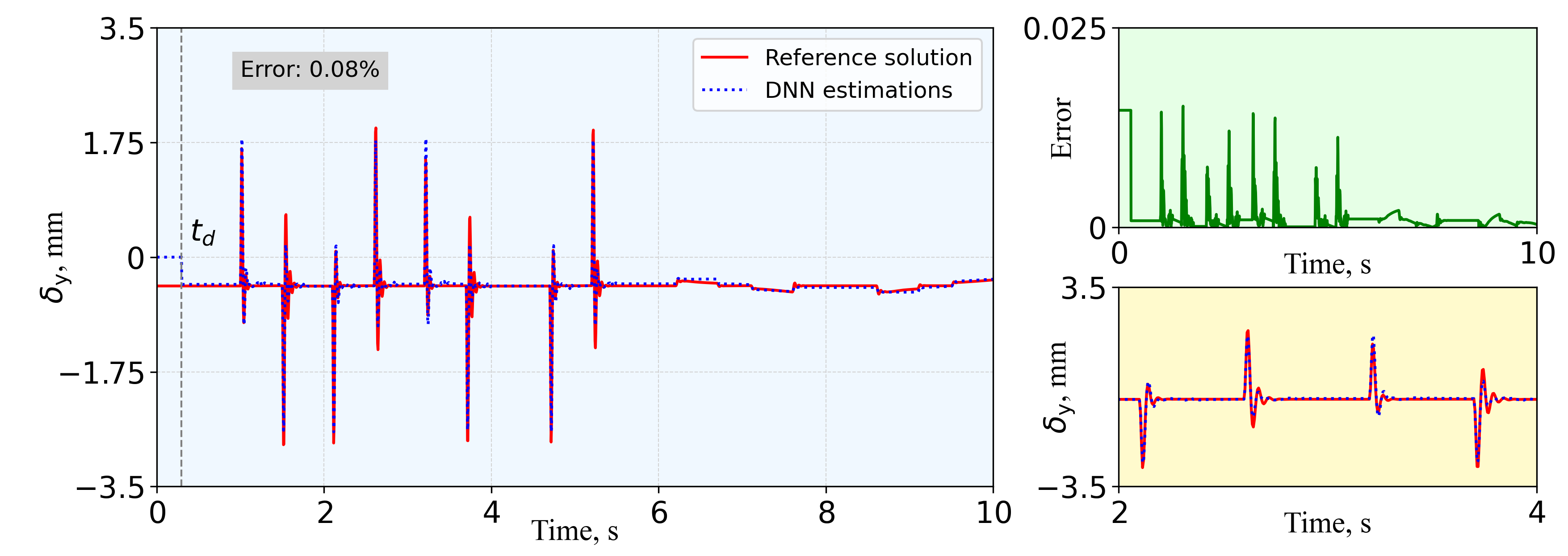}
        \label{Unseendata1}}
        \vfill
        \subfigure[\textbf{Payload 50~{kg}:} Estimating $\delta_{\mathrm{y}}$ using 'TLSLTLSLT' network with five sensors.]{\includegraphics[width=0.75\textwidth]{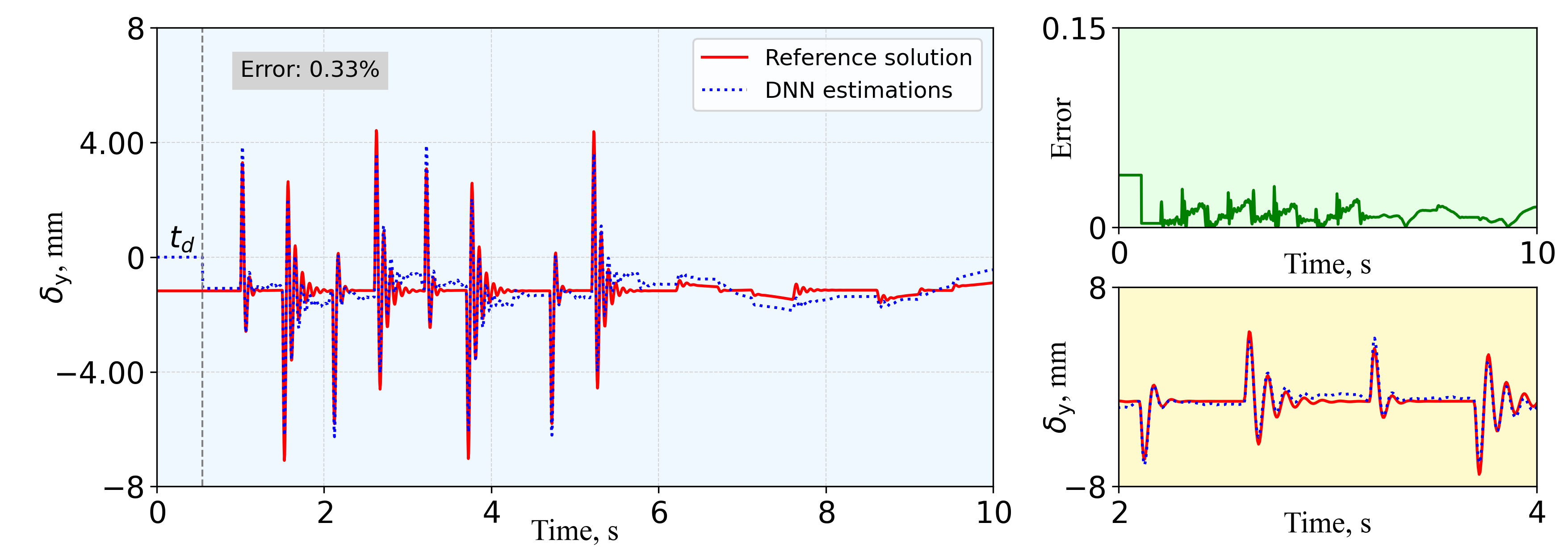}
        \label{Unseendata2}}
        \vfill
        \subfigure[\textbf{Payload 100~{kg}:} Estimating $\delta_{\mathrm{y}}$ using 'TLSLTLSLT' network with two sensors. ]{\includegraphics[width=0.75\textwidth]{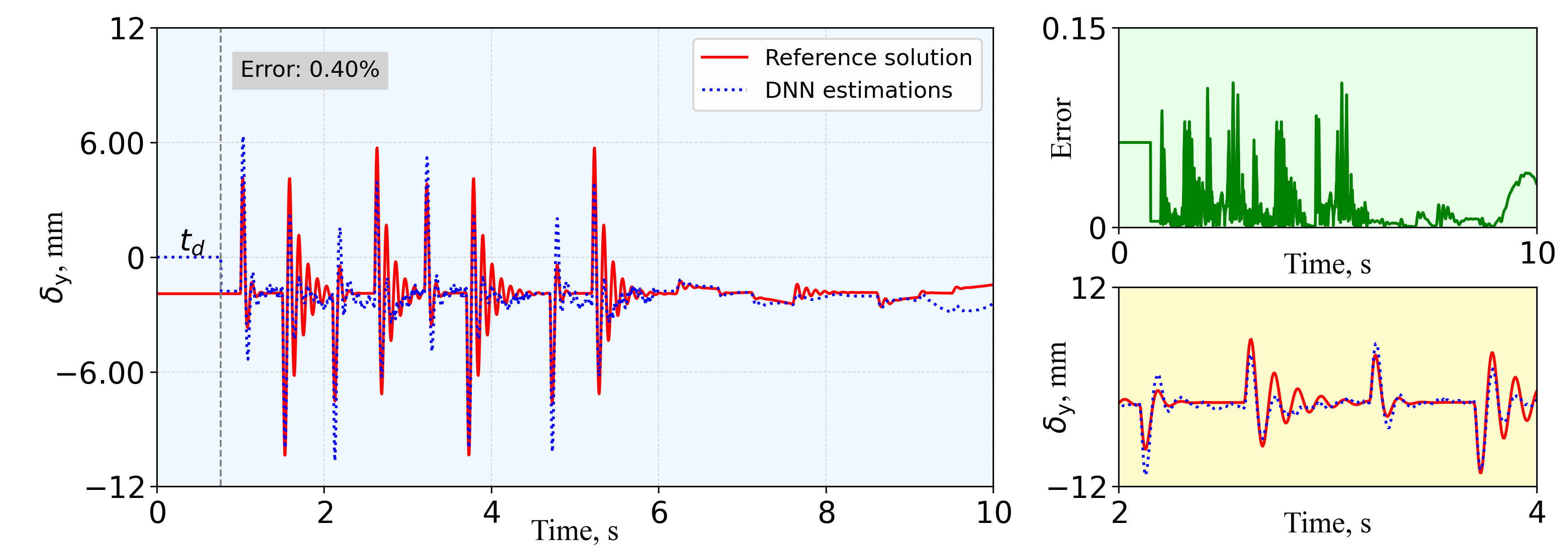} 
        \label{Unseendata3}}
    \caption{{\textbf{Estimation accuracy}—SLIDE-neural networks estimating structural deflection in a hydraulically actuated flexible boom, verifying various payloads and sensor combinations. Mean-absolute-percentage error~(MAPE) and mean-absolute error~(MAE) demonstrate the estimation accuracy with respect to the reference solutions.}  }
    \label{fig:allplots}
\end{figure}
Fig.~(\ref{Unseendata1}) demonstrates the SLIDE-neural network model estimation of $\delta_{\mathrm{y}}$ during 10 s dynamic simulation. The actuator control signal during simulation is presented in~Fig.~(\ref{EvaluationControl}).  The variation in control signal also affects $\delta_{\mathrm{y}}$ in the reference solution. 

For no payload, the DNN model makes no estimation at $\SLIDEWindow$, 0.3~\text{s}, due to the unavailability of $\InputLayer$ for this period. After $\SLIDEWindow$, the DNN model starts making single-step estimations with 0.08\% MAPE. 
The error with respect to the reference solution is further explained in mean-absolute-error $(\protect \darkgreenline)$. The accuracy of DNN estimations can be further seen in zoomed-in plot. The DNN model estimations are exactly following the uncertain and non-linear patterns $\delta_{\mathrm{y}}$.
\begin{figure}[h]
\centering
\includegraphics[width=0.925\textwidth]{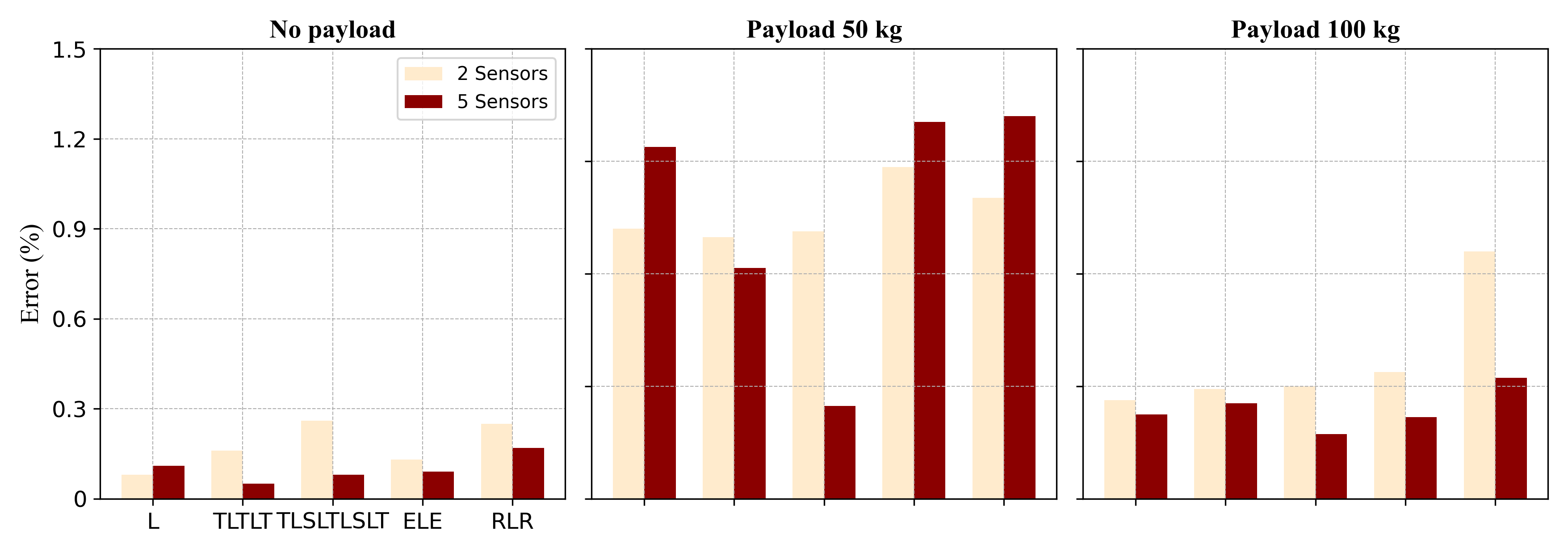}
\caption{{\textbf{Estimation summary} -- A summary of SLIDE-neural networks performance in estimating structural deflection}}
\label{fig:Summary}
\end{figure}
The estimation of $\delta_{\mathrm{y}}$ for $50$~\text{kg} and $100$~\text{kg}~are described in Fig.~(\ref{Unseendata2}) and Fig.~(\ref{Unseendata3}).  The maximum structural deflection of 11.25~\text{mm} occurs in payload 100~\text{kg}. The SLIDE windows for these cases are $0.55$~\text{s} and $0.77$~\text{s}. These results were obtained from the hidden layer 'TLSLTLSLT'. This case demonstrates that two sensors—$U$ and $s$—are enough for single-step DNN model estimations. Both models provides good accuracy with 0.33\% and 0.40\% MAPE for the $50$~\text{kg} and $100$~\text{kg} cases, respectively. 

The zoomed-in plot reveals that the DNN model follows the patterns of $\delta_{\mathrm{y}}$ from the reference solutions between 2--4~\text{s}. This demonstrates the robustness of the SLIDE approach to estimating the targets over longer simulations. Fig.~(\ref{fig:Summary}) describes the summary of SLIDE-neural networks performance when estimating structural deflection for $0$~\text{kg}, $50$~\text{kg} and $100$~\text{kg} payloads. The maximum MAPE occurs with $50$~\text{kg} payload using 'RLR', and it is below 1.5\%. The two sensors case presents accurate estimation results for all payloads and hidden layer combinations. It establishes that the SLIDE-neural network models can provide accurate estimations with actuator position and control signal data. 

\subsection{Computational efficiency}
Fig.~(\ref{ComputationalEfficiency}) shows the computational efficiency benefits of SLIDE-neural networks compared to flexible multibody simulation for $\delta_{\mathrm{y}}$.~Python's function \texttt{timeit} was used to determine the relative run time of SLIDE-neural network estimation and simulation solutions. CUDA takes 0.2 milliseconds to initialize the system GPUs. As Fig.~(\ref{SpeedUp}) also shows, the relative computational run time of the neural networks starts from $0.2~10^{-3}$ s for creating the equivalent data to the simulation batches.~The relative computational run time of the neural networks when running simulation batches is faster than the simulations. 
\begin{figure}[h]
    \centering
    \subfigure[Relative computational run time.  ]{
        \includegraphics[width=0.482\textwidth]{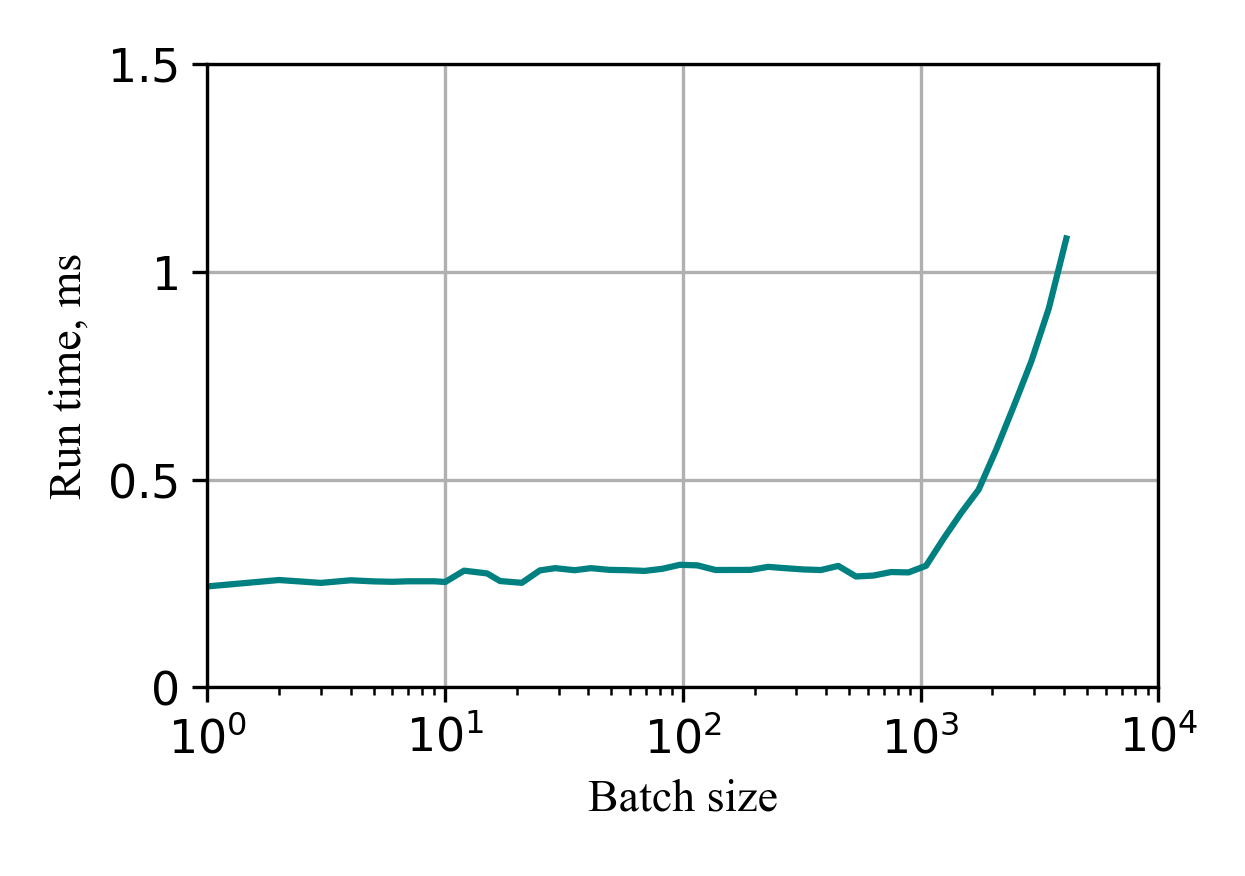}
        \label{SpeedUp}}
        \subfigure[Speed-up factor of SLIDE-neural networks. ]{
        \includegraphics[width=0.482\textwidth]{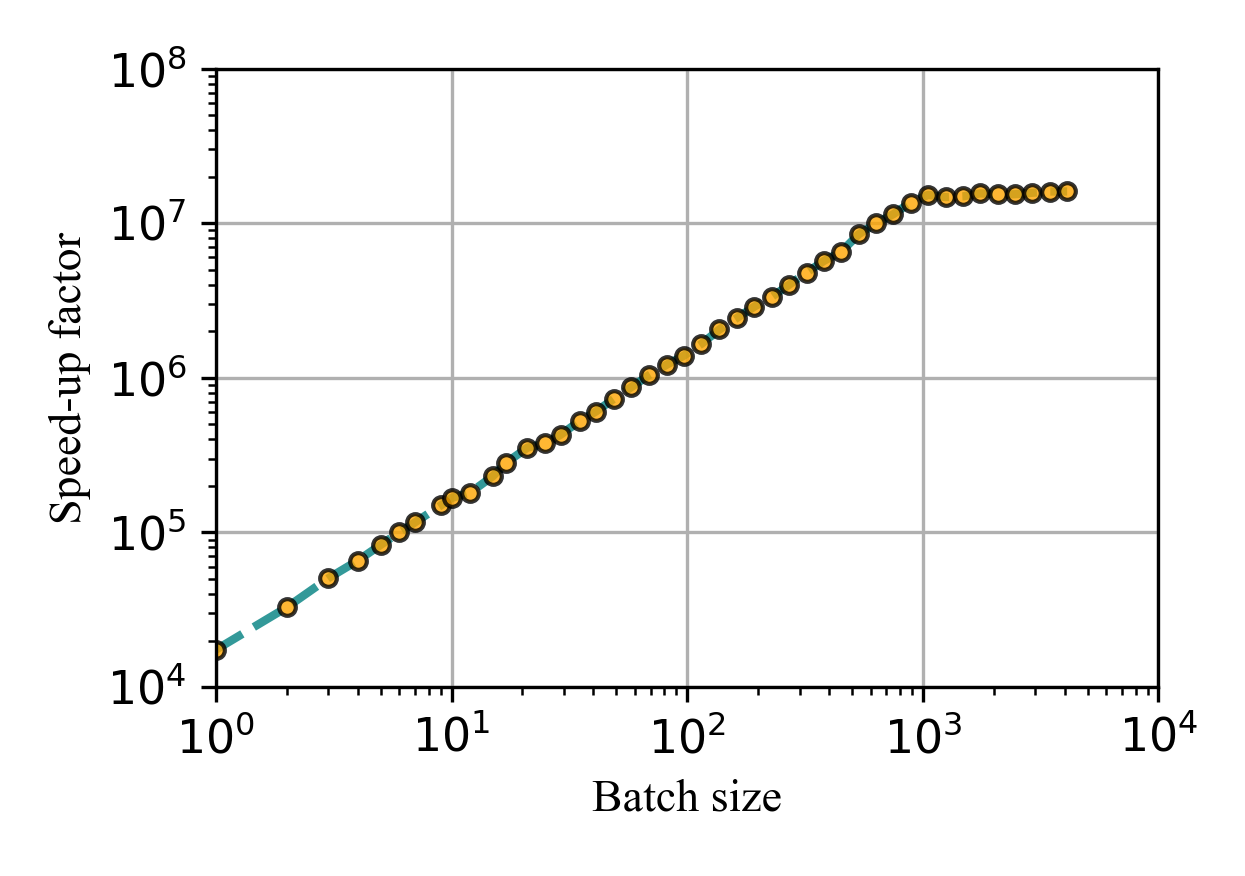}
        \label{RunTime}}
    \caption{\textbf{Computational efficiency}--Relative computational run time and speed-up factor of SLIDE-neural networks compared to the flexible multibody simulation of case example in running simulation batches for $\delta_{\mathrm{y}}$.  }
    \label{ComputationalEfficiency}
\end{figure}

Note that the simulations solutions are faster in the Exudyn framework. The SLIDE approach will provide more computational efficiency as compared to FE~\cite{uyar2023implementation} simulations. Further, Fig.~(\ref{ComputationalEfficiency}) demonstrates the computational speed-up factor, $S$, of neural network run time, $t_{\mathrm{NN}}$, with respect to the simulation solutions run time, $t_{\mathrm{sim}}$. It is measured as~\cite{manzl2024slide}
\begin{equation} \label{SpeedupEq}
    S = \frac{t_{\mathrm{sim}}}{t_{\mathrm{NN}}} \frac{n_{\mathrm{out}}}{n_{\mathrm{in}}},
\end{equation}
where ${n_{\mathrm{in}}}$ and ${n_{\mathrm{out}}}$ are the input number of simulations and output number of steps.
\section{Conclusion}
\label{Conc}
A real-time structural deflection estimation framework using novel SLIDE-neural network model has been proposed. The framework learns the behavior of structural deflections at arbitrary location on the full-scale {3D} FE model of a system within the SLIDE window size. Effectiveness can be confirmed as follows.
\begin{itemize}
\item {The robustness of this approach was tested with various payloads and sensor combinations. The SLIDE-trained networks accelerate the deflection estimation solutions for the hydraulically actuated three-dimensional flexible systems by a factor of $10^7$ as compared to the reference flexible multibody simulations batches.} 

\item {An algorithm was developed to generate training data for a hydraulically actuated flexible multibody system using randomized initial configurations, while also determining $\SLIDEWindow$ without the EOMs. The approximation of $\SLIDEWindow$ with the system’s linearized EOMs is explained in~\cite{manzl2024slide}. However, the $\SLIDEWindow$ computing method introduced here can also be applied to the experimental data.} 

\item{The findings show that a reasonable estimate of structural deflection can be achieved with fewer than five sensors, including the actuator position and control signal. The estimation accuracy of various SLIDE-neural networks was compared when estimating structural deflections. The maximum MAPE occurred in the case of 'RLR', which is below 1.50\%. } 
\end{itemize}
Unlike physics-inspired networks~\cite{roehrl2020modeling, lutter2023combining, khadim2024simulation}, SLIDE networks learn the behaviors, \textit{i.e.}, the physical dynamics of flexible systems using the SLIDE window when defining the input-output relationships.~The findings demonstrate the potential of the approach for various industrial applications when solving robotic manipulations, control, structural health monitoring, and automation problems. Joint friction and contact forces were not considered. These also contribute to the uncertainty of robotic manipulation and control~\cite{10438059}. Furthermore, the dynamics of lightweight structures can be studied using the SLIDE approach and control methods. As documented in the literature, these problem have been studied in control systems using approximated dynamic models~\cite{sarkhel2023robust,sun2018fuzzy}, FE~\cite{uyar2023implementation}, and direct measurement methods~\cite{jensen2022online}. 

Because of nonlinearity, the detailed FE models may not be applicable to industrial systems. However, the proposed approach offers an efficient tool to study flexible structures with complex geometries. Further, the SLIDE-driven approach would require less data in real-time estimation tasks for industrial systems, allowing sensors with less frequency. The SLIDE-based ML approach could have potential applications across various fields including engineering, medicine, and business. It could be employed in scenarios where data exhibits damping, because this would require fewer target values. The data generation tool developed here offers a versatile solution for hydraulically actuated systems, effectively facilitating research in ML, control, and structural health monitoring across various FE structures.
\section*{Acknowledgment}
This work was supported by the Business Finland (SANTTU-Oulu 8896/31/2021 and SANTTU-LUT 8859/31/2021). We want to thank the open-source projects Exudyn~\cite{gerstmayr2023exudyn}, PyTorch~\cite{2024_Ansel_Pytorch2} and Numpy~\cite{harris2020array}. The computational results presented here have been achieved in partly using the LEO HPC\footnote{High Performance Computing resources at the University of Innsbruck, \url{https://www.uibk.ac.at/en/mechatronik/mekt/research/research-profile/}} infrastructure of the University of Innsbruck, as well as Puhti\footnote{Puhti supercomputer, CSC - IT Center for Science, Finland, \url{https://www.csc.fi/en/puhti}} and LUMI\footnote{LUMI supercomputer, part of the EuroHPC Joint Undertaking and hosted by CSC - IT Center for Science, Finland, \url{https://www.lumi-supercomputer.eu}} provided by CSC.

\section*{Conflict of interest}
The authors declare that they have no conflict of interest.


\bibliography{mybibfile}

\begin{thebibliography}{10}
\expandafter\ifx\csname url\endcsname\relax
  \def\url#1{\texttt{#1}}\fi
\expandafter\ifx\csname urlprefix\endcsname\relax\def\urlprefix{URL }\fi
\expandafter\ifx\csname href\endcsname\relax
  \def\href#1#2{#2} \def\path#1{#1}\fi

\bibitem{bissadu2024society}
K.~D. Bissadu, S.~Sonko, G.~Hossain, Society 5.0 enabled agriculture: Drivers, enabling technologies, architectures, opportunities, and challenges, Information Processing in Agriculture (2024).
\newblock \href {https://doi.org/10.1016/j.inpa.2024.04.003} {\path{doi:10.1016/j.inpa.2024.04.003}}.

\bibitem{khadim2020targeting}
Q.~Khadim, E.-P. Kaikko, E.~Puolatie, A.~Mikkola, Targeting the user experience in the development of mobile machinery using real-time multibody simulation, Advances in Mechanical Engineering 12~(6) (2020) 1687814020923176.
\newblock \href {https://doi.org/10.1177/1687814020923176} {\path{doi:10.1177/1687814020923176}}.

\bibitem{billard2019trends}
A.~Billard, D.~Kragic, Trends and challenges in robot manipulation, Science 364~(6446) (2019) eaat8414.
\newblock \href {https://doi.org/10.1126/science.aat8414} {\path{doi:10.1126/science.aat8414}}.

\bibitem{sanchez2020innovation}
F.~S{\'a}nchez, P.~Hartlieb, Innovation in the mining industry: Technological trends and a case study of the challenges of disruptive innovation, Mining, Metallurgy \& Exploration 37~(5) (2020) 1385--1399.
\newblock \href {https://doi.org/10.1007/s42461-020-00262-1} {\path{doi:10.1007/s42461-020-00262-1}}.

\bibitem{mann2023benign}
K.~Mann, L.~P{\"u}ttmann, Benign effects of automation: New evidence from patent texts, Review of Economics and Statistics 105~(3) (2023) 562--579.
\newblock \href {https://doi.org/10.1162/rest_a_01083} {\path{doi:10.1162/rest_a_01083}}.

\bibitem{physicsbasedDT}
E.~Kurvinen, A.~Kutvonen, J.~Ukko, Q.~Khadim, Y.~S. Hagh, et~al., Physics-based digital twins merging with machines: Cases of mobile log crane and rotating machine, IEEE Access 10 (2022) 45962--45978.
\newblock \href {https://doi.org/10.1109/ACCESS.2022.3170430} {\path{doi:10.1109/ACCESS.2022.3170430}}.

\bibitem{haggerty2023control}
D.~A. Haggerty, M.~J. Banks, E.~Kamenar, A.~B. Cao, P.~C. Curtis, I.~Mezi{\'c}, E.~W. Hawkes, Control of soft robots with inertial dynamics, Science robotics 8~(81) (2023) eadd6864.
\newblock \href {https://doi.org/10.1126/scirobotics.add6864} {\path{doi:10.1126/scirobotics.add6864}}.

\bibitem{wagg2010nonlinear}
D.~Wagg, S.~Neild, Nonlinear vibration with control: for flexible and adaptive structures, Springer, 2010.
\newblock \href {https://doi.org/10.1007/978-90-481-2837-2_2} {\path{doi:10.1007/978-90-481-2837-2_2}}.

\bibitem{sarkhel2023robust}
P.~Sarkhel, M.~K. Dikshit, V.~K. Pathak, K.~K. Saxena, C.~Prakash, D.~Buddhi, Robust deflection control and analysis of a fishing rod-type flexible robotic manipulator for collaborative robotics, Robotics and Autonomous Systems 159 (2023) 104293.
\newblock \href {https://doi.org/10.1016/j.robot.2022.104293} {\path{doi:10.1016/j.robot.2022.104293}}.

\bibitem{10438059}
B.~Li, X.~Li, H.~Gao, F.-Y. Wang, Advances in flexible robotic manipulator systems—part i: Overview and dynamics modeling methods, IEEE/ASME Transactions on Mechatronics 29~(2) (2024) 1100--1110.
\newblock \href {https://doi.org/10.1109/TMECH.2024.3359067} {\path{doi:10.1109/TMECH.2024.3359067}}.

\bibitem{cui2020trajectory}
L.~Cui, H.~Wang, W.~Chen, Trajectory planning of a spatial flexible manipulator for vibration suppression, Robotics and Autonomous Systems 123 (2020) 103316.
\newblock \href {https://doi.org/10.1016/j.robot.2019.103316} {\path{doi:10.1016/j.robot.2019.103316}}.

\bibitem{lajunen2018overview}
A.~Lajunen, P.~Sainio, L.~Laurila, J.~Pippuri-M{\"a}kel{\"a}inen, K.~Tammi, Overview of powertrain electrification and future scenarios for non-road mobile machinery, Energies 11~(5) (2018) 1184.
\newblock \href {https://doi.org/10.3390/en11051184} {\path{doi:10.3390/en11051184}}.

\bibitem{czerwinski2021current}
F.~Czerwinski, Current trends in automotive lightweighting strategies and materials, Materials 14~(21) (2021) 6631.
\newblock \href {https://doi.org/10.3390/ma14216631} {\path{doi:10.3390/ma14216631}}.

\bibitem{rigatos2018robotic}
G.~Rigatos, K.~Busawon, Robotic manipulators and vehicles: control, estimation and filtering, Vol. 152, Springer, 2018.
\newblock \href {https://doi.org/10.1007/978-3-319-77851-8} {\path{doi:10.1007/978-3-319-77851-8}}.

\bibitem{uyar2023implementation}
M.~Uyar, L.~Malgaca, Implementation of active and passive vibration control of flexible smart composite manipulators with genetic algorithm, Arabian Journal for Science and Engineering 48~(3) (2023) 3843--3862.
\newblock \href {https://doi.org/10.1007/s13369-022-07279-2} {\path{doi:10.1007/s13369-022-07279-2}}.

\bibitem{avci2021review}
O.~Avci, O.~Abdeljaber, S.~Kiranyaz, M.~Hussein, M.~Gabbouj, D.~J. Inman, A review of vibration-based damage detection in civil structures: From traditional methods to machine learning and deep learning applications, Mechanical systems and signal processing 147 (2021) 107077.
\newblock \href {https://doi.org/10.1016/j.ymssp.2020.107077} {\path{doi:10.1016/j.ymssp.2020.107077}}.

\bibitem{malekloo2022machine}
A.~Malekloo, E.~Ozer, M.~AlHamaydeh, M.~Girolami, Machine learning and structural health monitoring overview with emerging technology and high-dimensional data source highlights, Structural Health Monitoring 21~(4) (2022) 1906--1955.
\newblock \href {https://doi.org/10.1177/14759217211036880} {\path{doi:10.1177/14759217211036880}}.

\bibitem{dwivedy2006dynamic}
S.~K. Dwivedy, P.~Eberhard, Dynamic analysis of flexible manipulators, a literature review, Mechanism and machine theory 41~(7) (2006) 749--777.
\newblock \href {https://doi.org/10.1016/j.mechmachtheory.2006.01.014} {\path{doi:10.1016/j.mechmachtheory.2006.01.014}}.

\bibitem{lee2020critical}
T.~S. Lee, E.~A. Alandoli, A critical review of modelling methods for flexible and rigid link manipulators, Journal of the Brazilian Society of Mechanical Sciences and Engineering 42 (2020) 1--14.
\newblock \href {https://doi.org/10.1007/s40430-020-02602-0} {\path{doi:10.1007/s40430-020-02602-0}}.

\bibitem{aldakheel2021feed}
F.~Aldakheel, R.~Satari, P.~Wriggers, Feed-forward neural networks for failure mechanics problems, Applied Sciences 11~(14) (2021) 6483.
\newblock \href {https://doi.org/10.3390/app11146483} {\path{doi:10.3390/app11146483}}.

\bibitem{jensen2022online}
K.~J. Jensen, M.~K. Ebbesen, M.~R. Hansen, Online deflection compensation of a flexible hydraulic loader crane using neural networks and pressure feedback, Robotics 11~(2) (2022) 34.
\newblock \href {https://doi.org/10.3390/robotics11020034} {\path{doi:10.3390/robotics11020034}}.

\bibitem{rouvinen1997deflection}
A.~Rouvinen, H.~Handroos, Deflection compensation of a flexible hydraulic manipulator utilizing neural networks, Mechatronics 7~(4) (1997) 355--368.
\newblock \href {https://doi.org/10.1016/S0957-4158(97)00009-3} {\path{doi:10.1016/S0957-4158(97)00009-3}}.

\bibitem{pezeshki2023state}
H.~Pezeshki, H.~Adeli, D.~Pavlou, S.~C. Siriwardane, State of the art in structural health monitoring of offshore and marine structures, in: Proceedings of the Institution of Civil Engineers-Maritime Engineering, Vol. 176, Thomas Telford Ltd, 2023, pp. 89--108.
\newblock \href {https://doi.org/10.1680/jmaen.2022.027} {\path{doi:10.1680/jmaen.2022.027}}.

\bibitem{fernandez2021long}
I.~Fernandez, C.~G. Berrocal, R.~Rempling, Long-term performance of distributed optical fiber sensors embedded in reinforced concrete beams under sustained deflection and cyclic loading, Sensors 21~(19) (2021) 6338.
\newblock \href {https://doi.org/10.3390/s21196338} {\path{doi:10.3390/s21196338}}.

\bibitem{grundkotter2022precision}
E.~Grundk{\"o}tter, J.~Melbert, Precision blade deflection measurement system using wireless inertial sensor nodes, Wind Energy 25~(3) (2022) 432--449.
\newblock \href {https://doi.org/10.1002/we.2680} {\path{doi:10.1002/we.2680}}.

\bibitem{kot2021recent}
P.~Kot, M.~Muradov, M.~Gkantou, G.~S. Kamaris, K.~Hashim, D.~Yeboah, Recent advancements in non-destructive testing techniques for structural health monitoring, Applied Sciences 11~(6) (2021) 2750.
\newblock \href {https://doi.org/10.3390/app11062750} {\path{doi:10.3390/app11062750}}.

\bibitem{chen2023experimental}
Y.~Chen, D.~T. Griffith, Experimental and numerical full-field displacement and strain characterization of wind turbine blade using a {3D} scanning laser doppler vibrometer, Optics \& Laser Technology 158 (2023) 108869.
\newblock \href {https://doi.org/10.1016/j.optlastec.2022.108869} {\path{doi:10.1016/j.optlastec.2022.108869}}.

\bibitem{rasmussen2022non}
S.~Rasmussen, J.~A. Krarup, G.~Hildebrand, Non-contact deflection measurement at high speed, in: Bearing Capacity Of Roads Volume 1, CRC Press, 2022, pp. 53--60.
\newblock \href {https://doi.org/10.1201/9781003078814-8} {\path{doi:10.1201/9781003078814-8}}.

\bibitem{sun2018fuzzy}
C.~Sun, H.~Gao, W.~He, Y.~Yu, Fuzzy neural network control of a flexible robotic manipulator using assumed mode method, IEEE transactions on neural networks and learning systems 29~(11) (2018) 5214--5227.
\newblock \href {https://doi.org/10.1109/TNNLS.2017.2743103} {\path{doi:10.1109/TNNLS.2017.2743103}}.

\bibitem{shi2023dynamics}
M.~Shi, B.~Rong, J.~Liang, W.~Zhao, H.~Pan, Dynamics analysis and vibration suppression of a spatial rigid-flexible link manipulator based on transfer matrix method of multibody system, Nonlinear Dynamics 111~(2) (2023) 1139--1159.
\newblock \href {https://doi.org/10.1007/s11071-022-07921-6} {\path{doi:10.1007/s11071-022-07921-6}}.

\bibitem{lochan2017robust}
K.~Lochan, B.~K. Roy, B.~Subudhi, Robust tip trajectory synchronisation between assumed modes modelled two-link flexible manipulators using second-order pid terminal smc, Robotics and Autonomous Systems 97 (2017) 108--124.
\newblock \href {https://doi.org/10.1016/j.robot.2017.08.008} {\path{doi:10.1016/j.robot.2017.08.008}}.

\bibitem{meng2021motion}
Q.~Meng, X.~Lai, Z.~Yan, C.-Y. Su, M.~Wu, Motion planning and adaptive neural tracking control of an uncertain two-link rigid--flexible manipulator with vibration amplitude constraint, IEEE Transactions on Neural Networks and Learning Systems 33~(8) (2021) 3814--3828.
\newblock \href {https://doi.org/10.1109/TNNLS.2021.3054611} {\path{doi:10.1109/TNNLS.2021.3054611}}.

\bibitem{shabana2020dynamics}
A.~A.~Shabana, \href{www.cambridge.org/9781107042650}{Dynamics of Multibody Systems}, Cambridge university press, Cambridge, 2020.
\newline\urlprefix\url{www.cambridge.org/9781107042650}

\bibitem{zwolfer2019co}
A.~Zw{\"o}lfer, J.~Gerstmayr, Co-rotational formulations for {3D} flexible multibody systems: a nodal-based approach, Contributions to Advanced Dynamics and Continuum Mechanics (2019) 243--263\href {https://doi.org/10.1007/978-3-030-21251-3_14} {\path{doi:10.1007/978-3-030-21251-3_14}}.

\bibitem{zwolfer2020concise}
A.~Zw{\"o}lfer, J.~Gerstmayr, A concise nodal-based derivation of the floating frame of reference formulation for displacement-based solid finite elements: Avoiding inertia shape integrals, Multibody System Dynamics 49~(3) (2020) 291--313.
\newblock \href {https://doi.org/10.1007/s11044-019-09716-x} {\path{doi:10.1007/s11044-019-09716-x}}.

\bibitem{zwolfer2021nodal}
A.~Zw{\"o}lfer, J.~Gerstmayr, The nodal-based floating frame of reference formulation with modal reduction: How to calculate the invariants without a lumped mass approximation, Acta Mechanica 232 (2021) 835--851.
\newblock \href {https://doi.org/https://doi.org/10.1007/s00707-020-02886-2} {\path{doi:https://doi.org/10.1007/s00707-020-02886-2}}.

\bibitem{2012_krizhevsky_AlexnetPaper}
A.~Krizhevsky, I.~Sutskever, G.~E. Hinton, Imagenet classification with deep convolutional neural networks, Advances in neural information processing systems 60~(6) (2012) 84--90.
\newblock \href {https://doi.org/10.1145/3065386} {\path{doi:10.1145/3065386}}.

\bibitem{2016_He_deepResidualLearningForImageRecognition}
K.~He, X.~Zhang, S.~Ren, J.~Sun, Deep residual learning for image recognition, in: Proceedings of the IEEE conference on computer vision and pattern recognition, 2016, pp. 770--778.
\newblock \href {https://doi.org/10.48550/arXiv.1512.03385} {\path{doi:10.48550/arXiv.1512.03385}}.

\bibitem{2013_Mnih_PlayingAtari_DQN}
V.~Mnih, K.~Kavukcuoglu, D.~Silver, A.~Graves, I.~Antonoglou, D.~Wierstra, M.~Riedmiller, Playing atari with deep reinforcement learning, arXiv preprint:1312.5602 (2013).
\newblock \href {https://doi.org/10.48550/arXiv.1312.5602} {\path{doi:10.48550/arXiv.1312.5602}}.

\bibitem{2017_Vaswani_attentionIsAllYouNeed}
A.~Vaswani, N.~Shazeer, N.~Parmar, J.~Uszkoreit, L.~Jones, A.~N. Gomez, {\L}.~Kaiser, I.~Polosukhin, Attention is all you need, Advances in neural information processing systems 30 (2017).
\newblock \href {https://doi.org/10.5555/3295222.3295349} {\path{doi:10.5555/3295222.3295349}}.

\bibitem{2021_Cai_PINNsForFluidDynamics}
S.~Cai, Z.~Mao, Z.~Wang, M.~Yin, G.~E. Karniadakis, Physics-informed neural networks ({PINN}s) for fluid mechanics: A review, Acta Mechanica Sinica 37~(12) (2021) 1727--1738.
\newblock \href {https://doi.org/10.1007/s10409-021-01148-1} {\path{doi:10.1007/s10409-021-01148-1}}.

\bibitem{2021_Cai_PINNsHeatTransfer}
S.~Cai, Z.~Wang, S.~Wang, P.~Perdikaris, G.~E. Karniadakis, Physics-informed neural networks for heat transfer problems, Journal of Heat Transfer 143~(6) (2021) 060801.
\newblock \href {https://doi.org/10.1115/1.4050542} {\path{doi:10.1115/1.4050542}}.

\bibitem{hashemiMultibodyDynamicsControl2023}
A.~Hashemi, G.~Orzechowski, A.~Mikkola, J.~McPhee, Multibody dynamics and control using machine learning, Multibody System Dynamics 58~(3) (2023) 397--431.
\newblock \href {https://doi.org/10.1007/s11044-023-09884-x} {\path{doi:10.1007/s11044-023-09884-x}}.

\bibitem{gerstmayrMultibodyModelsGenerated2024}
J.~Gerstmayr, P.~Manzl, M.~Pieber, Multibody {Models} {Generated} from {Natural} {Language}, Multibody System Dynamics 62~(2) (2024) 249--271.
\newblock \href {https://doi.org/10.1007/s11044-023-09962-0} {\path{doi:10.1007/s11044-023-09962-0}}.

\bibitem{goRapidlyTrainedDNN2024}
M.-S. Go, Y.-B. Kim, J.-H. Park, J.~H. Lim, J.-G. Kim, A rapidly trained {DNN} model for real-time flexible multibody dynamics simulations with a fixed-time increment, Engineering with Computers 40~(5) (2024) 3131--3156.
\newblock \href {https://doi.org/10.1007/s00366-024-01962-8} {\path{doi:10.1007/s00366-024-01962-8}}.

\bibitem{1989_Hornik_FFNareUniversalFunApproximators}
K.~Hornik, M.~Stinchcombe, H.~White, Multilayer feedforward networks are universal approximators, Neural networks 2~(5) (1989) 359--366.
\newblock \href {https://doi.org/10.1016/0893-6080(89)90020-8} {\path{doi:10.1016/0893-6080(89)90020-8}}.

\bibitem{omisore2021kinematics}
O.~M. Omisore, L.~Wang, Kinematics constraint modeling for flexible robots based on deep learning, in: 2021 43rd Annual International Conference of the IEEE Engineering in Medicine \& Biology Society (EMBC), IEEE, 2021, pp. 4940--4943.
\newblock \href {https://doi.org/10.1109/EMBC46164.2021.9630418} {\path{doi:10.1109/EMBC46164.2021.9630418}}.

\bibitem{lutter2023combining}
M.~Lutter, J.~Peters, Combining physics and deep learning to learn continuous-time dynamics models, The International Journal of Robotics Research 42~(3) (2023) 83--107.
\newblock \href {https://doi.org/10.1177/02783649231169492} {\path{doi:10.1177/02783649231169492}}.

\bibitem{khadim2024simulation}
Q.~Khadim, E.~Kurvinen, A.~Mikkola, G.~Orzechowski, Simulation-driven universal surrogates of coupled mechanical systems: Real-time simulation of a forestry crane, Journal of Computational and Nonlinear Dynamics 19~(7) (2024) 071003.
\newblock \href {https://doi.org/10.1115/1.4065015} {\path{doi:10.1115/1.4065015}}.

\bibitem{slimak2025machine}
T.~Slimak, A.~Zw{\"o}lfer, B.~Todorov, D.~Rixen, A machine learning approach to simulate flexible body dynamics, Multibody System Dynamics (2025) 1--27\href {https://doi.org/10.1007/s11044-024-10049-7} {\path{doi:10.1007/s11044-024-10049-7}}.

\bibitem{manzl2024slide}
P.~Manzl, A.~Humer, Q.~Khadim, J.~Gerstmayr, Slide: A machine-learning based method for forced dynamic response estimation of multibody systems, arXiv preprint:2409.18272 Under Review (2024).
\newblock \href {https://doi.org/10.48550/arXiv.2409.18272} {\path{doi:10.48550/arXiv.2409.18272}}.

\bibitem{2021_Angeli_DeepLearningMinimalCoordinates}
A.~Angeli, W.~Desmet, F.~Naets, Deep learning for model order reduction of multibody systems to minimal coordinates, Computer Methods in Applied Mechanics and Engineering 373 (2021) 113517.
\newblock \href {https://doi.org/https://doi.org/10.1016/j.cma.2020.113517} {\path{doi:https://doi.org/10.1016/j.cma.2020.113517}}.

\bibitem{2021_Choi_dataDrivenSimulation_DeepNerualNetworks}
H.-S. Choi, J.~An, S.~Han, J.-G. Kim, J.-Y. Jung, J.~Choi, G.~Orzechowski, A.~Mikkola, J.~H. Choi, Data-driven simulation for general-purpose multibody dynamics using deep neural networks, Multibody System Dynamics 51 (2021) 419--454.
\newblock \href {https://doi.org/10.1007/s11044-020-09772-8} {\path{doi:10.1007/s11044-020-09772-8}}.

\bibitem{2024_Slimak_overviewDesignConsiderationDataDrivenTimeStepping}
T.~Slimak, A.~Zw{\"o}lfer, B.~Todorov, D.~J. Rixen, Overview of design considerations for data-driven time-stepping schemes applied to nonlinear mechanical systems, Journal of Computational and Nonlinear Dynamics 19~(7) (2024).
\newblock \href {https://doi.org/10.1115/1.4065728} {\path{doi:10.1115/1.4065728}}.

\bibitem{2021_Han_DNNFlexible}
S.~Han, H.-S. Choi, J.~Choi, J.~H. Choi, J.-G. Kim, A {DNN}-based data-driven modeling employing coarse sample data for real-time flexible multibody dynamics simulations, Computer Methods in Applied Mechanics and Engineering 373 (2021) 113480.
\newblock \href {https://doi.org/10.1016/j.cma.2020.113480} {\path{doi:10.1016/j.cma.2020.113480}}.

\bibitem{roehrl2020modeling}
M.~A. Roehrl, T.~A. Runkler, V.~Brandtstetter, M.~Tokic, S.~Obermayer, Modeling system dynamics with physics-informed neural networks based on lagrangian mechanics, IFAC-PapersOnLine 53~(2) (2020) 9195--9200.
\newblock \href {https://doi.org/10.1016/j.ifacol.2020.12.2182} {\path{doi:10.1016/j.ifacol.2020.12.2182}}.

\bibitem{bergstra2012random}
J.~Bergstra, Y.~Bengio, Random search for hyper-parameter optimization, Journal of machine learning research 13~(2) (2012) 281–305.
\newblock \href {https://doi.org/10.5555/2188385.2188395} {\path{doi:10.5555/2188385.2188395}}.

\bibitem{kingma2014adam}
D.~P. Kingma, J.~Ba, Adam: A method for stochastic optimization, arXiv preprint:1412.6980 (2014).
\newblock \href {https://doi.org/https://doi.org/10.48550/arXiv.1412.6980} {\path{doi:https://doi.org/10.48550/arXiv.1412.6980}}.

\bibitem{Johannes2024}
J.~Gerstmayr, S.~Holzinger, A.~Zw{\"o}lfer, From {3D} solid finite elements to reduced flexible multibody bodies with constraint interfaces: A holistic approach, Journal of Sound and Vibration Under Review (2025).

\bibitem{brown2016continuous}
P.~Brown, J.~McPhee, A continuous velocity-based friction model for dynamics and control with physically meaningful parameters, Journal of Computational and Nonlinear Dynamics 11~(5) (2016) 054502.
\newblock \href {https://doi.org/10.1115/1.4033658} {\path{doi:10.1115/1.4033658}}.

\bibitem{watton2009fundamentals}
J.~Watton, Fluid Power systems: Modeling, Simulation, Analog, and Microcomputer Control, Prentice Hall, New York, 1989.
\newblock \href {https://doi.org/10.5555/73931} {\path{doi:10.5555/73931}}.

\bibitem{gerstmayr2023exudyn}
J.~Gerstmayr, Exudyn--~{A C++ based Python package for flexible multibody systems}, Multibody System Dynamics 60~(4) (2023) 533--561.
\newblock \href {https://doi.org/10.1007/s11044-023-09937-1} {\path{doi:10.1007/s11044-023-09937-1}}.

\bibitem{KHADIM2023105405}
Q.~Khadim, Y.~S. Hagh, D.~Jiang, L.~Pyrhönen, S.~Jaiswal, V.~Zhidchenko, X.~Yu, E.~Kurvinen, H.~Handroos, A.~Mikkola, Experimental investigation into the state estimation of a forestry crane using the unscented kalman filter and a multiphysics model, Mechanism and Machine Theory 189 (2023) 105405.
\newblock \href {https://doi.org/10.1016/j.mechmachtheory.2023.105405} {\path{doi:10.1016/j.mechmachtheory.2023.105405}}.

\bibitem{harris2020array}
C.~R. Harris, K.~J. Millman, S.~J. Van Der~Walt, R.~Gommers, P.~Virtanen, D.~Cournapeau, E.~Wieser, J.~Taylor, S.~Berg, N.~J. Smith, et~al., Array programming with numpy, Nature 585~(7825) (2020) 357--362.
\newblock \href {https://doi.org/10.1038/s41586-020-2649-2} {\path{doi:10.1038/s41586-020-2649-2}}.

\bibitem{2024_Ansel_Pytorch2}
J.~Ansel, E.~Yang, H.~He, N.~Gimelshein, A.~Jain, M.~Voznesensky, B.~Bao, P.~Bell, D.~Berard, E.~Burovski, et~al., Pytorch 2: Faster machine learning through dynamic python bytecode transformation and graph compilation, in: Proceedings of the 29th ACM International Conference on Architectural Support for Programming Languages and Operating Systems, Volume 2, 2024, pp. 929--947.
\newblock \href {https://doi.org/10.1145/3620665.3640366} {\path{doi:10.1145/3620665.3640366}}.

\end{thebibliography}


@article{bissadu2024society,
  title={Society 5.0 enabled agriculture: Drivers, enabling technologies, architectures, opportunities, and challenges},
  author={Bissadu, Kossi Dodzi and 
          Sonko, Salleh and 
          Hossain, Gahangir},
  journal={Information Processing in Agriculture},
  doi = {10.1016/j.inpa.2024.04.003},
  year={2024},
  publisher={Elsevier}
}

@article{khadim2020targeting,
  title={Targeting the user experience in the development of mobile machinery using real-time multibody simulation},
  author={Khadim, Qasim and Kaikko, Esa-Pekka and Puolatie, Eero and Mikkola, Aki},
  journal={Advances in Mechanical Engineering},
  volume={12},
  number={6},
  pages={1687814020923176},
  DOI  = {10.1177/1687814020923176},
  year={2020},
  publisher={SAGE Publications Sage UK: London, England}
}

@article{billard2019trends,
  title={Trends and challenges in robot manipulation},
  author={Billard, Aude and Kragic, Danica},
  journal={Science},
  volume={364},
  number={6446},
  pages={eaat8414},
  year={2019},
  DOI = {10.1126/science.aat8414},
  publisher={American Association for the Advancement of Science}
}


@article{sanchez2020innovation,
  title={Innovation in the mining industry: Technological trends and a case study of the challenges of disruptive innovation},
  author={S{\'a}nchez, Felipe and 
            Hartlieb, Philipp},
  journal={Mining, Metallurgy \& Exploration},
  volume={37},
  number={5},
  pages={1385--1399},
  DOI = {10.1007/s42461-020-00262-1}, 
  year={2020},
  publisher={Springer}
}

@article{lajunen2018overview,
  title={Overview of powertrain electrification and future scenarios for non-road mobile machinery},
  author={Lajunen, Antti and 
      Sainio, Panu and 
      Laurila, Lasse and 
      Pippuri-M{\"a}kel{\"a}inen, Jenni 
                and Tammi, Kari},
  journal={Energies},
  volume={11},
  number={5},
  pages={1184},
  year={2018},
  DOI = {10.3390/en11051184},
  publisher={MDPI}
}




@article{mann2023benign,
  title={Benign effects of automation: New evidence from patent texts},
  author={Mann, Katja and P{\"u}ttmann, Lukas},
  journal={Review of Economics and Statistics},
  volume={105},
  number={3},
  pages={562--579},
  DOI = {10.1162/rest_a_01083},
  year={2023},
  publisher={MIT Press One Rogers Street, Cambridge, MA 02142-1209, USA journals-info~…}
}


@article{haggerty2023control,
  title={Control of soft robots with inertial dynamics},
  author={Haggerty, David A and Banks, Michael J and Kamenar, Ervin and Cao, Alan B and Curtis, Patrick C and Mezi{\'c}, Igor and Hawkes, Elliot W},
  journal={Science robotics},
  volume={8},
  number={81},
  pages={eadd6864},
  year={2023},
  doi = {10.1126/scirobotics.add6864},
  publisher={American Association for the Advancement of Science}
}


@article{physicsbasedDT,
  author={Kurvinen, Emil and 
          Kutvonen, Antero and Ukko, Juhani and 
          Khadim, Qasim and Hagh, Yashar Shabbouei and others},
  journal={IEEE Access}, 
  title={Physics-Based Digital Twins Merging With Machines: Cases of Mobile Log Crane and Rotating Machine}, 
  year={2022},
  volume={10},
  number={},
  pages={45962-45978},
  keywords={Kalman filters;Finite element analysis;Computational modeling;Real-time systems;Mathematical models;Digital twin;Predictive models;Multibody simulation;finite element method;Kalman filter;state estimation;parameter estimation;physics-based simulation},
  doi={10.1109/ACCESS.2022.3170430}}

@article{sarkhel2023robust,
  title={Robust deflection control and analysis of a fishing rod-type flexible robotic manipulator for collaborative robotics},
  author={Sarkhel, Prasenjit and 
          Dikshit, Mithilesh K and 
          Pathak, Vimal Kumar and 
          Saxena, Kuldeep K and 
          Prakash, C and Buddhi, Dharam},
  journal={Robotics and Autonomous Systems},
  volume={159},
  pages={104293},
  year={2023},
  DOI = {10.1016/j.robot.2022.104293},
  publisher={Elsevier}
}


@ARTICLE{10438059,
  author={Li, Bai and Li, Xinyuan and Gao, Hejia and Wang, Fei-Yue},
  journal={IEEE/ASME Transactions on Mechatronics}, 
  title={Advances in Flexible Robotic Manipulator Systems—Part I: Overview and Dynamics Modeling Methods}, 
  year={2024},
  volume={29},
  number={2},
  pages={1100-1110},
  keywords={Mathematical models;Surveys;Manipulators;Solid modeling;Manipulator dynamics;Aerodynamics;Planning;Artificial intelligence (AI);dynamic modeling;flexibility;flexible robotic manipulator (FRM);vibration suppression},
  doi={10.1109/TMECH.2024.3359067}}

@article{cui2020trajectory,
  title={Trajectory planning of a spatial flexible manipulator for vibration suppression},
  author={Cui, Leilei and Wang, Hesheng and Chen, Weidong},
  journal={Robotics and Autonomous Systems},
  volume={123},
  pages={103316},
 DOI = {10.1016/j.robot.2019.103316},
  year={2020},
  publisher={Elsevier}
}

@article{shi2023dynamics,
  title={Dynamics analysis and vibration suppression of a spatial rigid-flexible link manipulator based on transfer matrix method of multibody system},
  author={Shi, Mingming and Rong, Bao and Liang, Jing and Zhao, Wenlong and Pan, Hongtao},
  journal={Nonlinear Dynamics},
  volume={111},
  number={2},
  pages={1139--1159},
  doi ={10.1007/s11071-022-07921-6},
  year={2023},
  publisher={Springer}
}

@article{lochan2017robust,
  title={Robust tip trajectory synchronisation between assumed modes modelled two-link flexible manipulators using second-order PID terminal SMC},
  author={Lochan, Kshetrimayum and Roy, Binoy Krishna and Subudhi, B},
  journal={Robotics and Autonomous Systems},
  volume={97},
  pages={108--124},
  year={2017},
doi = { 10.1016/j.robot.2017.08.008},
  publisher={Elsevier}
}

@inproceedings{omisore2021kinematics,
  title={Kinematics Constraint Modeling for Flexible Robots based on Deep Learning},
  author={Omisore, Olatunji Mumini and Wang, Lei},
  booktitle={2021 43rd Annual International Conference of the IEEE Engineering in Medicine \& Biology Society (EMBC)},
  pages={4940--4943},
  year={2021},
  doi ={10.1109/EMBC46164.2021.9630418},
  organization={IEEE}
}

@article{czerwinski2021current,
  title={Current trends in automotive lightweighting strategies and materials},
  author={Czerwinski, Frank},
  journal={Materials},
  volume={14},
  number={21},
  pages={6631},
  year={2021},
  doi = {10.3390/ma14216631 },
  publisher={MDPI}
}

@article{roehrl2020modeling,
  title={Modeling system dynamics with physics-informed neural networks based on Lagrangian mechanics},
  author={Roehrl, Manuel A and Runkler, Thomas A and Brandtstetter, Veronika and Tokic, Michel and Obermayer, Stefan},
  journal={IFAC-PapersOnLine},
  doi ={10.1016/j.ifacol.2020.12.2182}, 
  volume={53},
  number={2},
  pages={9195--9200},
  year={2020},
  publisher={Elsevier}
}

@article{lutter2023combining,
  title={Combining physics and deep learning to learn continuous-time dynamics models},
  author={Lutter, Michael and Peters, Jan},
  journal={The International Journal of Robotics Research},
  volume={42},
  number={3},
  pages={83--107},
  year={2023},
  DOI ={10.1177/02783649231169492},
  publisher={SAGE Publications Sage UK: London, England}
}

@article{sayahkarajy2018mode,
  title={Mode shape analysis, modal linearization, and control of an elastic two-link manipulator based on the normal modes},
  author={Sayahkarajy, Mostafa},
  journal={Applied mathematical modelling},
  volume={59},
  pages={546--570},
  doi ={10.1016/j.apm.2018.02.003},
  year={2018},
  publisher={Elsevier}
}

@article{slimak2025machine,
  title={A machine learning approach to simulate flexible body dynamics},
  author={Slimak, Tomas and Zw{\"o}lfer, Andreas and Todorov, Bojidar and Rixen, Daniel},
  journal={Multibody System Dynamics},
  pages={1--27},
  year={2025},
  doi = {10.1007/s11044-024-10049-7},
  publisher={Springer}
}

@article{KHADIM2023105405,
title = {Experimental investigation into the state estimation of a forestry crane using the unscented Kalman filter and a multiphysics model},
journal = {Mechanism and Machine Theory},
volume = {189},
pages = {105405},
year = {2023},
issn = {0094-114X},
author = {Qasim Khadim and Yashar Shabbouei Hagh and Dezhi Jiang and Lauri Pyrhönen and Suraj Jaiswal and Victor Zhidchenko and Xinxin Yu and Emil Kurvinen and Heikki Handroos and Aki Mikkola},
doi ={10.1016/j.mechmachtheory.2023.105405},
keywords = {State observer, Multiphysics model, Unscented Kalman filter, Accurate and smooth estimation, Multibody dynamics, Hydraulics},
}


@article{gerstmayr2023exudyn,
  title={Exudyn--~{A C++ based Python package for flexible multibody systems}},
  author={Gerstmayr, Johannes},
journal={Multibody System Dynamics},
  volume={60},
  number={4},
  pages={533--561},
DOI = {10.1007/s11044-023-09937-1},
  year={2023},
publisher={Springer}
}


@article{manzl2024slide,
  title={SLIDE: A machine-learning based method for forced dynamic response estimation of multibody systems},
  author={Manzl, Peter and Humer, Alexander and Khadim, Qasim and Gerstmayr, Johannes},
  journal={arXiv preprint:2409.18272},
volume={Under Review},
doi = {10.48550/arXiv.2409.18272},
  year={2024}
}


@article{2021_Han_DNNFlexible,
  title={A {DNN}-based data-driven modeling employing coarse sample data for real-time flexible multibody dynamics simulations},
  author={Han, Seongji and Choi, Hee-Sun and Choi, Juhwan and Choi, Jin Hwan and Kim, Jin-Gyun},
  journal={Computer Methods in Applied Mechanics and Engineering},
  volume={373},
  pages={113480},
  doi = {10.1016/j.cma.2020.113480},
  year={2021},
  publisher={Elsevier}
}



@article{2024_Slimak_overviewDesignConsiderationDataDrivenTimeStepping,
  title={Overview of Design Considerations for Data-Driven Time-Stepping Schemes Applied to Nonlinear Mechanical Systems},
  author={Slimak, Tomas and Zw{\"o}lfer, Andreas and Todorov, Bojidar and Rixen, Daniel J},
  journal={Journal of Computational and Nonlinear Dynamics},
  volume={19},
  number={7},
  year={2024},
  doi = {10.1115/1.4065728},
  publisher={American Society of Mechanical Engineers Digital Collection}
}

@article{pan2022sensor,
  title={Sensor placement and seismic response reconstruction for structural health monitoring using a deep neural network},
  author={Pan, Yuxin and Ventura, Carlos E and Li, Teng},
  journal={Bulletin of Earthquake Engineering},
  volume={20},
  number={9},
  pages={4513--4532},
  doi = {10.1007/s10518-021-01266-y},
  year={2022},
  publisher={Springer}
}

@inproceedings{2024_Ansel_Pytorch2,
  title={Pytorch 2: Faster machine learning through dynamic python bytecode transformation and graph compilation},
  author={Ansel, Jason and Yang, Edward and He, Horace and Gimelshein, Natalia and Jain, Animesh and Voznesensky, Michael and Bao, Bin and Bell, Peter and Berard, David and Burovski, Evgeni and others},
  booktitle={Proceedings of the 29th ACM International Conference on Architectural Support for Programming Languages and Operating Systems, Volume 2},
  pages={929--947},
  year={2024}, 
	doi={10.1145/3620665.3640366}
}


@book{oliphant2006guide,
  title={Guide to numpy},
  author={Oliphant, Travis E and others},
  volume={1},
  year={2006},
  publisher={Trelgol Publishing USA}
}

@misc{mining2024,
  title = {Mining Market Report},
  author = {The Business Research Company},
  year = {2024},
  url = {https://www.thebusinessresearchcompany.com/report/mining-and-oil-and-gas-field-machinery-manufacturing-global-market-report},
  note = {Accessed: 2024-10-16}
}

@misc{forestry2024,
  title = {Forestry Machinery Market Report},
  author = {The Business Research Company},
  year = {2024},
  url = {https://www.thebusinessresearchcompany.com/report/forestry-machinery-global-market-report},
  note = {Accessed: 2024-10-16}
}

@misc{agriculture2024,
  title = {Agriculture Machinery Market Report},
  author = {The Business Research Company},
  year = {2024},
  url = {https://www.thebusinessresearchcompany.com/report/agriculture-machinery-global-market-report},
  note = {Accessed: 2024-10-16}
}

@misc{automotive2024,
  title = {Autonomous Construction Equipment Market},
  author = {The Business Research Company},
  year = {2024},
  url = {https://www.thebusinessresearchcompany.com/report/autonomous-construction-equipment-market},
  note = {Accessed: 2024-10-16}
}

@misc{transportation2024,
  title = {Third-Party Logistics (3PL) Market Report},
  author = {The Business Research Company},
  year = {2024},
  url = {https://www.thebusinessresearchcompany.com/report/third-party-logistics-3pl-global-market-report},
  note = {Accessed: 2024-10-16}
}

@misc{railways2024,
  title = {Railways System Market Report},
  author = {Expert Market Research},
  year = {2024},
  url = {https://www.expertmarketresearch.com/reports/railway-system-market-report},
  note = {Accessed: 2024-10-16}
}


@article{bergstra2012random,
  title={Random search for hyper-parameter optimization},
  author={Bergstra, James and Bengio, Yoshua},
  journal={Journal of machine learning research},
  volume={13},
  number={2},
pages = {281–305},
doi = {10.5555/2188385.2188395},
  year={2012}
}

@article{zwolfer2021nodal,
  title={The nodal-based floating frame of reference formulation with modal reduction: How to calculate the invariants without a lumped mass approximation},
  author={Zw{\"o}lfer, Andreas and Gerstmayr, Johannes},
  journal={Acta Mechanica},
  volume={232},
  pages={835--851},
  year={2021},
  DOI = {https://doi.org/10.1007/s00707-020-02886-2},
publisher={Springer}
}

@article{2017_Vaswani_attentionIsAllYouNeed,
  title={Attention is all you need},
  author={Vaswani, Ashish and Shazeer, Noam and Parmar, Niki and Uszkoreit, Jakob and Jones, Llion and Gomez, Aidan N and Kaiser, {\L}ukasz and Polosukhin, Illia},
  journal={Advances in neural information processing systems},
  volume={30},
  year={2017},
  DOI ={10.5555/3295222.3295349}
}

@article{meng2021motion,
  title={Motion planning and adaptive neural tracking control of an uncertain two-link rigid--flexible manipulator with vibration amplitude constraint},
  author={Meng, Qingxin and Lai, Xuzhi and Yan, Ze and Su, Chun-Yi and Wu, Min},
  journal={IEEE Transactions on Neural Networks and Learning Systems},
  volume={33},
  number={8},
  pages={3814--3828},
  year={2021},
  publisher={IEEE},
doi = {10.1109/TNNLS.2021.3054611},
}

@article{Johannes2024,
  title={From {3D} solid finite elements to reduced flexible
multibody bodies with constraint interfaces: A
holistic approach},
  author={ Gerstmayr, Johannes and Holzinger, Stefan and  Zw{\"o}lfer, Andreas},
  journal={Journal of Sound and Vibration},
  volume={Under Review},
  pages={},
  year={2025},
  DOI = {},
publisher={Springer}
}

@article{kingma2014adam,
  title={Adam: A method for stochastic optimization},
  author={Kingma, Diederik P and Ba, Jimmy},
  journal={arXiv preprint:1412.6980},
  volume={},
  pages = {},
  DOI   = {
https://doi.org/10.48550/arXiv.1412.6980
},
year={2014}
}

@BOOK{watton2009fundamentals,
  title 	= {Fluid Power systems: Modeling, Simulation, Analog, and Microcomputer Control},
  publisher = {Prentice Hall},
  year 		= {1989},
  author 	= {Watton, John},
  address 	= {New York},
  Doi 		= {10.5555/73931},
  }


@article{brown2016continuous,
  title={A continuous velocity-based friction model for dynamics and control with physically meaningful parameters},
  author={Brown, Peter and McPhee, John},
  journal={Journal of Computational and Nonlinear Dynamics},
  volume={11},
  number={5},
  pages={054502},
   DOI = {10.1115/1.4033658 },
  year={2016},
  publisher={American Society of Mechanical Engineers}
}

@BOOK{shabana2020dynamics,
  title 	= {Dynamics of Multibody Systems},
  publisher = {Cambridge university press},
  year 		= {2020},
  author 	= {A. Shabana, Ahmed},
  address 	= {Cambridge},
  URL 		={www.cambridge.org/9781107042650},
  }

@manual{FunctionBay2024,
  author    = {{FunctionBay, Inc.}},
  title     = {{Recurdyn Online Help}},
  year      = {2024},
  note      = {Accessed on: November 04, 2024},
  howpublished = {\url{https://functionbay.com/documentation/onlinehelp/default.htm}}
}

@manual{MSCSoftware2024,
  author    = {{MSC Software Corporation}},
  title     = {{Theory of Flexible Bodies}},
  year      = {2024},
  note      = {Adams/Flex documentation. Accessed on: November 04, 2024},
  howpublished = {\url{https://simcompanion.mscsoftware.com/}}
}

@article{zwolfer2019co,
  title={Co-rotational formulations for {3D} flexible multibody systems: a nodal-based approach},
  author={Zw{\"o}lfer, Andreas and Gerstmayr, Johannes},
  journal={Contributions to Advanced Dynamics and Continuum Mechanics},
  pages={243--263},
  year={2019},
  DOI = {10.1007/978-3-030-21251-3_14},
  publisher={Springer}
}



@article{zwolfer2020concise,
  title={A concise nodal-based derivation of the floating frame of reference formulation for displacement-based solid finite elements: Avoiding inertia shape integrals},
  author={Zw{\"o}lfer, Andreas and Gerstmayr, Johannes},
  journal={Multibody System Dynamics},
  volume={49},
  number={3},
  pages={291--313},
  year={2020},
  publisher={Springer},
  DOI = {10.1007/s11044-019-09716-x},
}

@article{jensen2022online,
  title={Online Deflection Compensation of a Flexible Hydraulic Loader Crane Using Neural Networks and Pressure Feedback},
  author={Jensen, Konrad Johan and Ebbesen, Morten Kjeld and Hansen, Michael Rygaard},
  journal={Robotics},
  volume={11},
  number={2},
  pages={34},
  year={2022},
  doi = {10.3390/robotics11020034},
  publisher={MDPI}
}

@article{khadim2024simulation,
    author = {Khadim, Qasim and Kurvinen, Emil and Mikkola, Aki and Orzechowski, Grzegorz},
    title = {Simulation-Driven Universal Surrogates of Coupled Mechanical Systems: Real-Time Simulation of a Forestry Crane},
    journal = {Journal of Computational and Nonlinear Dynamics},
    volume = {19},
    number = {7},
    pages = {071003},
    year = {2024},
    month = {05},
doi = {10.1115/1.4065015},
  publisher={American Society of Mechanical Engineers}
}

@article{jensen2021development,
  title={Development of {3D} Anti-Swing Control for Hydraulic Knuckle Boom Crane},
  author={Jensen, Konrad J and Ebbesen, Morten K and Hansen, Michael R},
  year={2021}
}



@article{2012_krizhevsky_AlexnetPaper,
  title={Imagenet classification with deep convolutional neural networks},
  author={Krizhevsky, Alex and Sutskever, Ilya and Hinton, Geoffrey E},
  journal={Advances in neural information processing systems},
  volume={60},
  number={6},
  pages={84--90},
  doi = {10.1145/3065386},
  year={2012}
}

@inproceedings{2016_He_deepResidualLearningForImageRecognition,
  title={Deep residual learning for image recognition},
  author={He, Kaiming and Zhang, Xiangyu and Ren, Shaoqing and Sun, Jian},
  booktitle={Proceedings of the IEEE conference on computer vision and pattern recognition},
  pages={770--778},
  doi = {10.48550/arXiv.1512.03385
},
  year={2016}
}

@article{2013_Mnih_PlayingAtari_DQN,
  title={Playing atari with deep reinforcement learning},
  author={Mnih, Volodymyr and Kavukcuoglu, Koray and Silver, David and Graves, Alex and Antonoglou, Ioannis and Wierstra, Daan and Riedmiller, Martin},
  journal={arXiv preprint:1312.5602},
  doi = {10.48550/arXiv.1312.5602},
  year={2013}
}

@article{2021_Cai_PINNsForFluidDynamics,
  title={Physics-informed neural networks ({PINN}s) for fluid mechanics: A review},
  author={Cai, Shengze and Mao, Zhiping and Wang, Zhicheng and Yin, Minglang and Karniadakis, George Em},
  journal={Acta Mechanica Sinica},
  volume={37},
  number={12},
  pages={1727--1738},
  doi = {10.1007/s10409-021-01148-1},
  year={2021},
  publisher={Springer}
}


@article{2021_Cai_PINNsHeatTransfer,
  title={Physics-informed neural networks for heat transfer problems},
  author={Cai, Shengze and Wang, Zhicheng and Wang, Sifan and Perdikaris, Paris and Karniadakis, George Em},
  journal={Journal of Heat Transfer},
  volume={143},
  number={6},
  pages={060801},
  doi = {10.1115/1.4050542},
  year={2021},
  publisher={American Society of Mechanical Engineers}
}

@article{2021_Angeli_DeepLearningMinimalCoordinates,
title = {Deep learning for model order reduction of multibody systems to minimal coordinates},
journal = {Computer Methods in Applied Mechanics and Engineering},
volume = {373},
pages = {113517},
year = {2021},
issn = {0045-7825},
doi = {https://doi.org/10.1016/j.cma.2020.113517},
author = {Andrea Angeli and Wim Desmet and Frank Naets},
keywords = {Multibody dynamics, Minimal coordinates, Deep learning, Model order reduction}
}


@article{2021_Choi_dataDrivenSimulation_DeepNerualNetworks,
  title={Data-driven simulation for general-purpose multibody dynamics using Deep Neural Networks},
  author={Choi, Hee-Sun and An, Junmo and Han, Seongji and Kim, Jin-Gyun and Jung, Jae-Yoon and Choi, Juhwan and Orzechowski, Grzegorz and Mikkola, Aki and Choi, Jin Hwan},
  journal={Multibody System Dynamics},
  volume={51},
  pages={419--454},
  year={2021},
  doi= {10.1007/s11044-020-09772-8},
  publisher={Springer}
}

@article{1989_Hornik_FFNareUniversalFunApproximators,
  title={Multilayer feedforward networks are universal approximators},
  author={Hornik, Kurt and Stinchcombe, Maxwell and White, Halbert},
  journal={Neural networks},
  volume={2},
  number={5},
  pages={359--366},
  doi = {10.1016/0893-6080(89)90020-8},
  year={1989},
  publisher={Elsevier}
}

@article{dwivedy2006dynamic,
  title={Dynamic analysis of flexible manipulators, a literature review},
  author={Dwivedy, Santosha Kumar and Eberhard, Peter},
  journal={Mechanism and machine theory},
  volume={41},
  number={7},
  pages={749--777},
  year={2006},
  doi = {10.1016/j.mechmachtheory.2006.01.014},
  publisher={Elsevier}
}

@article{arkouli2021towards,
  title={Towards accurate robot modelling of flexible robotic manipulators},
  author={Arkouli, Z and Aivaliotis, P and Makris, S},
  journal={Procedia CIRP},
  volume={97},
  pages={497--501},
  doi ={10.1016/j.procir.2020.07.009},
  year={2021},
  publisher={Elsevier}
}

@article{avci2021review,
  title={A review of vibration-based damage detection in civil structures: From traditional methods to Machine Learning and Deep Learning applications},
  author={Avci, Onur and Abdeljaber, Osama and Kiranyaz, Serkan and Hussein, Mohammed and Gabbouj, Moncef and Inman, Daniel J},
  journal={Mechanical systems and signal processing},
  volume={147},
  pages={107077},
  year={2021},
   doi = {10.1016/j.ymssp.2020.107077},
  publisher={Elsevier}
}

@article{aldakheel2021feed,
  title={Feed-forward neural networks for failure mechanics problems},
  author={Aldakheel, Fadi and Satari, Ramish and Wriggers, Peter},
  journal={Applied Sciences},
  volume={11},
  number={14},
  pages={6483},
  year={2021},
  doi ={10.3390/app11146483}, 
  publisher={MDPI}
}


@article{malekloo2022machine,
  title={Machine learning and structural health monitoring overview with emerging technology and high-dimensional data source highlights},
  author={Malekloo, Arman and Ozer, Ekin and AlHamaydeh, Mohammad and Girolami, Mark},
  journal={Structural Health Monitoring},
  volume={21},
  number={4},
  pages={1906--1955},
  doi ={10.1177/14759217211036880},
  year={2022},
  publisher={SAGE Publications Sage UK: London, England}
}

@article{lee2020critical,
  title={A critical review of modelling methods for flexible and rigid link manipulators},
  author={Lee, Tian Soon and Alandoli, Esmail Ali},
  journal={Journal of the Brazilian Society of Mechanical Sciences and Engineering},
  volume={42},
  pages={1--14},
  year={2020},
  doi = {10.1007/s40430-020-02602-0},
  publisher={Springer}
}

@article{sun2018fuzzy,
  title={Fuzzy neural network control of a flexible robotic manipulator using assumed mode method},
  author={Sun, Changyin and Gao, Hejia and He, Wei and Yu, Yao},
  journal={IEEE transactions on neural networks and learning systems},
  volume={29},
  number={11},
  pages={5214--5227},
  year={2018},
  doi = {10.1109/TNNLS.2017.2743103},
  publisher={IEEE}
}


@article{uyar2023implementation,
  title={Implementation of active and passive vibration control of flexible smart composite manipulators with genetic algorithm},
  author={Uyar, M and Malgaca, LEVENT},
  journal={Arabian Journal for Science and Engineering},
  volume={48},
  number={3},
  pages={3843--3862},
  year={2023},
  doi = {10.1007/s13369-022-07279-2},
  publisher={Springer}
}

@book{rigatos2018robotic,
  title={Robotic manipulators and vehicles: control, estimation and filtering},
  author={Rigatos, Gerasimos and Busawon, Krishna},
  volume={152},
  year={2018},
  doi ={10.1007/978-3-319-77851-8},  
  publisher={Springer}
}

@book{wagg2010nonlinear,
  title={Nonlinear vibration with control: for flexible and adaptive structures},
  author={Wagg, David and Neild, Simon},
  year={2010},
  doi= {10.1007/978-90-481-2837-2_2},
  publisher={Springer}
}

@inproceedings{pezeshki2023state,
  title={State of the art in structural health monitoring of offshore and marine structures},
  author={Pezeshki, Hadi and Adeli, Hojjat and Pavlou, Dimitrios and Siriwardane, Sudath C},
  booktitle={Proceedings of the Institution of Civil Engineers-Maritime Engineering},
  volume={176},
  number={2},
  pages={89--108},
  doi ={10.1680/jmaen.2022.027},
year={2023},
  organization={Thomas Telford Ltd}
}

@article{fernandez2021long,
  title={Long-term performance of distributed optical fiber sensors embedded in reinforced concrete beams under sustained deflection and cyclic loading},
  author={Fernandez, Ignasi and Berrocal, Carlos G and Rempling, Rasmus},
  journal={Sensors},
  volume={21},
  number={19},
  pages={6338},
  doi = {10.3390/s21196338},
  year={2021},
  publisher={MDPI}
}


@article{grundkotter2022precision,
  title={Precision blade deflection measurement system using wireless inertial sensor nodes},
  author={Grundk{\"o}tter, Eike and Melbert, Joachim},
  journal={Wind Energy},
  volume={25},
  number={3},
  pages={432--449},
  doi = {10.1002/we.2680},
  year={2022},
  publisher={Wiley Online Library}
}

@article{kot2021recent,
  title={Recent advancements in non-destructive testing techniques for structural health monitoring},
  author={Kot, Patryk and Muradov, Magomed and Gkantou, Michaela and Kamaris, George S and Hashim, Khalid and Yeboah, David},
  journal={Applied Sciences},
  volume={11},
  number={6},
  DOI = {10.3390/app11062750},
  pages={2750},
  year={2021},
  publisher={MDPI}
}


@article{chen2023experimental,
  title={Experimental and numerical full-field displacement and strain characterization of wind turbine blade using a {3D} Scanning Laser Doppler Vibrometer},
  author={Chen, Yuanchang and Griffith, D Todd},
  journal={Optics \& Laser Technology},
  volume={158},
  doi = {10.1016/j.optlastec.2022.108869},
  pages={108869},
  year={2023},
  publisher={Elsevier}
}

@incollection{rasmussen2022non,
  title={Non-contact deflection measurement at high speed},
  author={Rasmussen, S{\o}ren and Krarup, J{\o}rgen A and Hildebrand, Gregers},
  booktitle={Bearing Capacity Of Roads Volume 1},
  pages={53--60},
  doi = {10.1201/9781003078814-8},
  year={2022},
  publisher={CRC Press}
}

@article{rouvinen1997deflection,
  title={Deflection compensation of a flexible hydraulic manipulator utilizing neural networks},
  author={Rouvinen, Asko and Handroos, Heikki},
  journal={Mechatronics},
  volume={7},
  number={4},
  pages={355--368},
  year={1997},
doi = {10.1016/S0957-4158(97)00009-3},
  publisher={Elsevier}
}

@article{li2022multibody,
  title={Multibody system dynamic analysis and payload swing control of tower crane},
  author={Li, Kun and Liu, Manlan and Yu, Zuqing and Lan, Peng and Lu, Nianli},
  journal={Proceedings of the Institution of Mechanical Engineers, Part K: Journal of Multi-body Dynamics},
  volume={236},
  number={3},
  pages={407--421},
  year={2022},
  publisher={SAGE Publications Sage UK: London, England}
}

@article{soderena2024pathways,
  title={Pathways for CO2 regulation in NRMM},
  author={S{\"o}derena, Petri and Pihlatie, Mikko and Nylund, Nils-Olof},
  year={2024},
  journal={VTT Research Report},
  publisher={VTT Technical Research Centre of Finland}
}

@article{hagan2022non,
  title={Non-road mobile machinery emissions and regulations: A review},
  author={Hagan, Rita and Markey, Emma and Clancy, Jerry and Keating, Mark and Donnelly, Aoife and O’Connor, David J and Morrison, Liam and McGillicuddy, Eoin J},
  journal={Air},
  volume={1},
  number={1},
  pages={14--36},
  year={2022},
  publisher={MDPI}
}


@article{hashemiMultibodyDynamicsControl2023,
	title = {Multibody dynamics and control using machine learning},
	volume = {58},
	issn = {1573-272X},
	doi = {10.1007/s11044-023-09884-x},
	language = {en},
	number = {3},
	urldate = {2023-06-05},
	journal = {Multibody System Dynamics},
	author = {Hashemi, Arash and Orzechowski, Grzegorz and Mikkola, Aki and McPhee, John},
	year = {2023},
	pages = {397--431}
}


@article{gerstmayrMultibodyModelsGenerated2024,
	title = {Multibody {Models} {Generated} from {Natural} {Language}},
	volume = {62},
	issn = {1573-272X},
	doi = {10.1007/s11044-023-09962-0},
	number = {2},
	journal = {Multibody System Dynamics},
	author = {Gerstmayr, Johannes and Manzl, Peter and Pieber, Michael},
	year = {2024},
	keywords = {Automotive Engineering, Large language models, Modeling, Natural language, Text-based simulation},
	pages = {249--271}
}


@article{goRapidlyTrainedDNN2024,
	title = {A rapidly trained {DNN} model for real-time flexible multibody dynamics simulations with a fixed-time increment},
	volume = {40},
	issn = {1435-5663},
	doi = {10.1007/s00366-024-01962-8},
	language = {en},
	number = {5},
	urldate = {2024-12-02},
	journal = {Engineering with Computers},
	author = {Go, Myeong-Seok and Kim, Young-Bae and Park, Jeong-Hoon and Lim, Jae Hyuk and Kim, Jin-Gyun},
	year = {2024},
	pages = {3131--3156},
}


@article{harris2020array,
  title={Array programming with NumPy},
  author={Harris, Charles R and Millman, K Jarrod and Van Der Walt, St{\'e}fan J and Gommers, Ralf and Virtanen, Pauli and Cournapeau, David and Wieser, Eric and Taylor, Julian and Berg, Sebastian and Smith, Nathaniel J and others},
  journal={Nature},
  volume={585},
  number={7825},
  pages={357--362},
  year={2020},
  doi= {10.1038/s41586-020-2649-2},
  publisher={Nature Publishing Group UK London}
}

\appendix

\section*{Supplementary S1: Training losses with different units}
Fig.~(\ref{TrainingUnits}) demonstrates the superior training performance of single-step SLIDE-neural network models, evaluated with different number of units in the hidden layers.
\begin{figure}[h]
    \centering
    \subfigure[Training loss with $\SLIDEWindow$ units. ]{
\includegraphics[width=0.4\textwidth]{Result_5sensors_1280Samples_0kg.png}
\label{Training_td}}
\hfill
\subfigure[Training loss with $2 \times \SLIDEWindow$ units. ]{
\includegraphics[width=0.4\textwidth]{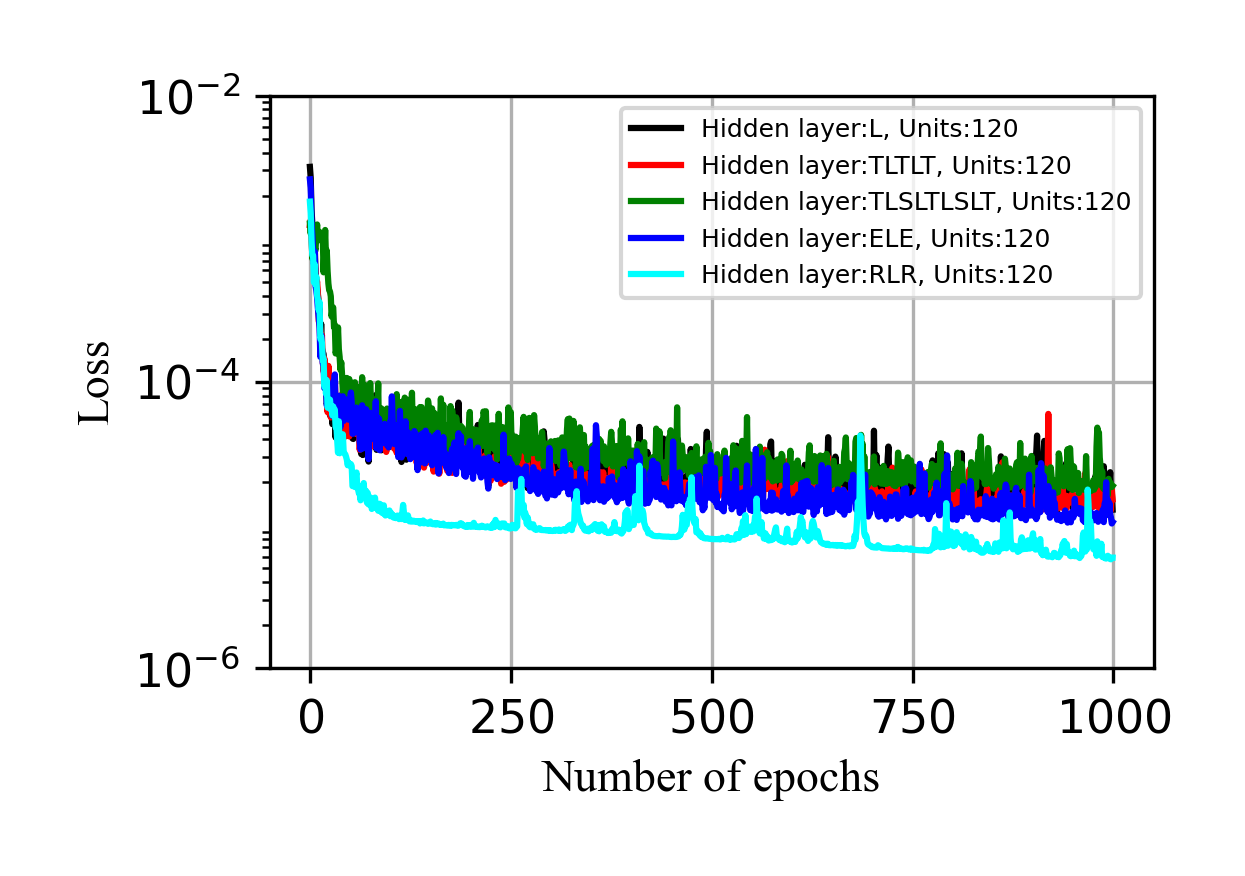}
\label{Training_2td}}
\vfill
\subfigure[Training loss with $3 \times \SLIDEWindow$ units. ]{
\includegraphics[width=0.4\textwidth]{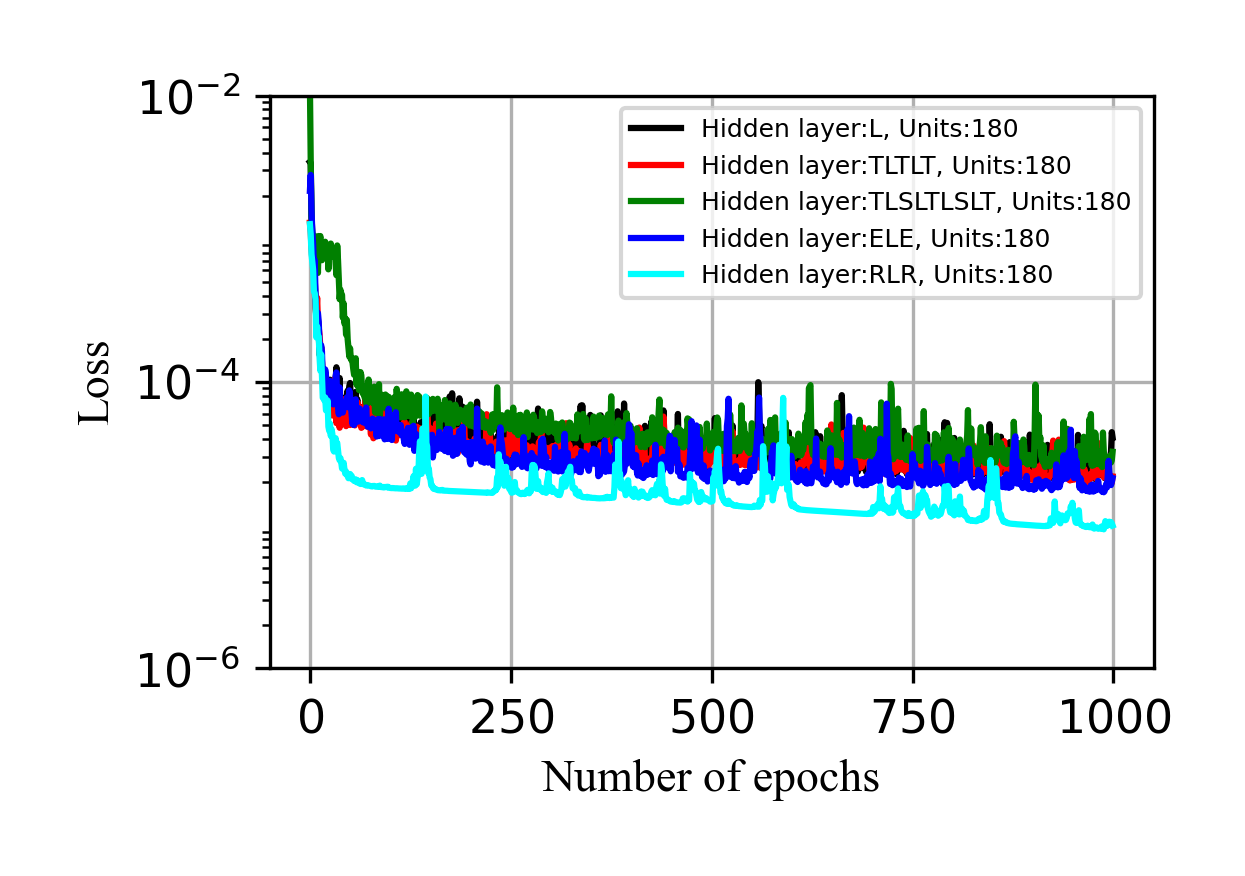}
\label{Training_3td}}
\hfill
\subfigure[Training loss with $200$ units. ]{
\includegraphics[width=0.4\textwidth]{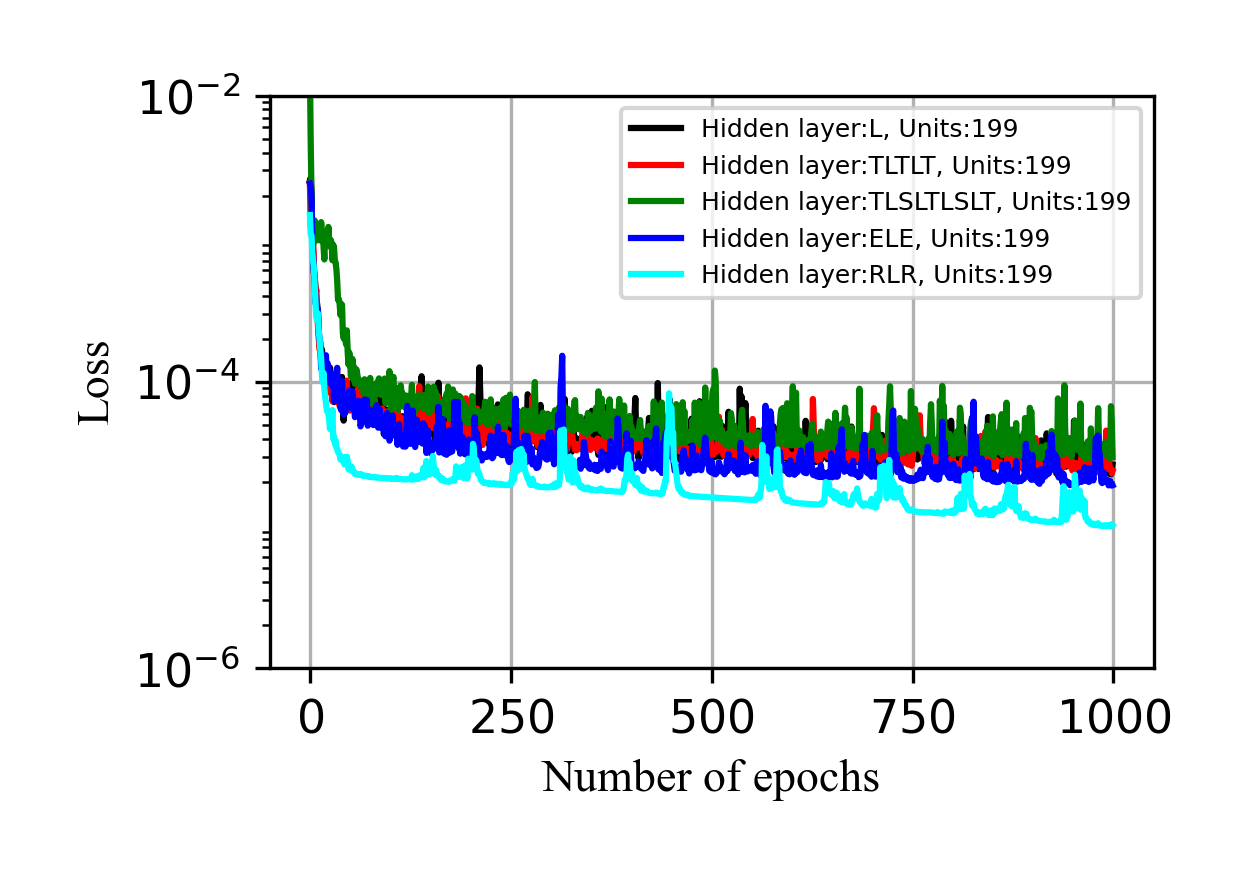}
\label{Training_200}}
\caption{\textbf{Training performance}— Comparing the mean squared error (MSE) loss of single-step SLIDE-neural network models, evaluated with different number of units in the hidden layers.}
    \label{TrainingUnits}
\end{figure}
The neural networks require more data in the supervised learning with $2 \times \SLIDEWindow$, $3 \times \SLIDEWindow$ and $200$ units in the hidden layer. As shown in Fig.~(\ref{Training_2td})--Fig.~(\ref{Training_200}), the supervised learning of neural networks demonstrate overfitting. However, the SLIDE-network arrangement in Fig.~(\ref{Training_td}) shows stable supervised learning with relatively less training data and standard PyTorch parameters.

\section*{Supplementary S2: Computing SLIDE window}
For no and 50~\text{kg} payload cases, the computing of ${\SLIDEWindow}^{*}$ is shown in Fig.~(\ref{TestDef_loads}) with the control signal, described in Sec.~(\ref{ComputeSlide}). 
\begin{figure}[h]
    \centering
    \subfigure[The SLIDE window is ${\SLIDEWindow}^{*}=0.3~\text{s}$ with no payload. ]{
        \includegraphics[width=0.482\textwidth]{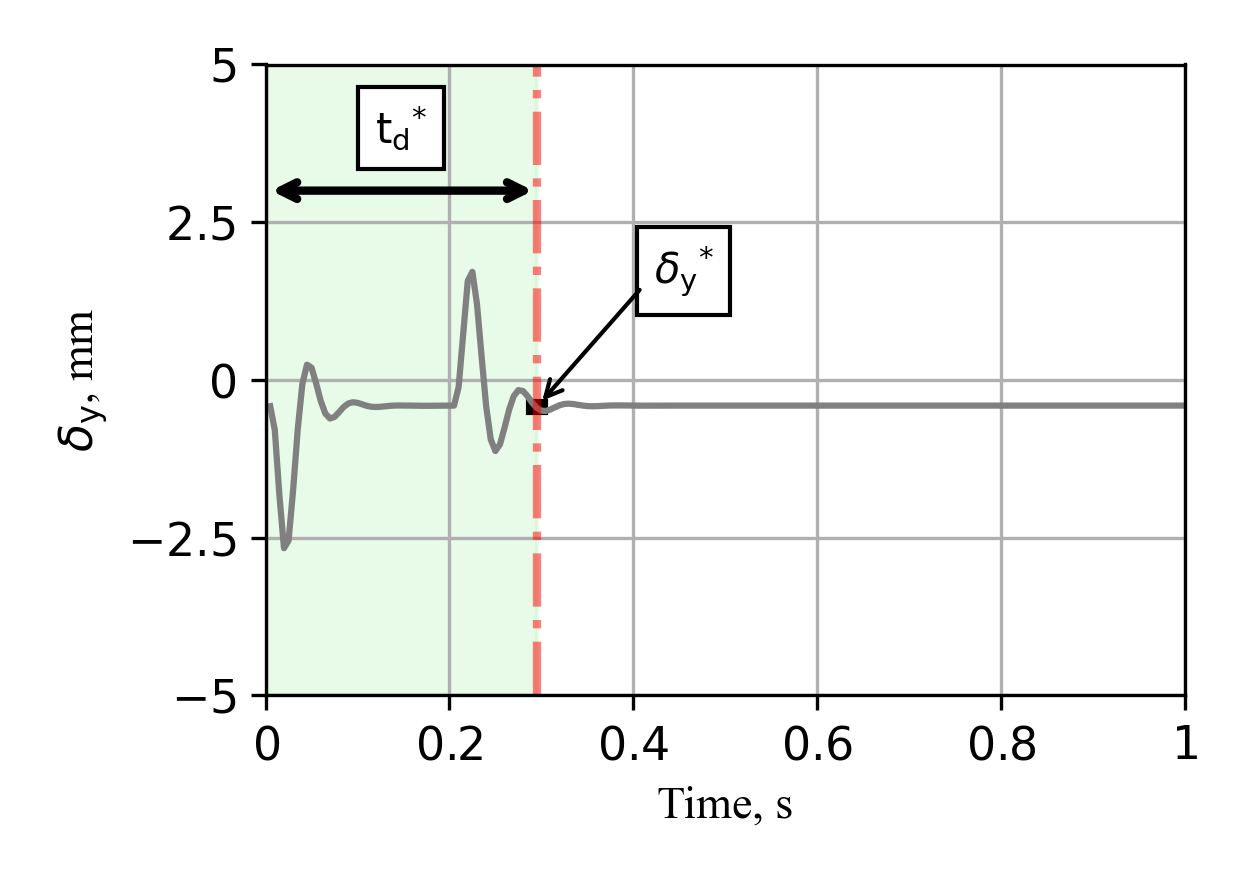}
        \label{TestDef_0kg}}
        \subfigure[The SLIDE window is ${\SLIDEWindow}^{*}=0.55~\text{s}$ with 50~\text{kg}. ]{
        \includegraphics[width=0.482\textwidth]{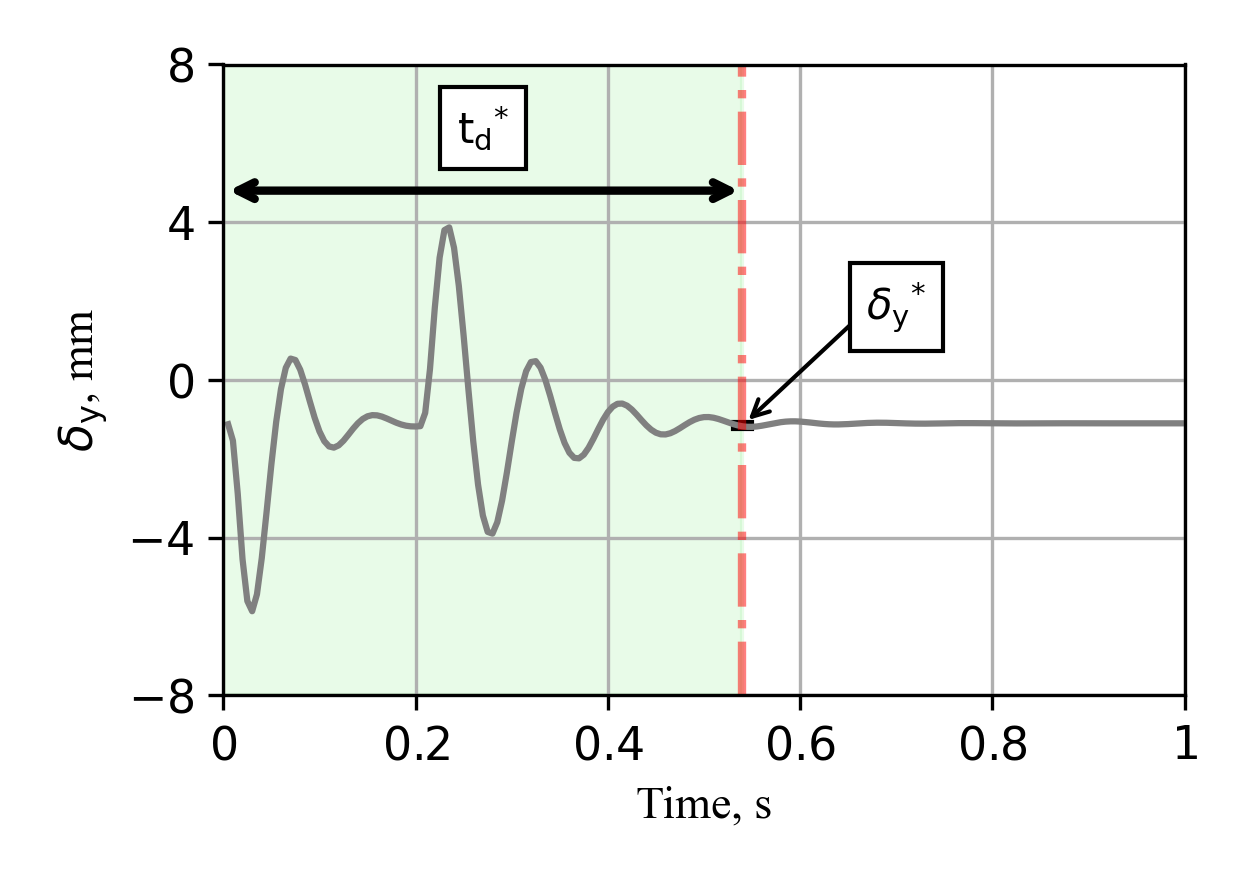}
        \label{TestDef_50kg}}
    \caption{\textbf{Computing SLIDE window}--${\SLIDEWindow}^{*}$ is computed with no payload and 50~\text{kg} payload in the flexible boom.  }
    \label{TestDef_loads}
\end{figure}
Rayleigh damping $\mathbf{D} = 3.35~\times~{10}^{-3}\text{Ns}\text{m}^{-1}$ is employed in modeling the flexible boom, which effects the steady-state behavior of structural deflection under the forced excitation.

\begin{figure}[h]
\centering
\includegraphics[width=0.5\textwidth]{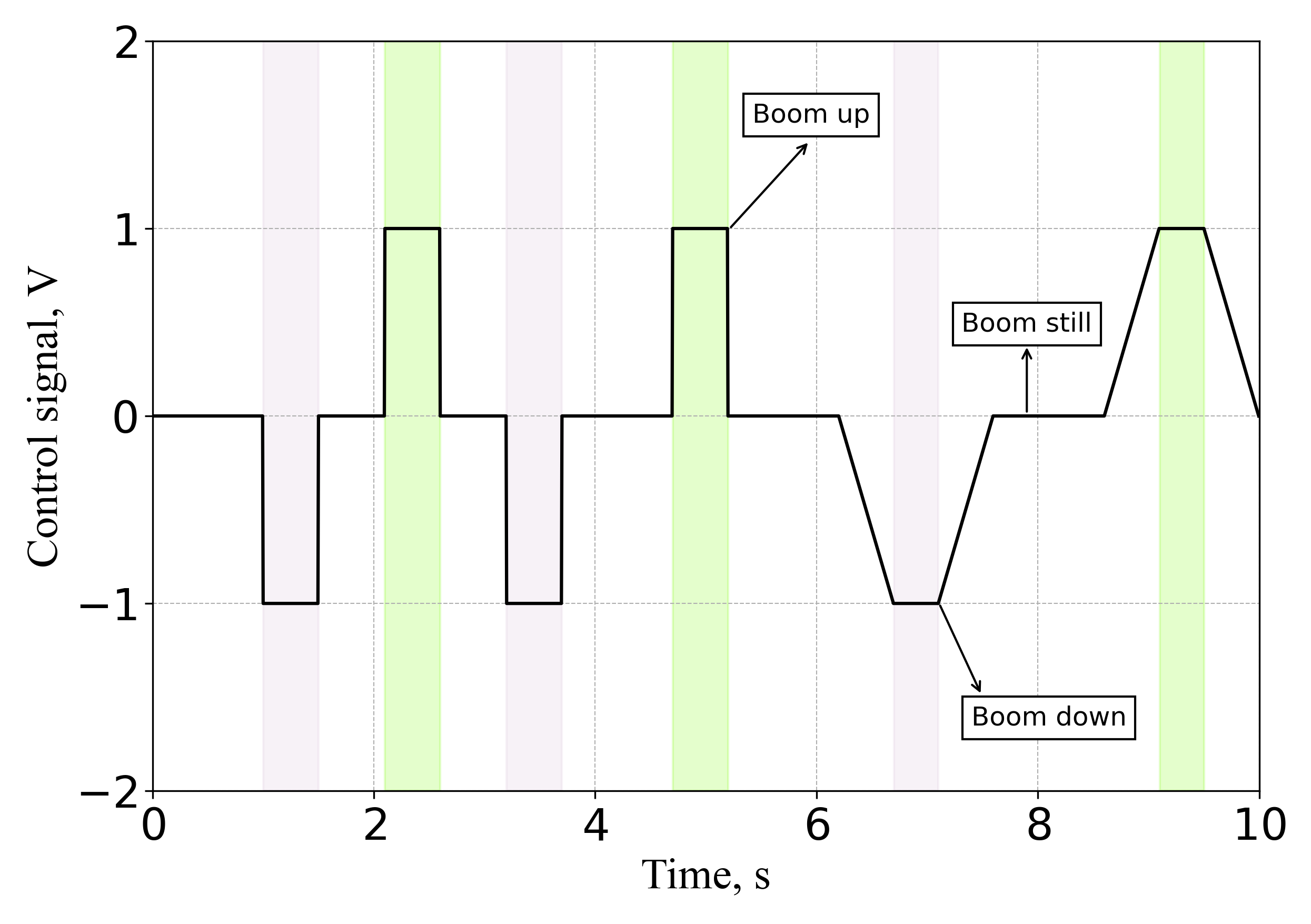}
\caption{Control valve signal in the working cycle during the evaluation phase.}
\label{EvaluationControl}
\end{figure}
\section*{Supplementary S4: valuation working cycle}
In the evaluation phase, the hydraulically actuated flexible boom is actuated with the control signal shown in Fig.~(\ref{EvaluationControl}). It ensures the movement of flexible boom up and down during the working cycle, accounting for the variation of $\delta_{\mathrm{y}}$ both with and without payloads. This control signal was not provided during the supervised learning of SLIDE-neural network.

\section*{Supplementary S3: Multi-steps estimation}
The estimations made by multi-steps SLIDE-networks are described in Fig.~(\ref{fig:MultistepEstimations}) with no payload, payload 50~\text{kg} and payload 100~\text{kg}.
For multi-steps, 10, 15 and 30 steps have been estimated forward in $\SLIDEWindow$, which requires the arrangement of $\InputLayer$ and  $\OutputLayer$ accordingly to Eq.~\ref{inputVec} and Eq.~\ref{outputVec}. Standard PyTorch parameters, described in Table~\ref{NN_Parameters} have been used in the supervised learning of networks.
\begin{figure}[h]
\centering
\includegraphics[width=0.8\textwidth]{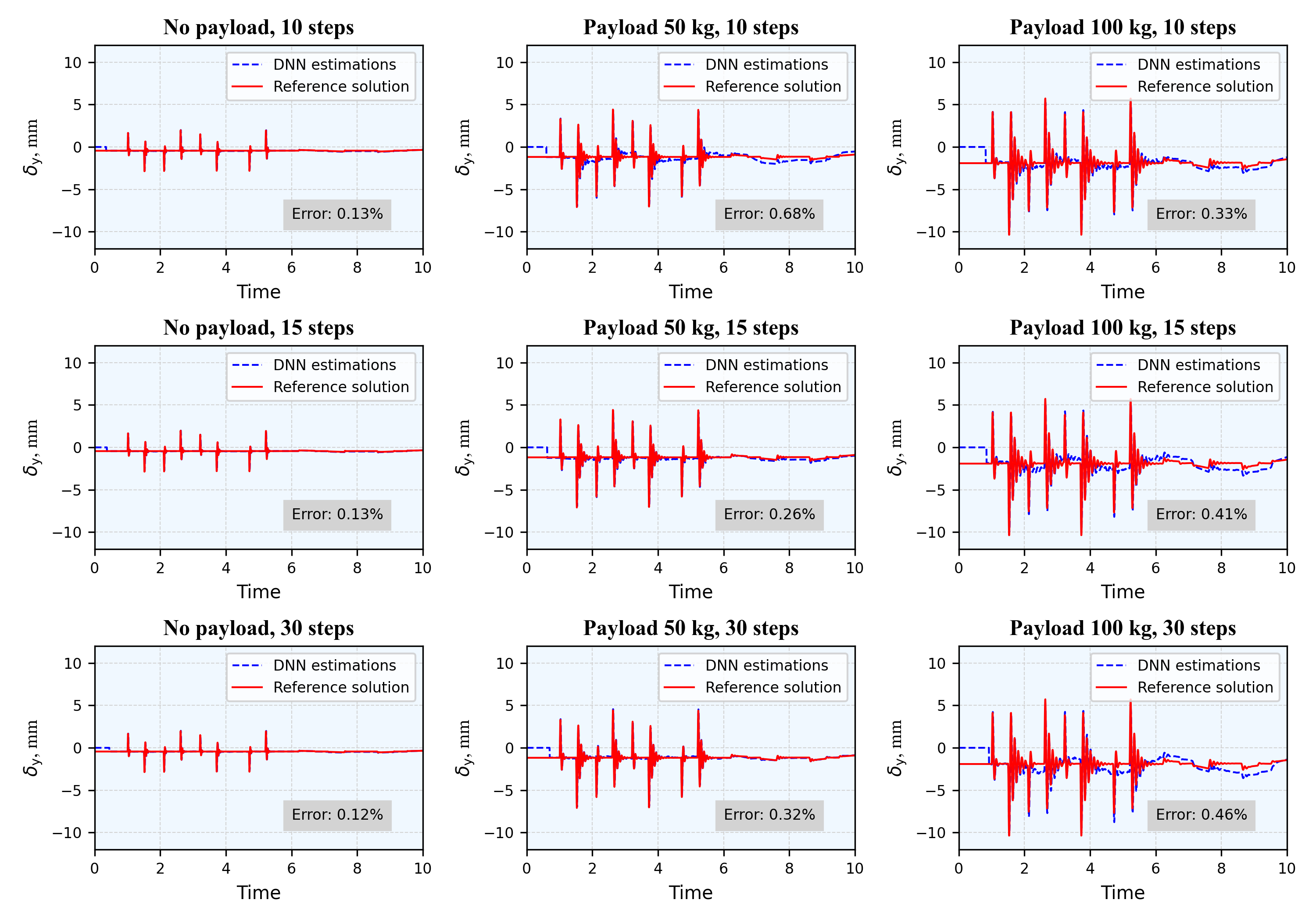}
\caption{{\textbf{Multi-steps estimation}—A summary of SLIDE-neural networks performance in multi-steps estimations of structural deflection with different payloads forward in $\SLIDEWindow$.}}
\label{fig:MultistepEstimations}
\end{figure}
In evaluation phase, SLIDE-L network is making multi-steps estimations for all cases in Fig.~(\ref{fig:MultistepEstimations}). The MAPE in estimating the structural deflection increases as the nonlinearity of the problem elevates with payloads 50~\text{kg} and 100~\text{kg}. This nonlinearity can be due to highly nonlinear dynamics and discontinuities. On other hand, no payload case demonstrates pretty stable estimations with 'L' hidden layer.

\end{document}